\documentclass[12pt]{article}
\usepackage{amssymb}
\usepackage{axodraw}
\makeatletter
\renewcommand\section{\@startsection {section}{1}{\z@}%
                                  {-3.5ex \@plus -1ex \@minus -.2ex}%nn
                                    {2.3ex \@plus.2ex}%
                                    {\normalfont\large\bfseries}}
\renewcommand\subsection{\@startsection{subsection}{2}{\z@}%
                                    {-3.25ex\@plus -1ex \@minus -.2ex}%
                                      {1.5ex \@plus .2ex}%
                                      {\normalfont\bfseries}}
\renewcommand\subsubsection{\@startsection{subsubsection}{3}{\z@}%
                                    {-3.25ex\@plus -1ex \@minus -.2ex}%
                                      {1.5ex \@plus .2ex}%
                                      {\normalfont\itshape}}
\makeatother

\newcommand{\Letter}{
     \setlength{\textwidth}{7in}
     \setlength{\textheight}{9.5in}
     \hoffset=-0.85in
     \voffset=-1in }

\Letter

\newcommand{\non}{\nonumber \\}

\newcommand{\alp}{\alpha}     \newcommand{\bet}{\beta}
\newcommand{\gam}{\gamma}     \newcommand{\del}{\delta}
\newcommand{\eps}{\epsilon}   
      \renewcommand{\th}{\theta}
 
     \newcommand{\lam}{\lambda}
   \newcommand{\sig}{\sigma}

\newcommand{\Gam}{\Gamma}     
     \newcommand{\Lam}{\Lambda}

\newcommand{\cA}{{\cal A}}    \newcommand{\cB}{{\cal B}}

    \newcommand{\cJ}{{\cal J}}
    
\newcommand{\cM}{{\cal M}}    \newcommand{\cN}{{\cal N}}

\newcommand{\ZZ}{\mathbb{Z}}
\newcommand{\CC}{\mathbb{C}}
\newcommand{\PP}{\mathbb{P}}
\newcommand{\RR}{\mathbb{R}}

\newcommand{\inv}{^{-1}}

\newcommand{\rar}{\rightarrow}

\newcommand{\one}{1\!\!1\,\,}
\newcommand{\im}{{\rm i}}

\newcommand{\bra}[1]{\langle #1|}

\newcommand{\hlf}{\frac{1}{2}}
\newcommand{\ove}[1]{\frac{1}{#1}}

\newcommand{\btau}{\bar{\tau}}
\newcommand{\brho}{\bar{\rho}}

\newcommand{\bz}{\bar{z}}

\newcommand{\mh}{\mathfrak{h}}
\newcommand{\mod}{~\mbox{mod}~}
\begin{document}
\bibliographystyle{utphys}

\thispagestyle{empty}
\begin{flushright}
\parbox[t]{2in}{CU-TP-1173\\
ITFA-2006-53\\
hep-th/0612228}
\end{flushright}

\vspace*{0.5in}

\begin{center}
{\large \bf
Crosscaps in Gepner Models and the Moduli space of $T^2$
Orientifolds
}

\vspace*{0.5in}
{Brandon Bates${}^{1}$, Charles Doran${}^{2}$,
and Koenraad Schalm${}^{1,3}$}\\[.3in]
{\em ${}^1$ Institute for Strings, Cosmology and Astroparticle
   Physics\\
Department of Physics\\
Columbia University, New York, NY 10027\\[.1in]
${}^{2,*}$ Department of Mathematics\\ University of Washington, Seattle,
WA 98195 \\[.05in]
\& Department of Mathematics\\
Columbia University, New York, NY 10027\\[.1in]
${}^{3,*}$ Institute for Theoretical Physics\\
University of Amsterdam\\
Valckenierstraat 65, Amsterdam, 1018XE
}
\end{center}

\vspace*{0.5in}

\begin{center}
{\bf
Abstract}
\end{center}
We study $T^2$ orientifolds and their moduli space in detail. 
Geometrical insight into the
involutive automorphisms of $T^2$ allows a straightforward derivation
of the moduli space of orientifolded $T^2$'s. Using $c=3$ Gepner models, we
compare the explicit worldsheet sigma model of an
orientifolded $T^2$ compactification with the CFT results. 
In doing so, we derive half-supersymmetry preserving
crosscap coefficients for generic unoriented Gepner models using simple
current techniques to construct the charges and tensions of Calabi-Yau
orientifold planes. For $T^2$s we are able to identify the O-plane
charge directly as the number of
fixed points of the involution; this number plays an important role
throughout our analysis. At several points we make connections with the
mathematical literature on real elliptic curves. We conclude with a preliminary extension of these results to
elliptically fibered K3s.

\vfill

\hrulefill\hspace*{4in}

{\footnotesize
Email addresses: \parbox[t]{5.5in}{bdbates@phys.columbia.edu,
   doran@math.washington.edu, schalm@science.uva.nl}

${}^*$ Current address.}

\newpage

\section{Introduction}

Orientifolds are mysterious beasts, whose taxonomic classification is
starting to come of age. Orientifolds are an important
aspect in the study of string theory vacua. As we have become recently
aware, the majority of (supersymmetric)
vacua are probably of the orientifold type \cite{Douglas:2003um}.
Generically orientifold compactifications will contain orientifold
planes: a non-perturbative `object' in string theory which
characteristically can carry negative tension and charge.
All compactifications can in fact be organized in
superselection sectors determined by their orientifold plane content.

Of phenomenological interest are those string compactifications with
$\cN=1$ $d=4$ supersymmetry. Next to heterotic Calabi-Yau
compactifications and M-theory on $G_2$ manifolds, type II Calabi-Yau
compactifications with D-branes (type I Calabi-Yau
compactifications)
provide a new class.  This class has shown promising signs of
not only being able to describe cosmological string theories with a
period of
slow-roll inflation,
\cite{Kachru:2003sx,Kachru:2003aw,Blanco-Pillado:2004ns,Burgess:2004kv},
but also contain models with SM-model-like
spectra \cite{Dijkstra:2004ym}. With the gauge sector descending from
D-branes,
these compactifications are `brane-world' scenarios. Type I
compactifications must obey a consistency condition --- tadpole
cancellation --- which is Gauss's law that the total charge in the
internal space vanishes, and supersymmetry dictates that the only
compatible `negative' charge object is the orientifold plane. Hence
all supersymmetric type II brane compactifications must be orientifolds.
For the study of string vacua it is of
great interest to know what the possible orientifold planes are for
a Calabi-Yau manifold and their charges with respect to internal
gauge-fields.

The mystery of orientifold planes lies in their properties. 
They can behave
as negative tension objects at long range, yet do not violate any
GR energy theorems \cite{Johnson:1999qt,Marolf:2002np}.
Perturbatively they have no
moduli, which leads to the question whether orientifolds are objects at all.
Non-perturbatively they are thought to be a
either a (non-dynamical) condensation of  D-branes
\cite{Sen:1997gv,Sen:1997kw,Sen:1996xx} or resolve into
a smooth geometry \cite{Seiberg:1996nz}. Particularly these last
results suggest that
orientifold planes are perhaps remnants of purely quantum-geometrical
characteristics with no true classical analogue, rather in the same
way that fermions are intrinsically quantum objects.

Indeed the main obstacle in our understanding of orientifold planes is
that they lack an intrinsic classical geometric description. D-branes
are the quantum-geometric version of
vector-bundles. Orientifolds, however, as we now know them, are
intrinsically defined in perturbative string theory by modding out by
a worldsheet parity transformation (times a spacetime
involution). This projection will generically also remove a number of
geometric
moduli. In particular, toroidal orientifold compactifications are known
where
the true large volume limit is absent \cite{Polchinski:1996ry}.
Consistent with the suggestion
above, a classical geometric version of such a
compactification does not seem to
exist. For others the large volume limit does exist.
This study partly seeks whether some characteristics of
orientifold planes may be geometrically determined, especially
those characteristics which are relevant to the construction of phenomenologically
viable string vacua, namely the location of the planes, their charges and
tensions, the
effect and relation to the moduli space of the oriented parent theory;
and how
O-planes behave throughout the moduli space.
Because orientifold
compactifications are intrinsically defined at the
worldsheet level, we will use worldsheet CFT methods to try to extract this information. The systematics of
building consistent unoriented CFTs and specifically rational CFTs
(RCFTs) are
by now well understood \cite{thesis}. To solve the associated
algebraic constraints in practice is computationally involved and may
generically only be possible mechanically
\cite{Dijkstra:2004ym,Dijkstra:2004cc,Blumenhagen:2005mu}.
  A search for a geometrical
understanding is motivated by the expectation that geometrical insight
will allow analytic insight into consistent orientifold
compactifications.

The Gepner construction is
the most well-known example of an RCFT description of a Calabi-Yau
compactification at a special point in its moduli space.
In principle the construction of
unoriented Gepner models has been known for some time (see
e.g. \cite{Blumenhagen:1998tj});
the emphasis on the geometrical aspects is more
recent
\cite{Acharya:2002ag,Blumenhagen:2002wn,Diaconescu:2003dq,Misra:2003xx,Misra:2003zv,Misra:2004yp,Govindarajan:2003vv,Govindarajan:2003vp,Brunner:2003zm,Brunner:2004zd,Brunner:2002em,Blumenhagen:2003su,Blumenhagen:2004cg,Aldazabal:2003ub}
In particular, the two articles by Brunner and  Hori, and by Brunner,
Hori, Ha and Walcher,
1) show that O-planes are located at fixed
points of (anti-)holomorphic isometries, 2) argue that the moduli
space is unobstructed and computed the local dimension, and 3) in
specific examples study the geometry and moduli-space of consistent CY-3fold
orientifolds. In these examples the charges and
tensions of the O-planes are known and canceled by the
addition of appropriate D-branes.  
We study here the simplest non-trivial
example in detail: a CY-1fold or 2-torus compactification. The
advantage is that throughout the torus moduli-space we have
an additional description, aside from the abstract Gepner
description and the gauged linear sigma model description. The
worldsheet sigma model covers the
full moduli space rather than a subspace.  
Furthermore, our approach will differ from the methods of Brunner
et al in
one subtle detail. Their orbifold CFT results apply to any
Gepner model, but they are not extendible to generic RCFTs.
In general one
ought to use simple current techniques (see e.g. \cite{Fuchs:2000gv,walcher}. We do so
here. Simple current techniques for Gepner orientifolds were
recently used in \cite{Blumenhagen:2004qu} and underlie the search for
Standard-Model like brane-world compactifications in
\cite{Dijkstra:2004ym,Dijkstra:2004cc}.
We review both the Gepner construction and the
unoriented simple current extended RCFTs in section 3; details are
provided in appendix A. 
In section 4 we will use these results to
determine the O-plane properties of $c=3$ Gepner models
corresponding to
2-torus compactifications, 
Finally in section 5 we compare the results with the
exact worldsheet sigma model description. 
We conclude with a
discussion and outlook on how our torus results may provide insight in
elliptically fibered CYs and K3s in particular. In terms of the general
aim of this study we will succeed in giving a geometrical description
of the charge/tension of
A-type O-planes as the number of fixed points of the associated
involution. 
This is a direct extension of the results by
  Brunner and Hori. This number of fixed points plays a guiding role
  throughout. Classically it is an invariant. Indeed in section 2
We begin with a simple geometrical
analysis of orientifolds of $T^2$s by analyzing for which points in
the moduli space involutions exist. We will then show how in the moduli space of unoriented
$T^2$s this number of fixed points is a topological obstruction. We
conclude in section 6 with an outlook how the results for $T^2$
orientifolds may be applied to elliptically fibered families and K3s
in particular. Curiously we will show here that the number of fixed
points of the $T^2$ fiber can change in the family. In such fibered
surfaces, it is therefore no longer an invariant. 

A final result of this article is to connect various mathematical
results on real elliptic curves
\cite{Alling-Greenleaf,Alling,Duval,Huisman:2001} with the physics of orientifold compactifications.

\section{The geometry of $T^2$ Orientifolds}
\label{sec:geom-t2-orient}
\setcounter{equation}{0}

Geometrically the most pronounced characteristic of orientifolds is their localization
at fixed points of involutive spacetime automorphisms. Given a
manifold
and its involutive automorphisms, we can decide to restrict our attention to
physical states invariant under the action of the involutive automorphisms. If
we combine the action of the involutive automorphism with a
worldsheet orientation reversal,
the fixed point locus of the automorphism is an
orientifold plane.
String theory, however, tells us that
these orientifold planes carry both tension and charges under various
$p$-form potentials on the manifold. To understand these charges from
a purely geometric construction is one of the main aims of the recent
work on orientifold constructions. A proper
understanding will facilitate the construction of consistent orientifold
compactifications in particular on Calabi-Yau manifolds (see for example,
\cite{Dijkstra:2004ym,Dijkstra:2004cc}).

To illustrate the power of geometrical insight, we
will deduce in this section properties of $T^2$
orientifolds purely from geometry. The remainder of this article will be
devoted to a review and construction of $T^2$ orientifolds from 
first principles. They will naturally confirm the
geometrical results. We will clearly see the benefits and shortcomings
of both methods. The geometrical approach yields us the full moduli
space of $T^2$ orientifolds with little effort. However, the charges
and tensions of the orientifold planes require some work. Fortunately
for $T^2$s it is straightforward and we will show that the O-plane
charge is simply related to the number of fixed points of the
involution. 
The
string theory approach yields the converse. The charges are explicitly
computable but the moduli space is less clear. By an argument of
Kapranov and Oh 
\cite{Brunner:2003zm} (page 112) the moduli-space of orientifolds
is unobstructed, so than a perturbative analysis in string theory
should suffice to find the moduli space as well. For $T^2$s this does
not seem to be the case, as we will show below. In section \ref{sec:disc-outl-towards} we do
find evidence that this obstruction is
lifted when the $T^2$ is non-trivially fibered over some base.

\subsection{$T^2$ Orientifolds}

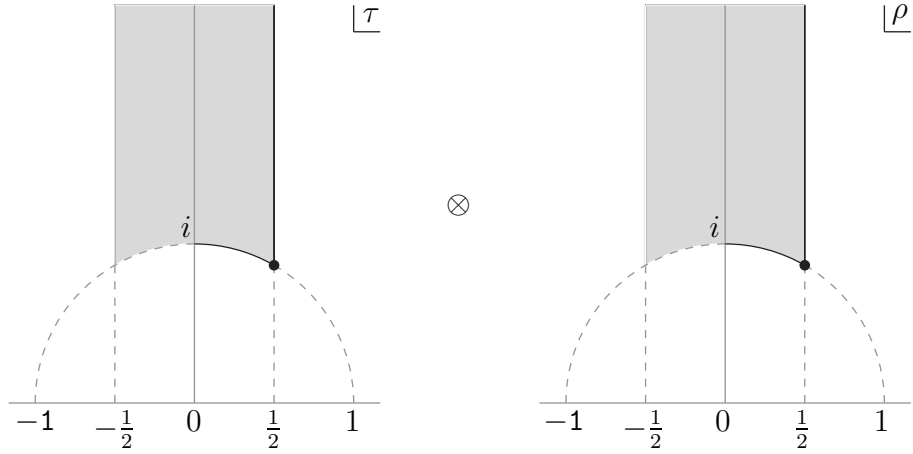
\begin{figure}[tbp]
   \begin{center}
     \begin{picture}(300,150)(0,0)
       % Filling
       \GBox(20,0)(80,150){0.85}
       \GBox(220,0)(280,150){0.85}
       \SetColor{White}
       \Line(20,150)(80,150)
       \Line(20,51.96)(20,150)
       \BCirc(50,0){60}
       \Line(220,150)(280,150)
       \Line(220,51.96)(220,150)
       \BCirc(250,0){60}
       \SetColor{Black}
       % Axes
       \SetColor{Gray}
       \Line(-20,0)(120,0)
       \Line(180,0)(320,0)
       \Line(50,0)(50,150)
       \Line(250,0)(250,150)
       \SetColor{Black}
       % Fundamental Domain
       %   Vertical boundaries
       %\Line(20,51.96)(20,150)
       \Line(80,51.96)(80,150)
       %\Line(220,51.96)(220,150)
       \Line(280,51.96)(280,150)
       %  Unit Circle
       \CArc(50,0)(60,60,90)
       \CArc(250,0)(60,60,90)
      %  Background
       \SetColor{Gray}
       \DashCArc(50,0)(60,0,60){3}
       \DashCArc(50,0)(60,90,180){3}
       \DashLine(20,0)(20,51.96){3}
       \DashLine(80,0)(80,51.96){3}
       \DashCArc(250,0)(60,0,60){3}
       \DashCArc(250,0)(60,90,180){3}
       \DashLine(220,0)(220,51.96){3}
       \DashLine(280,0)(280,51.96){3}
       \SetColor{Black}
       %  Vertices
%      \BCirc(20,51.96){2}
       \Vertex(80,51.96){2}
%      \BCirc(220,51.96){2}
       \Vertex(280,51.96){2}
        % Text
       \Text(150,75)[]{$\otimes$}
       \Text(50,-3)[t]{0}
       \Text(110,-3)[t]{$1$}
       \Text(80,-3)[t]{{$\hlf$}}
       \Text(49,63)[br]{$i$}
       \Text(20,-3)[t]{$-\hlf$}
       \Text(-10,-3)[t]{$-${\tt 1}}
       \Text(250,-3)[t]{0}
       \Text(310,-3)[t]{$1$}
       \Text(280,-3)[t]{{$\hlf$}}
       \Text(249,63)[br]{$i$}
       \Text(220,-3)[t]{$-\hlf$}
       \Text(190,-3)[t]{$-${\tt 1}}
       % Legend
       \Line(110,140)(110,150)
       \Line(110,140)(120,140)
       \Text(117,147)[]{$\tau$}
       \Line(310,140)(310,150)
       \Line(310,140)(320,140)
       \Text(317,147)[]{$\rho$}
     \end{picture}
%}
   \end{center}
   \caption{{\em The moduli space of $T^2$ compactifications: 
The
       boundary of the fundamental domain is included when bold.}  }
   \label{fig:1}
\end{figure}

The complete geometrical data of a $T^2$ compactification is given by
two complex numbers $\tau$ and $\rho$. Consider the torus as
$\RR^2$ modded out by a lattice $\Lambda$ generated by two vectors
$e_1,e_2$. Scaling and rotating $e_1$ to the unit vector $(1,0)$ and reflecting
across the horizontal axis if necessary, the
complex number $\tau$, $\mbox{Im } \tau > 0$ is the lattice
vector $e_2$ in $\CC \simeq \RR^2$.
The natural lattice structure implies that
\begin{equation}
\label{eq:7}
\tau \rar \frac{a\tau+b}{c\tau+d}~,~~
\left(\matrix{a &b \cr c &d}\right) \in PSL(2,\ZZ)~,
\end{equation}
describes the same torus.
Geometrically $\tau$ classifies the complex structure of
the torus. The other complex number $\rho$ is the complexified K\"{a}hler
class: $\rho = \int_{T_2} B+\im J$. Here $J \sim \sqrt{G}dx \wedge dy$
is the K\"{a}hler form
parametrizing the size of the torus, and $B$ is the NS-NS
two-form.  T-duality combined
   with the gauge symmetry $B \rar B+1$ implies that
\begin{equation}
\label{eq:8}
  \rho \rar \frac{a\rho+b}{c\rho+d}~,~~
\left(\matrix{a &b \cr c &d}\right) \in PSL(2,\ZZ)~,
\end{equation}
also describes the same torus. The quantum moduli space of an elliptic
curve is thus
given by two copies of the fundamental domain of $PSL(2,\ZZ)$ (see Fig.
\ref{fig:1}) subject to three global $\ZZ_2$ symmetries: mirror
symmetry
which exchanges $\tau$ with $\rho$: $(\tau,\rho) \rar (\rho,\tau)$,
spacetime parity which sends $(\tau,\rho)$ to minus their complex
conjugates: $(\tau,\rho) \rar (-\btau,-\brho)$, and worldsheet parity
which sends $\rho$ to minus its complex conjugate: $(\tau,\rho)
\rar (\tau,-\bar{\rho})$ \cite{Gukov:2002nw}.

To consider orientifold compactifications we need to identify the
involutive automorphisms of a $T^2$. These
are easily derived and can also be found in the mathematical literature
  \cite{Alling-Greenleaf,Alling}. Any automorphism of $T^2$ will, up to
a lattice
vector,
also be an automorphism of the covering space $\RR^2$.
Using complex coordinates $z=x+\im y$, these divide into
holomorphic, $z \rar \gam z+\del$, and anti-holomorphic,
  $z \rar \alp\bar{z} + \beta $, automorphisms. 
The $\RR^2$ automorphisms are involutions of the torus
$T^2=\RR^2/\Lam$, iff (1)
acting once, they return the same torus (i.e. map the lattice vectors to
other lattice
vectors) and (2) acting twice they return the same point on the
torus. The latter requirement tells us that for
$\gam=1,~2\del=n\tau+m$,
or $\gam=-1,~\del \in \CC$ we obtain holomorphic involutions, and for
$|\alp|^2 =1,~\alp\bar{\beta}+\beta =n\tau+m$ we have antiholomorphic
involutions. Without loss of generality, the shifts $\beta \in
\CC/\Lam$ can be limited to those within the torus; the unit norm
constraint, $|\alp|^2=1$, then implies that $\beta$ is half a lattice
vector: $\beta = (n\tau+m)/2$ with $n,
m \in \{0,1 \}$, if $|\beta|^2 = |n\tau+m-\beta|^2$ is to have a
solution. For reasons we will explain in
section 3 the latter, i.e.
anti-holomorphic involutions, are known as
type A and holomorphic ones as type B. Moreover, mirror symmetry maps
type A involutions into type B and vice versa.

For anti-holomorphic type A involutions, the first requirement is
equivalent to demanding that
\begin{eqnarray}
   \label{eq:4}
     \alp \cdot 1  = a\tau+b +\beta~,~~ a,b \in \ZZ, \\
\label{eq:4b}
     \alp \cdot \bar{\tau}   = c\tau+d +\beta~,~~ c,d \in \ZZ.
\end{eqnarray}
 Equation (\ref{eq:4}) states
   that $\alp$ is a lattice vector. Suppose $|\tau|^2 >1$. The condition
that $\alp$ has norm
   one, then uniquely fixes $\alp=\pm 1$; all other lattice vectors
   have norm larger than one. Furthermore, from
   equation (\ref{eq:4b}) we conclude that
$\bar{\tau}$ is then also
a lattice vector. This implies that $\tau+\bar{\tau} \in \ZZ$. For
   $\tau$ in the fundamental domain, FD, this has the  solutions
Re$(\tau)=0,\hlf$. To
derive the explicit involution, there are five natural
   cases to consider
   \begin{itemize}
   \item[(a)] $\tau = \im\tau_2$ with $\tau_2>1$ (and $\tau_2 \in \RR$).
Here $\alp =\pm 1$. {For both values of $\alp$ all shifts
   $\beta=0,\hlf,\frac{\im\tau_2}{2},\frac{\im\tau_2+1}{2}$ are
   compatible with the involution constraint $\alp\bar{\beta}+\beta=
   c\tau+d$.}
\item[(b)] $\tau=\im$. The constraint (\ref{eq:4}) that $\alp$ be a
   lattice vector has the solutions $\alp =\pm 1$ and $\alp=\pm \im$. The
   second constraint (\ref{eq:4b}) is always satisfied. {Again all
   shifts $\beta$ are allowed for $\alp=\pm 1$. For $\alp=\pm \im$ only
   the shifts $\beta = 0, \frac{\im+1}{2}$ are allowed.}
\item[(c)] $\tau=\exp (\im\theta)$ with $\pi/2>\theta>\pi/3$. The
constraint (\ref{eq:4}) that
   $\alp$ be a unit norm
lattice vector has solutions $\alp=\pm 1$ and $\alp =\pm
   \tau$. Now, however, the second constraint (\ref{eq:4b}) only allows
   the solution $\alp = \pm \tau$. {As in case (b) this allows the
   shifts $\beta =0, \frac{\tau+1}{2}$.}
\item[(d)] $\tau=\exp(\im\pi/3)$. The constraint (\ref{eq:4}) that
   $\alp$ be a unit norm lattice vector has the solutions $\alp=\pm 1$,
   $\alp = \pm \exp(\im\pi/3)$ and $\alp =\pm \exp(2\im\pi/3)$. All obey
   the second constraint (\ref{eq:4b}). {The allowed shifts
   are: $\beta =0,\hlf$ for $\alp =\pm 1$, $\beta =0,\frac{\tau+1}{2}$
   for $\alp=\pm \tau$, and $\beta =0,\frac{\tau}{2}$ for $\alp=\pm
(\tau-1)$.}
\item[(e)] $\tau= \hlf +\im\tau_2$ with $\tau_2 > \hlf\sqrt{3}$. Here
again $\alp=\pm 1$. {The only
   shifts compatible with the involution constraint are $\beta=0,\hlf$.}
\end{itemize}

Not all these involutions are independent. Involutions
$\sig(z)=\alp\bz+\beta$ which are
conjugate to each other by an automorphism of the covering space,
$g(z)\sig(z)g\inv(z)\simeq \sig'(z)$ with $g(z) =az+b$, $a\in \CC^*$ and $b\in \CC$, have
{\em identical} action on the torus. For such an automorphism $g(z)$ to take the period
parallelogram associated with $\Lambda$ into another, it must be area-preserving.
Thus $b$ is arbitrary, while $a$ must be  a square root of 1 in cases (a), (c), and (e),
a fourth root of 1 in case (b), and a sixth root of 1 in case (d).  Explicitly, we have therefore
\begin{equation}
\sigma'(z) = \alpha a^2 \overline{z} + a \beta + (b - \alpha a^2
\overline{b}) \ ,
\end{equation}
since $a/\overline{a} = a^2$.  Note also that if $\alpha a^2 = 1$ then
$b - \alpha a^2 \overline{b} = 2 i \mbox{Im}(b)$, whereas if $\alpha a^2 = -1$ then it
equals $2 \mbox{Re}(b)$.  A detailed analysis \cite{Alling} (particularly section 12.33)
of each of the cases (a)-(e) above using these observations results in a complete classification of the
inequivalent antiholomorphic involutions on a torus (see Table
\ref{table:involutions}). To illustrate the method, let us check case (a).
If $\alpha = 1$, then $\alpha a^2 = 1$, so $b - \alpha a^2 \overline{b} = 2 \im \mbox{Im}(b)$.
We can thus choose $\beta$ to be real, and since it lies in the period parallelogram, to lie
in the interval $[0,1)$.  But $\overline{\beta} + \beta \in \Lambda$,
so $\beta$ is either $0$ or $1/2$.  Similarly, if $\alpha = -1$, then $\alpha a^2 = -1$,
so $b - \alpha a2 \overline{b} = 2 \mbox{Re}(b)$.  We can thus choose $\beta$ to be purely imaginary
in the interval $[0,\im \tau_2)$. But $\overline{\beta} - \beta \in \Lambda$,
so $\beta$ is either $0$ or $\im \tau_2 / 2$. It is noteworthy that this same
classification was also found by Du Val \cite{Duval} by analyzing the
properties of elliptic functions on the lattices defined by these
tori. 

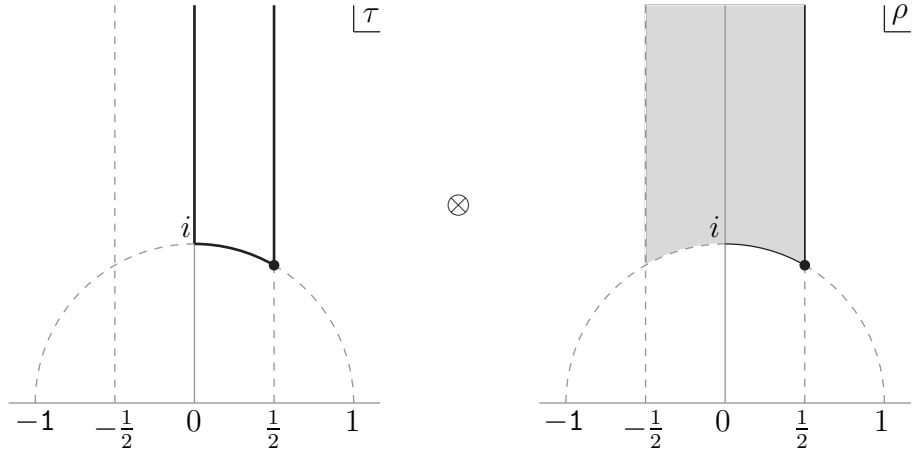
\begin{figure}[tbp]
   \begin{center}
     \begin{picture}(300,150)(0,0)
       % Filling
%      \GBox(20,0)(80,150){0.99}
       \GBox(220,0)(280,150){0.85}
       \SetColor{White}
       \Line(20,150)(80,150)
       \Line(20,51.96)(20,150)
       \BCirc(50,0){60}
       \Line(220,150)(280,150)
       \Line(220,51.96)(220,150)
       \BCirc(250,0){60}
       \SetColor{Black}
       % Axes
       \SetColor{Gray}
       \Line(-20,0)(120,0)
       \Line(180,0)(320,0)
       \Line(50,0)(50,150)
       \Line(250,0)(250,150)
       \SetColor{Black}
       % Fundamental Domain
       %   Vertical boundaries
       \SetWidth{1}
       %\Line(20,51.96)(20,150)
       \Line(80,51.96)(80,150)
       \Line(50,60)(50,150)
       %\Line(220,51.96)(220,150)
       \SetWidth{0.5}
       \Line(280,51.96)(280,150)
       %  Unit Circle
       \SetWidth{1}
       \CArc(50,0)(60,60,90)
       \SetWidth{0.5}
       \CArc(250,0)(60,60,90)
      %  Background
       \SetColor{Gray}
       \DashCArc(50,0)(60,0,60){3}
       \DashCArc(50,0)(60,90,180){3}
       \DashLine(20,0)(20,150){3}
       \DashLine(80,0)(80,51.96){3}
       \DashCArc(250,0)(60,0,60){3}
       \DashCArc(250,0)(60,90,180){3}
       \DashLine(220,0)(220,150){3}
       \DashLine(280,0)(280,51.96){3}
       \SetColor{Black}
       %  Vertices
%      \BCirc(20,51.96){2}
       \Vertex(80,51.96){2}
%      \BCirc(220,51.96){2}
       \Vertex(280,51.96){2}
        % Text
       \Text(150,75)[]{$\otimes$}
       \Text(50,-3)[t]{0}
       \Text(110,-3)[t]{$1$}
       \Text(80,-3)[t]{{$\hlf$}}
       \Text(49,63)[br]{$i$}
       \Text(20,-3)[t]{$-\hlf$}
       \Text(-10,-3)[t]{$-${\tt 1}}
       \Text(250,-3)[t]{0}
       \Text(310,-3)[t]{$1$}
       \Text(280,-3)[t]{{$\hlf$}}
       \Text(249,63)[br]{$i$}
       \Text(220,-3)[t]{$-\hlf$}
       \Text(190,-3)[t]{$-${\tt 1}}
       % Legend
       \Line(110,140)(110,150)
       \Line(110,140)(120,140)
       \Text(117,147)[]{$\tau$}
       \Line(310,140)(310,150)
       \Line(310,140)(320,140)
       \Text(317,147)[]{$\rho$}
     \end{picture}
%}
   \end{center}
   \caption{ {\em The ``moduli space'' for type A $T^2$
       orientifolds. In Figure \ref{fig:2a} we will show that the
       space is not connected but there exists a topological
       obstruction at $\tau =i$. Type B
       orientifolds are related by mirror symmetry whereby
the roles of $\tau$ and $\rho$ are interchanged.}}
   \label{fig:modulispace}
\end{figure}

These results identify the subclass of $T^2$ complex structures that
permit the existence of an anti-holomorphic involution.\footnote{This
  extends the results of \cite{Gukov:2002nw}.} This set is 
shown in Figure \ref{fig:modulispace}. We will see momentarily that
this $T^2$ orientifold ``moduli space''  consists of two disconnected
components. 
\begin{table}[ht]
\begin{center}
\begin{tabular}{||c|c|c|c|c|c|c||}
\hline \rule{0pt}{5mm} Case & $\tau$ & $J(\tau)$ & $\alp$
& $\beta$ & $s$ & Fixed pts \\[3pt]
\hline \hline \rule{0pt}{5mm}
(a) & $\im \tau_2 $ \ with $\tau_2>1$
& $J  >1$
& 1
& $0$ & $2$ & Im$(z)=0$; Im$(z)=\tau_2/2$ \\[3pt]
\cline{4-7} \rule{0pt}{5mm}  & &  & $-1$
& $0$ & $2$ & Re$(z)=0$; Re$(z)$=1/2 \\[3pt]
\cline{4-7} \rule{0pt}{5mm}  &  & & $1$ & $1/2$ & $0$ & \\[3pt]
\cline{4-7} \rule{0pt}{5mm}
& & & $-1$ & $\tau/2$ & $0$ & \\[3pt]
\hline \rule{0pt}{5mm} (b) & $\im$ & $1$ & $1 \sim -1$ & $0$ & $2$
& Im$(z)=0$; Im$(z)=1/2$ \\[3pt]
\cline{4-7} \rule{0pt}{5mm}
  & & & $\im \sim -\im$ & $0$ & $1$ & $z=re^{\im\pi/4}, r \in \RR$
\\[3pt]
  \cline{4-7} \rule{0pt}{5mm}
  & & & $1 \sim -1$ & $1/2$ & $0$ &\\[3pt]
  \hline \rule{0pt}{5mm}
  (c) & $e^{\im \theta}$ \ with $\pi/3 < \theta < \pi/2$ &  $(0,1)$ &
$\tau$
  & $0$ & $1$ & $z=re^{\im\theta/2}, r
\in \RR$ \\[3pt]
  \cline{4-7} \rule{0pt}{5mm}
  & & & $-\tau$ & $0$ & $1$ & $z=\im re^{\im\theta/2}, r
\in \RR$\\[3pt]
   \hline \rule{0pt}{5mm}
   (d) & $e^{\im \pi/3}$ & $0$ & $1 \sim e^{ \im 2\pi/3} \sim
   e^{ \im 4\pi/3}$ & $0$ & $1$ & Im$(z)=0,\sqrt{3}/4$\\[3pt]
   \cline{4-7} \rule{0pt}{5mm}
  & & & $e^{\im \pi/3} \sim -1 \sim
   e^{ \im 5\pi/3}$ & $0$ & $1$ &Re$(z)=0,1/2$\\[3pt]
   \hline \rule{0pt}{5mm}
    (e) & $\hlf + \im\tau_2 \ \mbox{with} \  \tau_2>\hlf\sqrt{3}$ &
$J<0$ & $1$
    & $0$ & $1$ & Im$(z)=0,\tau_2/2$\\[3pt]
    \cline{4-7} \rule{0pt}{5mm}
   &  & & -1 & 0 & 1 & Re$(z)=0,1/2$\\[3pt]
     \hline
\end{tabular}
\end{center} \caption{{\em Table of Anti-Holomorphic Involutions.}}
\label{table:involutions}
\end{table}

\bigskip

To see what happens to the K\"{a}hler moduli space, we recall
that the NS field $B \equiv$ Re$(\rho)$ is
odd under worldsheet parity. Under an antiholomorphic involution {\em
   combined} with worldsheet parity
$B_{z\bar{z}} \rar -B_{\bar{z}z}$ therefore survives. Clearly the
   volume is unchanged under any involution and hence the K\"{a}hler moduli
   space is unaffected by a type A orientifold projection (Figure
\ref{fig:modulispace}).  Vice versa
type A orientifolds exist for all values of $\rho$.

\bigskip

Mirror symmetry already tells us that for type B orientifolds the
roles of the K\"{a}hler and complex structure moduli are reversed. One can
also see directly that under a holomorphic involution all pairs of
lattice
vectors $(1,\tau)$ remain lattice vectors. To find the type B-parity
compatible
points in K\"{a}hler moduli
space, we
realize that the odd-worldsheet parity of $B_{NS}$ means that under
a holomorphic involution plus worldsheet parity, $\rho \rar
-\bar{\rho}$. Note, however, that this equation will generically bring
us outside the fundamental domain (it is not defined on $SL(2,\ZZ)$
equivalence classes). Including an arbitrary $SL(2,\ZZ)$ transformation
to bring us back, holomorphic involutions
thus exist for K\"{a}hler moduli for which the involution returns the
K\"{a}hler modulus up to an $SL(2,\ZZ)$ transformation, i.e. for those
$\rho$ which obey
\begin{eqnarray}
   \label{eq:3}
   -\bar{\rho} &=& \frac{a\rho+b}{c\rho+d}~.
\end{eqnarray}
To solve this equation,
recall that a convenient representation of $\rho$ is as the ratio
$\rho=\lam_1/\lam_2$ in terms of the components  $\left(\matrix{\lam_1 \cr
     \lam_2}\right)$ of the fundamental representation of
$SL(2,\ZZ)$. Eq. (\ref{eq:3}) is therefore equivalent to the two
equations
\begin{eqnarray}
   \label{eq:5}
   -\bar{\lam_1} &=& a\lam_1+b\lam_2 ~, \non
  -\bar{\lam_2} &=& c\lam_1+ d\lam_2~.
\end{eqnarray}
Dividing both sides by $\lam_2$,
\begin{eqnarray}
   \label{eq:6}
   - \frac{\bar{\lam_2}}{\lam_2} \frac{\lam_1}{\lam_2} &=& a
     \frac{\lam_1}{\lam_2}+b ~, \non
  - \frac{\bar{\lam_2}}{\lam_2} &=& c \frac{\lam_1}{\lam_2} + d~,
\end{eqnarray}
we recognize the equations~(\ref{eq:4}) and (\ref{eq:4b}) with $\alp=
-\bar{\lam_2}/{\lam_2}$ (and $\beta=0$). We therefore recover directly
the inference of mirror symmetry: that the the existence of
anti-holomorphic involutions in complex structure
moduli-space maps to the existence of holomorphic involutions in the 
K\"{a}hler moduli space.

\subsubsection{Fixed points, species and O-plane charges}

It is at the fixed points of the involution that orientifold-planes are
located. In the covering space $\RR^2$ the solutions to the
fixed-point equation are easily found. E.g.
 for anti-holomorphic involutions with zero shift they are the cycles
$z\bar{z}^{-1} = \alp$: lines through
the origin at an angle $\phi =\frac{1}{2\im}\ln(\alp)$.  Type A $T^2$
orientifolds therefore have O1-planes (we ignore the external
dimensions). Fixed points of anti-holomorphic
involutions are used to construct middle dimensional
special Lagrangian submanifolds which can support supersymmetric
D-branes. Reassuringly, this guarantees the trivial solution to the
tadpole equations with D-branes on top of the O-planes.

For holomorphic involutions fixed points are even dimensional
holomorphic submanifolds. For $z\rar
z$ the fixed submanifold is the torus itself: the O-plane is an
O2-plane. For $z \rar -z$ the fixed submanifold is the origin: the
O-plane is an O0-plane.

The complete set of fixed points for actual $T^2$ involutions
(a)-(e) is given in Table \ref{table:involutions}. 
In the classification Table \ref{table:involutions} we
follow
Alling and Greenleaf \cite{Alling-Greenleaf, Alling}
in their work on Klein surfaces
and real elliptic curves and attach to each nontrivial involution an
invariant called the {\em species}, $s$. This is the number of
connected components of the fixed point locus. In answer to one of the
motivating questions, we will
see this quantity re-appear on the CFT-side as the charge of the 
O-planes present in the Gepner construction.\footnote{Alling and
   Greenleaf imagine the fixed point locus as the boundary of some
   special submanifold. Thus for them $s=2$
corresponds to an annulus, $s=1$ to the M\"{o}bius strip, and
$s=0$ to the Klein bottle.}

The geometric confirmation that the species $s$ is related to the
O-plane charge arises from the
topological generating formula O-plane charge in terms of Hirzebruch
polynomials. This formula can be deduced from chiral anomaly
cancellation 
\cite{Brunner:2003zm,Morales:1998ux,Scrucca:1999uz}.
 In condensed
notation the (set of) RR-charge(s) 
of an O-plane is given by
\begin{eqnarray}
  \label{eq:9}
  Q = \frac{1}{2}\int_{fixed~point~locus} C \wedge \sqrt{ \frac{L(T/4)}{L(N/4)}}. 
\end{eqnarray}
Here $C= C^{(0)}+C^{(2)}+ \ldots$ denotes collectively the RR fields;
$L(T)$ and $L(N)$ are the Hirzebruch polynomials of (a quarter of) 
the tangent and
normal bundle to the O-plane localized at the fixed point. For a $T^2$
A-type orientifold it is a straightforward matter to show that the
unique RR-charge equals the species $s$. 
Brunner and Hori showed in generality that for A-type O-planes the
charge (\ref{eq:9}) equals half the self-intersection number of the
fixed-point locus  \cite{Brunner:2003zm}.
For $T^2$ this equals
$s$ by inspection.\footnote{In general the set of all possible O-plane charges is a subset of all
possible D-brane charges: one can think of the O-plane charges as a
set of vectors in the D-brane charge lattice. 
}

A closer inspection of Table \ref{table:involutions} then reveals that
the species $s$ is an obstruction in the naive ``moduli space'' of
A-type $T^2$ orientifolds. The true moduli space consists of two
disconnected pieces as the species number $s$ is not arbitrary for
each distinct anti-holomorphic involution. Rather involutions for
$\tau=i\tau_2$ only have $s=1$, whereas involutions for which
$|\tau|=1$ or $\tau=\hlf +i\tau_2$ can only have $s=0$ or $s=2$. The
species $s$ therefore serves as a superselection sector on the naive
``moduli space'' and obstructs continuous deformation from the branch
connected to large complex structure at $\tau=i\infty$ to the branch
connected to large complex structure at $\tau=\hlf+i\infty$. In the
mirror B-0 type orientifold there are therefore two disconnected large
volume $T^2$ orientifolds: one with the NS-NS field $B=0$ and one with
$B=\hlf$. This is directly analogous to Brunner and Hori's results on 
the moduli-space of the B-type
orientifold of the quintic with $\tau=i$ serving as the conifold
point \cite{Brunner:2003zm}. Similar to their study we will discuss in
section 6 indications that this obstruction is lifted when the $T^2$
orientifold is embedded in a larger family.

\subsection{The $J$-line moduli space}

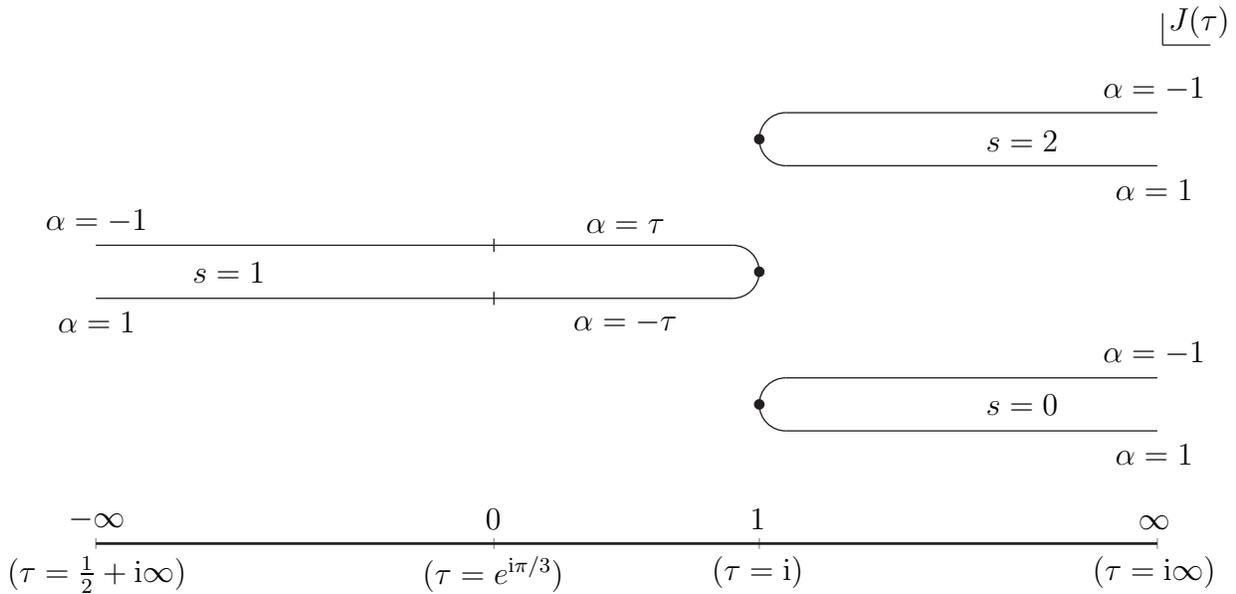
\begin{figure}[tbp]
   \begin{center}
     \begin{picture}(300,200)(-50,-50)
       % Axes
       \SetColor{Gray}
       \Line(-100,-52)(-100,-48)
       \Line(150,-52)(150,-48)
       \Line(300,-52)(300,-48)
       \Line(50,-52)(50,-48)
       \SetColor{Black}
       % Structure at 1
             % Degenerations
       \Vertex(150,52.5){2}
       \Vertex(150,102.5){2}
       \Vertex(150,2.5){2}
       \Line(50,40)(50,45)
       \Line(50,60)(50,65)
       \Line(-100,62.5)(140,62.5)
       \Line(-100,42.5)(140,42.5)
       \CArc(140,52.5)(10,270,90)
       \Line(160,112.5)(300,112.5)
       \Line(160,92.5)(300,92.5)
       \CArc(160,102.5)(10,90,270)
       \Line(160,12.5)(300,12.5)
       \Line(160,-7.5)(300,-7.5)
       \CArc(160,2.5)(10,90,270)

             % Text
       \Text(-50,52.25)[c]{$s=1$}

       \Text(-100,37.5)[t]{$\alpha=1$}
       \Text(-100,67.5)[b]{$\alpha=-1$}
       \Text(100,37.5)[t]{$\alpha=-\tau$}
       \Text(100,67.5)[b]{$\alpha=\tau$}
       \Text(300,87.5)[t]{$\alpha=1$}
       \Text(300,117.5)[b]{$\alpha=-1$}
       \Text(300,-12.5)[t]{$\alpha=1$}
       \Text(300,17.5)[b]{$\alpha=-1$}
       \Text(250,2.5)[c]{$s=0$}

       %\Text(250,87.5)[t]{$s=2$}
       \Text(250,102.25)[c]{$s=2$}
       % Text
       \Text(50,-45)[b]{0}
       \Text(150,-45)[b]{1}
       \Text(300,-45)[b]{$\infty$}
       \Text(-100,-45)[b]{$-\infty$}
       \Text(50,-55)[t]{$(\tau = e^{\im\pi/3})$}
       \Text(150,-55)[t]{$(\tau=\im)$}
       \Text(300,-55)[t]{$(\tau=\im\infty)$}
       \Text(-100,-55)[t]{$(\tau=\hlf+\im\infty)$}
       % Moduli Line
       \SetWidth{1}
       \Line(-100,-50)(300,-50)
       \SetWidth{0.5}
%      \Vertex(250,0){2}
%      \Vertex(50,0){2}
       % Legend
       \Line(302,138)(302,150)
       \Line(302,138)(320,138)
       \Text(317,147)[]{$J(\tau)$}
     \end{picture}
   \end{center}
   \caption{
{\em The complex structure (K\"{a}hler) moduli space type A(B)
        $T^2$ orientifolds in the $J(\tau)$-plane.}
}
   \label{fig:2a}
\end{figure}

We used a natural geometric description of the moduli-space of $T^2$s
as $\RR^2/\Lambda$.
For compactifications of phenomenological interest --- Calabi-Yau
orientifolds --- an intrinsic geometric description of the moduli
space is lacking. Rather, we only understand the complex structure
moduli space in terms of deformations of algebraic equations, and the
K\"{a}hler moduli space through the complex structure of the mirror
We can connect the results for $T^2$ compactifications
with this algebro-geometric description of the moduli space through
the $J$-function: the 1-1 map of the fundamental domain
of $PSL(2,\ZZ)$ to the complex plane: 
\begin{eqnarray}
   \label{eq:14}
   J(\tau) =
\frac{2^8}{24^3}\frac{(\theta_2^8(\tau)+\theta_3^8(\tau)+\theta_4^8(\tau))^3}
{(\theta_2(\tau)\theta_3(\tau)\theta_4(\tau))^8},
\end{eqnarray}
where the $\theta_i(\tau)$ are the Jacobi theta-functions. 
Not coincidentally, the
type
A(B) complex-structure (K\"{a}hler)
orientifold-moduli space projects precisely onto the real
$J$-line. Real values of $\tau$ manifestly admit the trivial
anti-holomorphic involution $\tau \leftrightarrow \bar{\tau}$ and the
1-1 correspondence between distinct tori and values of $J$ then almost
directly implies that the real $J$-line
parametrizes
the complex analytic 2-tori admitting
anti-holomorphic involutions.  However, it is important to note that
the real $J$-line is {\em not} the moduli
space for complex analytic 2-tori with anti-holomorphic
involutions. As the list of anti-holomorphic involutions (a)-(e)
shows, for fixed $\tau$ (i.e. $J$) there are multiple possible
choices
for the involution, and the moduli space for each is distinct
although its $J$-value is the same. As we show in Figure \ref{fig:2a},
the true moduli space is a {\em double cover} of the real $J$-line
\cite{Alling-Greenleaf,Alling}.
On the other hand, that the moduli space is parametrized by real
values of $J$ is significant in that it suggests that the
``orientifold moduli space'', as a subset of the parameter space
for our family of complex 2-tori, is literally the inverse image
of the real part of the $J$-line in parameter space.

As we already indicate in Fig. \ref{fig:2a}, the hidden multiplicity
inherent in the $J$-function can be resolved by two additional
invariants, one of which is the species invariant $s$.  Recall that
any nonsingular elliptic curve may be presented as a cubic
hypersurface in {\em Weierstrass form} via projective coordinates
$(X:Y:Z)$ where $g_2(\tau), g_3(\tau) \in {\mathbb{C}}$ as
\begin{equation}
\label{eq:19}
Y^2 Z = 4 X^3 - g_2 X Z^2 - g_3Z^3 \ ,~~~~~~ \{Y,X,Z\} \sim \{\lam Y,\lam
X,\lam Z\}~.
\end{equation}
Parametrizing the $\CC\PP^2$ spanned by $X,Y,Z$ in terms of $y=Y/Z$
  and $x=X/Z$, one recognizes the more conventional Weierstrass form
  $y^2 = W(g_2,g_3,x)$ in $\CC^2$.
In terms of the coefficients of the Weierstrass equation, the
$J$-function --- also known as the $J$-invariant of the elliptic curve
--- equals
\begin{eqnarray}
   \label{eq:13}
   J(\tau) = \frac{g_2^3}{g_2^3 - 27 g_3^2} \ .
\end{eqnarray}
Finite values of $J$ correspond to a unique
oriented nonsingular complex elliptic curve up to
isomorphism (ignoring K\"{a}hler structure).

For real elliptic curves, i.e. elliptic curves in Weierstrass form with
$g_2$ and $g_3$ real, treat $\infty$ as a real root of
$W(g_2,g_3,x)$. Then the number of real roots is $4$, $2$, or
$0$.  Half this number is the species value $s$ for the corresponding
``standard'' anti-holomorphic involution (i.e., the one inherited from
the ambient ${\mathbb{P}}^2$, fixing the solutions defined over
the reals).

The other invariant which, together with the species,
completely determines the pair of an analytic 2-torus and an
anti-holomorphic involution up to isomorphism, is
essentially a square root of $J(\tau) - 1$, or rather the sign of the
discriminant\cite[S17.50]{Alling}:
\begin{equation}
\Delta= g_2^3 - 27 g_3^2.
\label{eq:dis}
\end{equation}
More precisely,
consider the quartic
$$P(x) = A x^4 + 4 B x^3 + 6 C x^3 + 4 D x + E$$
and the associated Weierstrass invariants $g_2$ and $g_3$ with
$W(g_2,g_3,x) = P(x)$ (after Cayley and Boole \cite[S17.30]{Alling}) expressed as
$$g_2 = A E - 4 B D + 3 C^2 \ , \ g_3 = A C E + 2 B C D - A D^2 -
B^2 E - C^3 \ .$$  
If the discriminant, $\Delta $,
is positive, then the invariant is 
$$H = 3^{3/2} g_3/\sqrt{\Delta} \,$$ 
and if it is negative, then it is
$$H = 3^{3/2} g_3/i\sqrt{-\Delta} \ .$$  
In either case, $H^2 = J - 1$.
The function $H$ is Huisman's ``real J-invariant'' \cite{Huisman:2001}.

\subsubsection{A comparison to $GLSM$ results for projective spaces}

The Weierstrass description of the elliptic curve is, however, not a
natural one from the worldsheet
string. The most general description we know for string theories on
Calabi-Yau surfaces is as gauged $\cN=2$
linear sigma models (GLSM). The
target CY surface arises in algebro-geometric form from the minimum of
the GLSM superpotential as the zero locus
of a set of equations in weighted projective space.
Brunner, Hori, Hosomichi and Walcher classified the possible
anti-holomorphic involutions and their fixed
point loci (\cite{Brunner:2003zm} for unweighted  and
\cite{Brunner:2004zd} for weighted projective spaces).
There are three such algebraic descriptions of the torus. It is precisely
these algebraic descriptions in the
GLS model that make contact with the exact CFT description. At special
points in the moduli
space the IR limit of GLSM is given by a Gepner model. Following the
Gepner notation \footnote{The Gepner model $(k_1,\ldots,k_r)$
is obtained by tensoring $r$ $\cN=2$ minimal models of level $k_i$. This
corresponds to a superpotential $x_1^{k_1+2}+\cdots+x_r^{k_r+2}\in
W\PP_{H/(k_1+2),\dots, H/(k_r+2)}$
where $H$ is the least common multiple of the $k_i+2$. For the variety to
be Calabi-Yau the $k_i$ must satisfy, $-1+\sum_{i=0}^r
\frac{1}{k_i+2}=0$, which might
require the inclusion of trivial, $k_i=0$, minimal models
(e.g. \cite{Brunner:2004zd, Greene}).}
the three models are the (1,1,1) model corresponding to the variety
\begin{eqnarray}
   x_1^3+x_2^3+x^3_3 = 0,  ~~~~~ x_i \in \CC\PP^2
\end{eqnarray}
which describes a torus with $\tau=e^{\frac{\im \pi}{3}}$, the (2,2)
model with variety
\begin{equation}
x_1^4+x_2^4+x_3^2=0,~~~~~ x_i \in W\CC\PP_{1,1,2}
\end{equation}
which is a torus with $\tau=\im$ and the (1,4) model,
\begin{eqnarray}
   \label{eq:38}
   x_1^6+x_2^3+x_3^2 = 0, ~~~~~ x_i \in W\CC\PP_{1,2,3}
\end{eqnarray}
which has $\tau=e^{\frac{\im 2 \pi}{3}}\sim e^{\frac{\im \pi}{3}}$.

To analyze the antiholomorphic involutions, consider first the (1,1,1) 
representation of the
torus as an algebraic variety in unweighted projective space: the
Fermat cubic,
\begin{eqnarray}
   \label{eq:27}
   x_1^3+x_2^3+x^3_3 = 0,  ~~~~~ x_i \in \CC\PP^2.
\end{eqnarray}
The two distinct anti-holomorphic involutions are the canonical one,
$x_i \mapsto
\bar{x}_i$, and the permutation involution,
$(x_1,x_2,x_3) \mapsto
(\bar{x}_2,\bar{x}_1, \bar{x}_3)$).
Following Brunner and Hori \cite{Brunner:2003zm}, we therefore expect
to find two kinds of O1 planes at the fixed points corresponding to two
distinct
anti-holomorphic involutions. In order to make a connection to the
Weierstrass description,
\begin{eqnarray}
   \label{eq:28}
   Y^2Z = 4X^3-g_2XZ^2 -g_3Z^3,~~~~~\{X,Y,Z\} \in \CC\PP^2,
\end{eqnarray}
which allows a direct map to the lattice description through
the Weierstrass functions
\begin{eqnarray}
   \label{eq:29}
   J(\tau) &=& \frac{g_2^3}{g_2^3 - 27 g_3^2} \ ,\non
   X &=&
   \wp(z;\tau) \equiv \frac{1}{z^2}+ {\sum_{m,n=-\infty}^{\infty}}'
   \frac{1}{(z -(m\tau+n))^2}-\frac{1}{(m\tau+n)^2} ~,~\non
   Y &=& \frac{d}{dz}\wp(z;\tau),
\end{eqnarray}
we apply the change of variables,
\begin{eqnarray}
   \label{eq:30}
   x_1 = \frac{1}{6} Z+\frac{\sqrt{3}}{18}Y~,~x_2 = \frac{1}{6}Z
-\frac{\sqrt{3}}{18}Y~,~x_3 =
   -\ove{3}X.
\end{eqnarray}
This yields a Weierstrass equation with $g_2=0$ and
$g_3=1$. Thus
$J=0$, which is known to correspond to $\tau=e^{\frac{\im\pi}{3}}$. With
the Weierstrass correspondence between the algebraic Fermat cubic and the
lattice description in hand, we can match the anti-holomorphic involutions on
both sides. For $\alp= 1$ (and zero shift) the involution is
$\tau \rar \bar{\tau}$. As the Weierstrass $\wp$-function is a
holomorphic function in $\tau$ with real coefficients, we see that
this implies $X \rar \bar{X}$ and $Y \rar \bar{Y}$. The transformation
to the Fermat cubic is also holomorphic and thus the involution $z
\rar \bar{z}$ corresponds to the canonical involution $x_i \rar
\bar{x_i}$. For the involution $\tau \rar -\bar{\tau}$, the definition
of the $\wp$-function shows that $X \rar \bar{X}$, but $Y \rar
-\bar{Y}$. From the transformation rules (\ref{eq:30}), we immediately
see that this corresponds to $x_1+x_2 \rar \bar{x}_1+\bar{x}_2$,
$x_1-x_2 \rar -(\bar{x}_1-\bar{x}_2)$ and $x_3 \rar \bar{x}_3$ and hence
to the permutation involution.

For an arbitrary algebraic variety in weighted projective space it was shown in
\cite{Brunner:2004zd}
that the parity transformations are $x_i\rightarrow e^{\frac{\im 2\pi
m_i}{k_i+2}} \bar{x}_i$, up to permutation,
subject to the following equivalences: $m_i\sim m_i+2n_i,\forall~ n_i\in
\ZZ$ and $m_i\sim m_i+1$.
Thus for the  (2,2) Gepner model, corresponding to a variety
\begin{equation}
x_1^4+x_2^4+x_3^2=0, ~~~~x_i\in W\CC\PP_{1,1,2}
\end{equation}
we expect there to be the following
three types of involutions:
\begin{itemize}
\item[(1)]  $(x_1,x_2,x_3)\rightarrow (\bar{x}_1, \bar{x}_2, \bar{x}_3)$,
\item[(2)] $(x_1,x_2,x_3)\rightarrow (\bar{x}_1,  \bar{x}_2, -\bar{x}_3)$,
\item[(3)] $(x_1,x_2,x_3)\rightarrow ( \bar{x}_1,  \im \bar{x}_2,
-\bar{x}_3)$.
\end{itemize}
The fixed point locus of the first case is empty, as it described by the
solution set
of
\begin{equation}
a_1^4+a_2^4+a_3^2=0, ~~~~ a_i\in W\RR\PP_{1,1,2}
\end{equation}
which, as a sum of positive terms, cannot be equal to zero. According to our
classification this corresponds to a species $s=0$ type involution (see
Table \ref{table:involutions}).
The second case reduces to the requirement that
\begin{equation}
a_1^4+a_2^4=a_3^2, ~~~~ a_i\in W\RR\PP_{1,1,2}
\end{equation}
which appears as closed curves in the $a_1,a_2$ plane with $a_1$
intercept parametrized by $\sqrt{|a_3|}$.
There are actually two disconnected curves as $a_3$ and $-a_3$ yield the
same radius parameter.
This corresponds to the double fixed point locus in our table, i.e. an
$s=2$ type involution.

\begin{figure}[tpb]
\begin{center}\begin{picture}(300,200)(0,0)
\SetOffset(150,100)
\LinAxis(-150,00)(150,00)(4,5,-2,0,.5)
\LinAxis(00,-100)(00,100)(4,5,2,0,.5)
\SetScale{100.}\SetWidth{0.005}
\Curve{(-1.2,1.03615)
(-1.16,0.900355)
(-1.12,0.757311)
(-1.08,0.600407)
(-1.04,0.412139)
(-1,0)
}
\Curve{(1.2,1.03615)
(1.16,0.900355)
(1.12,0.757311)
(1.08,0.600407)
(1.04,0.412139)
(1,0)
}
\Curve{(-1.2,-1.03615)
(-1.16,-0.900355)
(-1.12,-0.757311)
(-1.08,-0.600407)
(-1.04,-0.412139)
(-1,-0)
}
\Curve{(1.2,-1.03615)
(1.16,-0.900355)
(1.12,-0.757311)
(1.08,-0.600407)
(1.04,-0.412139)
(1,-0)
}
\Vertex(-1,0){.012}\Text(-100,3)[rb]{$B^-$}
\Vertex(1,0){.012}\Text(100,3)[lb]{$B^+$}
\SetScale{1}
\end{picture}
\end{center}
\caption{{\em  The curve $a_1^4-a_2^4=a_3^2$ in the chart where $a_2\neq0$}}
\label{figure:disjoint}
\end{figure}
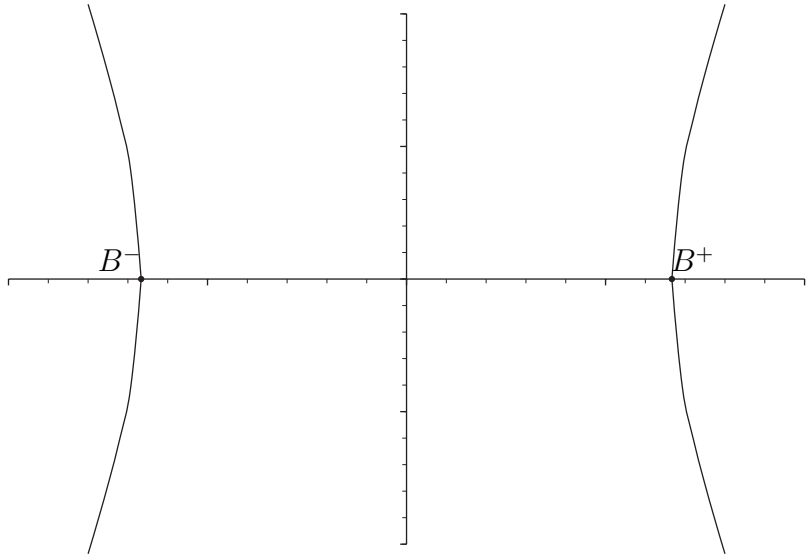

\begin{figure}[tbp]
\begin{center}
\begin{picture}(250,250)(0,0)
\SetOffset(100,100)
\LinAxis(-125,00)(125,00)(4,5,-2,0,.5)
\LinAxis(00,-125)(00,125)(4,5,2,0,.5)
\SetScale{100.}\SetWidth{0.005}
\Curve{(-1.,0.)
(-.995,0.140892)
(-.99,0.198504)
(-.98,0.27862)
(-.96,0.388141)
(-.94,0.468243)
(-.92,0.532548)
(-.9,0.58643)(-.8,0.768375)
(-.7,0.871722)(-.6,0.932952)
(-.5,0.968246)
(-.4,0.987117)
(-.3,0.995942)
(-.2,0.9992)(-.1,0.99995)(0,1)}
\Curve{(0.,1.)(0.1,0.99995)
(0.2,0.9992)(0.3,0.995942)
(0.4,0.987117)
(0.5,0.968246)
(0.6,0.932952)
(0.7,0.871722)
(0.8,0.768375)
(0.9,0.58643)
(.92,0.532548)
(.94,0.468243)
(.96,0.388141)
(.98,0.27862)
(.99,0.198504)
(.995,0.140892)
(1,0)}
\Curve{(-1.,0.)
(-.995,-0.140892)
(-.99,-0.198504)
(-.98,-0.27862)
(-.96,-0.388141)
(-.94,-0.468243)
(-.92,-0.532548)
(-.9,-0.58643)(-.8,-0.768375)
(-.7,-0.871722)(-.6,-0.932952)
(-.5,-0.968246)
(-.4,-0.987117)
(-.3,-0.995942)
(-.2,-0.9992)(-.1,-0.99995)(0,-1)}
\Curve{(0.,-1.)(0.1,-0.99995)
(0.2,-0.9992)(0.3,-0.995942)
(0.4,-0.987117)
(0.5,-0.968246)
(0.6,-0.932952)
(0.7,-0.871722)
(0.8,-0.768375)
(0.9,-0.58643)
(.92,-0.532548)
(.94,-0.468243)
(.96,-0.388141)
(.98,-0.27862)
(.99,-0.198504)
(.995,-0.140892)
(1,0)}

\Vertex(-1,0){.012} \Text(-100,3)[rb]{$B^-$}
\Vertex(1,0){.012} \Text(100,3)[lb]{$B^+$}
\SetScale{1}
\end{picture}
\end{center}
\caption{{\em The curve $a_1^4-a_2^4=a_3^2$ in the chart where $a_1\neq0$}}
\label{fig:connected}
\end{figure}
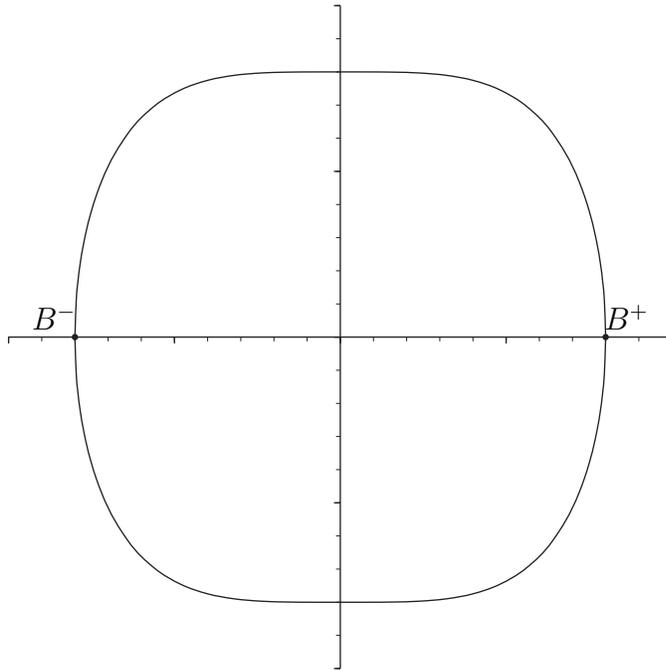

The third possibility reduces to
\begin{equation}
a_1^4-a_2^4=a_3^2, ~~~~ a_i\in W \RR\PP_{1,1,2},
\end{equation}
which resembles a hyperboloid with two disconnected branches. However, a
careful study of the locus,
chart by chart, reveals that the two branches connect through infinity.
For example,
in the $a_2\neq0$ chart one gets the expression
\begin{equation}
\left(\frac{a_1}{a_2}\right)^4-1=\left(\frac{a_3}{a_2}\right)^2,
\end{equation}
which has two branches which intersect the $\frac{a_3}{a_1}$ axis at the
points,
$B^{\pm}$, given by coordinates $(\frac{a_1}{a_2}=\pm
1,\frac{a_3}{a_2}=0)$ (see Figure \ref{figure:disjoint}). Although
these two points appear disjoint in this chart they appear connected in
the $a_1\neq 0$ chart, where
the equation reduces to
\begin{equation}
1-\left(\frac{a_2}{a_1}\right)^4=\left(\frac{a_3}{a_4}\right)^2,
\end{equation}
which describes a closed loop around the origin (see Figure
\ref{fig:connected}). The points $B^{\pm}$ now appear
in the opposite sides of the loop. Similarly, the chart $a_3\neq 0$ has
two apparently disconnected
branches whose points are connected in the $a_1\neq 0$
chart. Thus the third possibility has $s=1$.

We can express the (2,2) Gepner variety in Weierstrass form,
\begin{eqnarray}
   \label{eq:37}
   Z Y^2 = 4X^3 + X Z^2,
\end{eqnarray}
by applying the transformations
\begin{eqnarray}
   \label{eq:36}
   x_1= \left(X/Z+\hlf\right)~,~~
   x_2 = e^{\frac{\im\pi}{4}}\left(\hlf -X/Z\right)~,~~x_3= \im Y/Z.
\end{eqnarray}
Here $J=1$ which indeed corresponds to $\tau=\im$.
Recall that the number of roots of the Weierstrass equation over the
reals is twice the species number
for the {\em standard} antiholomorphic involution - i.e. just the barring of all
the coordinates. With this Weierstrass form we obtain 2 roots ($X=0, X=\infty$), hence
$s=1$. Conjugating the Weierstrass
variables induces the transformation
\begin{equation}
x_1\rightarrow \bar{x}_1,~~~x_2\rightarrow \im \bar{x}_2,~~~
x_3\rightarrow - \bar{x}_3,
\end{equation}
which is the third involution case we examined and showed to be of type
$s=1$.

Finally, the (1,4) model corresponds to the superpotential
\begin{eqnarray}
   \label{eq:sp14}
   x_1^6+x_2^3+x_3^2 = 0 ~,~ \in W\CC\PP_{1,2,3},
\end{eqnarray}
which, using the parities described in \cite{Brunner:2004zd}, has the
following
possible parity involutions:
\begin{itemize}
\item[(1)]  $(x_1,x_2,x_3)\rightarrow (\bar{x}_1, \bar{x}_2, \bar{x}_3)$,
\item[(2)] $(x_1,x_2,x_3)\rightarrow (\bar{x}_1, e^{\frac{\im 2 \pi}{3}}
\bar{x}_2, \bar{x}_3)~$,
\item[(3)] $(x_1,x_2,x_3)\rightarrow (e^{\frac{\im \pi}{3}} \bar{x}_1,
\bar{x}_2, \bar{x}_3)$,
\item[(4)] $(x_1,x_2,x_3)\rightarrow ( \bar{x}_1,  \bar{x}_2, -\bar{x}_3)~\sim( e^{\frac{\im \pi}{3}} \bar{x}_1, e^{\frac{\im 2\pi}{3}} \bar{x}_2, \bar{x}_3).$
\end{itemize}
The first and second cases have fixed point loci isomorphic to
\begin{equation}
a_1^6+a_2^3+a_3^2=0 ~,~ a_i\in W\RR\PP_{1,2,3}.
\end{equation}
This follows directly in case (1). The fixed point locus of transformation (2) is
isomorphic to this curve under the map $a_2\rightarrow -a_2$. The fixed point loci for
these cases are of species $s=1$. The cases (3) and (4) have fixed point loci that are isomorphic to
\begin{equation}
-a_1^6+a_2^3+a_3^2=0~,~ a_i\in W\RR\PP_{1,2,3}
\end{equation}
and hence are also of species $s=1$. The isomorphism relating case (3) to case (4) is also
the map $a_2\rightarrow -a_2$. It is easier to see this if one uses the second
presentation of the involution map for case (4).
Thus in all there are only two different
classes of involution for this
non-linear sigma model, both of which have $s=1$. This fits in nicely
with our results of Table \ref{table:involutions}.
As a final check, we can apply the transformation
$x_1 = \lam$, $x_2= -\lam^2 4^{1/3} X/Z$, $x_3 =
\lam^3 Y/Z$, to bring the elliptic curve to Weierstrass form
\begin{eqnarray}
   \label{eq:41b}
    Y^2 Z = 4X^3 -Z^3.
\end{eqnarray}
  We can see that the Weierstrass equation has 2 real roots at
$X=\infty$ and at $X=(1/3)^\frac{1}{3}$ so that the natural involution
also has $s=1$, as was to be expected.
Note that the Weierstrass equation has $J=0$. Hence this is the same
elliptic curve as the (1,1,1) model
even though the value $g_3=1$ is different.

The importance of the species $s$ in the analysis of involutions of
$T^2$ and their fixed points is self-evident. We will now show how the
species $s$ appears as the O-plane charge from a microscopic
worldsheet CFT analysis. We have clearly seen, however, how much an analysis solely
based on geometrical methods is able to tell us.

\section{Crosscaps in Calabi-Yau compactifications: review}
\setcounter{equation}{0}

We will use the equivalence of O-planes and worldsheet crosscaps to gain
an understanding of charges and tensions of orientifold planes.
First we will write down rational
world sheet CFTs which correspond to toroidal compactifications. In
particular we will
work with the so called ``Gepner'' model of the torus. The
orientifold(-planes) will then correspond to crosscaps in the CFT.
This section reviews how to construct unoriented RCFTS and how to read
off the O-plane data from the crosscap states. The knowledgeable reader
may skip to section \ref{sec:orient-plan-from} where we apply these
methods to $T^2$ compactifications as $c=3$ Gepner models. 

\subsection{Gepner models: a brief review}
A Gepner model is a $c=3n$, $\cN=2$ CFT  produced by tensoring several
$c_i=3k_i/(k_i+2)$ $\cN=2$ minimal models
and aligning the worldsheet supersymmetry currents. The resulting
$\cN=2$ theory, $\cA^{3n}$, can be
matched to a specific $d=2n$ algebraic variety. The full theory
requires the inclusion of the external $d=10-2n$ dimensional
dimensional spacetime which is modeled (via the bosonic string map 
\cite{Fuchs:2000gv,Gepner1989,Englert1985,Schellekens1987,LSW}) as
a $\cN=2$ WZW $D_{N-1,1}$ component to be tensored with $\cA^{3n}$ and again susy
aligned. The resulting theory $\cA^{ws}$ is finally GSO projected to produce a
consistent string compactification.\footnote{When studying closed oriented theories, a
   pre-GSO projection can be done on the $\cA^{3n}$ theory to
produce a more or less independent CFT theory corresponding to
the internal compactification space \cite{Fuchs:2000gv}. However, in
unoriented closed/open theories,
the supersymmetric properties of the theory in general and crosscap
states in particular cannot be
divorced from the ambient space
\cite{Govindarajan:2003vv,Huiszoon:2003ai,Govindarajan:2003vp}.}

The supersymmetry alignment of the different components is an example of {\em a
   simple current extension} of a CFT. 
A simple current extension is an extension
   of the chiral algebra $\cA \stackrel{J}{\subset} \cA^{ext}$
by integer conformal weight simple currents: primary fields $J$
whose fusion with any other
primary field, $i$, yields a {\em single} field $j=Ji$~\cite{simple,simple2}.
Under this extension primary
fields arrange themselves into orbits $[i] =
\left\{i,Ji,J^2i,\ldots\right\}$. The conformal weight of the orbit is well
defined modulo integers. The exact
conformal weight of the orbit is taken to be
the lowest of the weights of its representative elements \cite{Huiszoon:2003ai}.
Orbits with integer monodromy charge
$Q_J(i)\equiv h_i+h_J-h_{Ji}$ mod
$\ZZ$ under $J$, are the primaries of the extended chiral algebra
$\cA^{ext}$. Non-integer charged
fields are projected out.

The simple currents of an $c_i=3k_i/(k_i+2)$ $\cN=2$ minimal model are
reproduced in Table \ref{tab:sc} \cite{Fuchs:2000fd}.
\begin{table}[tbp]
   \centering
\[
   \begin{array}{|c|c|c|}
\hline
   \mbox{Current} & \mbox{Primary field label} & \mbox{Order} \\
\hline
    v & (0,0,2) & 2 \\
    s & (0,1,1) & 2k+4~{\rm if}~k \in 2\ZZ,~4k+8~{\rm otherwise} \\
    p = s^2v & (0,2,0) & k+2 \\
    f & (k,0,0)=(0,k+2,2) & 2 \\
\hline
   \end{array}
\]
   \caption{{\em Simple currents of $\cN=2$ minimal models. Recall that
   primary fields are labeled by $(\ell,m,s)$ with $0\leq \ell \leq k$,
   $-k-1 \leq m_{\rm mod ~2k+4} \leq k+1$, $ -1 \leq s_{\rm mod ~4} \leq
   2$ subject to field identification $(\ell,m,s) \simeq
   (k-\ell,m+k+2,s+2)$ (For a review see \cite{Fuchs:2000gv,Fuchs:2000fd}).}}
   \label{tab:sc}
\end{table}
Worldsheet supersymmetry is generated by
the `vector' simple current, $v_i$, of order 2. The alignment of the
worldsheet supersymmetries in a tensor product of $i=1,2,3,...,r$
minimal models is
thus enforced by
the group of currents generated by $w_{ij}$: the pairwise choice of
$v_i$'s. In our calculations we will prefer to work in two stages. First we
will align the minimal model tensor product(the ``Gepner'' part,
$\cA^{3n}$) and then align it with the spacetime part $D_{8-n,1}$
to yield the worldsheet theory, $\cA^{ws}$.

The final step in constructing the Gepner model is to
perform the GSO projection which will ensure spacetime supersymmetry.
This is done by extending the theory by the spectral flow operator of
$\cA^{ws}$, $s$, which is the product of the
spectral flow operators of the individual $\cN=2$ factors in the tensor
theory. The order of the group
generated by this theory depends on the structure of the
components. The relevant feature is that this current is even, as
described in \cite{Huiszoon:2003ai}.

In constructing simple current extensions 
there is an important subtlety that one must be aware of. When the
order of the current is even, there may exist fixed points,
i.e. fields $i=Ji$ that are invariant under the action of the
extension current $J$. In general one must resolve this degeneracy for
a complete description of the theory \cite{simple,simple2,Fuchs:2000cm}.
However, the conjecture is that 
the one known consistent solution for the crosscap
coefficients (see below) of orientifolded simple
current extensions needs only the trivial solution to the fixed point
resolution \cite{thesis,Fuchs:2000cm} and we can therefore ignore this subtlety.
 
\subsection{Building consistent $\cN=2$ unoriented CFTs}

The worldsheet supersymmetry alignment ensures that the Gepner model
will be $\cN=2$ superconformal. The bulk worldsheet theory symmetry
currents will
therefore contain a left- and right-moving copy of the $\cN=2$
superconformal algebra \cite{Greene}
\begin{eqnarray}
[L_n,L_m]&=&(n-m) L_{n+m}+\frac{c}{12}(n^3-n)\delta_{n+m}, \non
\{G^\pm_r,G^\mp_s\}&=&2L_{r+s}\pm(r-s)J_{r+s}\frac{c}{3}(r^2-\frac{1}{4})\delta_{r+s},\non
\{G^\pm_r,G^\pm_s\}&=&0,\non
{}[ L_n,G^\pm_r ]&=&(\frac{n}{2}-r)G^\pm_{n+r},\non
{}[J_n,J_m]&=&\frac{c}{3}n\delta_{n+m},\non
{}[L_n,J_m]&=&-mJ_{n+m},\non
{}[ J_n, G^\pm_r ] &=&\pm G^\pm_{n+r}.
\end{eqnarray}
Important for us are the $\ZZ_2\times U(1)$ {\em automorphisms} of the
$\cN=2$ superconformal algebra. The $U(1)$ is an inner automorphism
produced by conjugation with the $U(1)$ R-charge, while the $\ZZ_2$ is
an outer automorphism which flips the sign
of the R-charge.

These automorphisms, and in particular the involutive ones, play a
constitutive part in the method of {\em open descendants}: the
construction of consistent unoriented closed/open string theories from
a consistent, i.e. modular invariant closed string theory
\cite{Pradisi,sagnotti,sagnotti1996}.
Starting from the torus partition function, ${\cal{Z}}^{orient}=Z$, of
this oriented closed string theory one can project, or in a sense
``orbifold'',
   by the world sheet parity operator, $\Omega$.
The resulting partition function, ${\cal{Z}}=(Z+K)/2$, is the sum of the
torus contribution and an unoriented component, $K$, the Klein
bottle partition.
The Klein bottle can be described as a tube ending on two crosscaps.
Depending on how time is oriented, there are two ways of describing the
physics:
the ``open loop channel'', or the ``transverse closed string channel''.
In the open loop time runs azimuthally with open strings spanning the
length of the tube and ending on the crosscaps. The resulting boundary
condition
on an open string ending on the crosscap is  $\partial_\tau X(\sigma,
\tau)=-\partial_{\tau}X(\pi/2-\sigma,\tau)$. The open string partition
function can, by a conformal map aligning time with the cylinder axis,
also be seen as a tree level closed string exchange between the two
crosscaps.
The open string boundary condition is then reinterpreted as a gluing
condition at the crosscap state, $|C\rangle$,
\begin{eqnarray}
    \label{eq:crosscap_reflection}
    W_n -(-1)^{h_W+n}w[\tilde{W}_n]|C\rangle =0,
\end{eqnarray}
where $W_n$ is the mode expansion of a chiral field $W$ of weight
$h_W$ \cite{thesis}.
For the standard bosonic string $W_n$ would be a string
oscillator mode,
$\alpha_n$, and the weight would be unity. The symbol $w$ is used to denote
any possible automorphisms of the chiral algebra which could provide
additional
ways of matching the right and left moving algebras.

Generically the projected theory, $\tilde{\cal{Z}}=(Z+K)/2$, suffers
from UV divergences. These can be canceled by adding new ``twisted
states''. These come
from the open string sector, described by the Annulus partition function, $A$,
which is also in turn projected by $\Omega$. This projection yields the sum of the Annulus
and the M\"{o}bius partition functions: $(A+M)/2$. The annulus can be described as
the familiar open cylinder, while the M\"{o}bius can be envisioned as a
cylinder with a boundary
on one end and a crosscap on the other. In the direct channel picture we
can attach an open string to the boundary so that it satisfies
Dirichlet, $\partial_\tau X=0$,
or Neumann, $\partial_\sigma X=0$, boundary conditions. In the closed
string transverse channel this boundary condition can be interpreted as
the tree level exchange of
closed strings from a boundary state, $|B\rangle$,  with the boundary
gluing condition
\begin{eqnarray}
    \label{eq:boundary_reflection}
    W_n -(-1)^{h_W}w[\tilde{W}_n]|B\rangle =0.
\end{eqnarray}
For the standard bosonic string $w$ is the identity in the Neumann case, or the map
$\tilde{\alpha}_n\rightarrow - \tilde{\alpha}_n$ in the
Dirichlet case.

Formally the boundary and crosscap states can be expanded as linear combinations
of boundary, $|I\rangle\rangle_{B,w}$, and crosscap,
$|I\rangle\rangle_{C,W}$, Ishibashi states \cite{ishibashi}. These Ishibashi
state are solutions to the reflection conditions
(\ref{eq:crosscap_reflection}) and (\ref{eq:boundary_reflection})
that are (irreducible) representations of the conformal symmetry. In
this formal sense
the boundary and crosscap states are
\begin{eqnarray}
|B_a\rangle_w&=&\sum_I B_{aI}|I\rangle\rangle _{B,w},\non
|C\rangle_w&=&\sum_I \Gamma_I|I\rangle\rangle _{C,w}.
\end{eqnarray}
Here the boundary state acquires a boundary label, $a$, to distinguish
different boundaries. Explicitly,
the boundary Ishibashi states are given by \cite{thesis}
\begin{equation}
|I\rangle\rangle_{B,w}=\sum_s|s,I\rangle\otimes UV_w|s,I\rangle,
\end{equation}
where $I$ is a primary field of the theory and $s$ labels the states in that
module. The operator $U$ is antiunitary and
satisfies the commutation relation
\begin{equation}
U\tilde{W}_n=\tilde{W}_nU(-1)^{h_W},
\end{equation}
while $V_w$ is an ``intertwiner'' which maps a module $I$ to a module
$w(I)$ and satisfies the composition rule
\begin{equation}
V_w\circ W_n=w[W_n]\circ V_w~.
\end{equation}
The crosscap reflection condition (\ref{eq:crosscap_reflection}) is
similarly solved by the crosscap Ishibashi states,
\begin{equation}
|I\rangle\rangle_{C,w}=\sum_s|s,I\rangle\otimes (-1)^{\tilde{L}_0-h_I}
UV_w|s,I\rangle.
\end{equation}

The boundary coefficients, $B_{aI}$, and the crosscap coefficients,
$\Gamma_I$, must obey certain ``sewing''
constraints for consistency of the theory. These sewing constraints are very
difficult to solve in general. In practice one limits one's attention to
a subset of these: the positivity and integrality constraints.
Recall that the 1-loop open string partition contains information about
the spectrum
of the theory. Therefore the coefficients of the
characters, transformed back from the transverse channel, have to be
natural numbers. These conditions are already so
restrictive that only one set of solutions is known.
   For the boundary coefficients with trivial automorphism action 
one has the Cardy solution \cite{cardy},
\begin{equation}
B_{Ia}=\frac{S_{Ia}}{\sqrt{S_{I0}}},
\end{equation}
where $S$ is the representation of the modular generator,
$\tau\rightarrow -1/\tau$, on the conformal characters (irreducible
building blocks of the partition function). For the crosscap
coefficients (again with trivial automorphism action) there is the
Rome solution \cite{Rome,sagnotti},
\begin{equation}
\Gamma_I=\frac{P_{I0}}{\sqrt{S_{I0}}},
\end{equation}
where $P=\sqrt{T}S T^2 S \sqrt{T}$ and $T$ is the other generator of
the modular group $\tau\rightarrow \tau+1$.

These equations can
be generalized to the case of simple current extensions
\cite{thesis,Huiszoon:2003ai}.
For a theory whose algebra, $\cA^{ext}$,
is extended from $\cA$ by a simple current
group $G$ of order $N$, the boundary states are \cite{Fuchs:1999zi,Fuchs:1999xn},
\begin{eqnarray}
      \label{eq:D-expansion}
      |B_{[a]}\rangle&=&\frac{1}{\sqrt{N}}\sum_{J\in G}|B_{Ja}\rangle
\nonumber\\
      &=&\sum_I \sqrt{N}\frac{S_{Ia}}{\sqrt{S_{I0}}}|I\rangle\rangle_D,
\end{eqnarray}
where one uses the familiar S-matrix identity,
\begin{equation}
\label{eq:Sidentity}
S_{J i,j}=e^{\im 2 \pi Q_J(j)}S_{i,j}~,
\end{equation}
and that $Q_J(I)=0$, i.e. the Ishibashi states are labeled by `fields'
$I$ of $\cA$
that are integer charged under $J$. The crosscap is similarly extended
as \cite{klein}:
\begin{equation}
     \label{eq:23}
      |C\rangle_{\sigma}^{[K]} = \sum_{J \in G} \sigma(JK)
\frac{P^{\cA}_{KJ,I}}{\sqrt{S^{\cA}_{0,I}}}|I\rangle\rangle_C.
    \end{equation}
The symbols $\sig(JK)$ are signs, where $\sig(K)=1$ by
convention, subject to the following
constraint \cite{thesis}. Define $\beta_K(J) = \sig(JK)\eps_J(K)$, then
$\beta_K(J_1J_2)=\beta_K(J_1)\beta_K(J_2)$ and $\beta_K(0)=1$. These
constraints, valid for integer spin extensions, are equivalent to
the assumption that the crosscap should only couple to fields which are
GSO invariant (hence physical). The
field $K$ is a simple current in $G$, called the Klein bottle
current; it must obey that for each order 2 current $J_2 \in G$,
$Q_{J_2}(K)=0 \mod 1$
in order to be able to solve the sign constraints. Note that $K$ serves to
label
different crosscap states. The total number of boundary and crosscap
states for a cyclic, $\ZZ_N$, simple
current extension is therefore $N$ boundary states and $2\cdot N/2$
(signs times number of Klein bottles) in both cases
\cite{Huiszoon:2003ai}.
Experience indicates that for $K$ odd and not in $G$ we get different
crosscaps \cite{thesis}. The current extensions will play an important
role here in that they will be used
to generate the non-trivial boundary and crosscap coefficients in
addition to constructing the Gepner model.

\subsubsection{Boundary conditions:}

Boundary and crosscap states clearly intertwine the Left and Right
moving algebras. In particular the full $\cN=(2,2)$ supersymmetry
will be broken to a linear $\cN=2$ combination of the Left and Right supersymmetry algebras.
The different ways a boundary or crosscap state can preserve the $\cN=2$ susy is
thus given by the automorphisms of the $\cN=2$ algebra.  At the same time
we wish to preserve
$T|B\rangle = \bar{T}|B\rangle$ as the conserved charge
associated with the unbroken translations is (worldsheet) `energy'
rather than
momentum; hence there is no energy leakage across the boundary.
The full $U(1)\times \ZZ_2$ isomorphisms leave 
$T$ unchanged and can therefore be used to vary the gluing
conditions. 
The most general boundary conditions fall in two classes, A and B
distinguished by whether the $\ZZ_2$ acts trivially (type A) or  not
(type B).
\begin{eqnarray}
    \label{eq:32}
    A{\rm -type}:~~~&& G^{\pm}|B\rangle =
    e^{\pm \im\alpha_A}\bar{G}^{\mp}|B\rangle~, \non
&&J|B\rangle = \bar{J} |B\rangle~. \non
B{\rm -type}:~~~&& G^{\pm} |B\rangle =
   e^{\pm \im\alp_B}\bar{G}^{\pm} |B\rangle ~,\non
   &&J|B\rangle = -\bar{J} |B\rangle~.
\end{eqnarray}
Note that the A-type boundary conditions preserve the $U(1)_A \equiv
U(1)_{R,L}-U(1)_{R,R}$ symmetry, but break the $U(1)_V \equiv
U(1)_{R,L}+U(1)_{R,R}$ symmetry, whereas the B-type boundary
conditions do the reverse. There must therefore be a $U(1)_V$ {\em
   multiplet} of consistent A-type boundary conditions, and it is not
hard to see that these are precisely the boundary conditions
parametrized by $\alp_A$. One obtains these by a rotation by
$e^{\im\frac{\alp_A}{2} J_V}$:
\begin{eqnarray}
    \label{eq:17}
    |B\rangle_{\alp_A} = e^{\im\frac{\alp_A}{2} J_V}|B\rangle_0 :~~~~
     G^{\pm} |B\rangle_{\alp_A} &=& G^{\pm} e^{\im\frac{\alp_A}{2}
     J_V}|B\rangle_0 \non
    %%&=& e^{i\frac{\alp_A}{2}
      %%(J_V \mp 1)} G^{\pm} |B\rangle_{0} \non
   %%&=& e^{i\frac{\alp_A}{2}
     %% (J_V \mp 1)} G^{\mp} |B\rangle_{0} \non
%% &=& e^{i\frac{\alp_A}{2}(\mp 1\mp 1)} G^{\mp} e^{i\frac{\alp_A}{2}
    %%  J_V}|B\rangle_0 \non
&=& e^{\pm \im \alp_A} \bar{G}^{\mp} |B\rangle_{\alp_A}~.
\end{eqnarray}
Similarly, B-branes fall a $U(1)_A$ multiplet of boundary states,
    which preserve half the supersymmetry.\footnote{This is correct for
    superconformal theories. For a generic $N=2$ susy theory, however,
    $U(1)_A$ is broken at the quantum
    level. It is only preserved if $c_1(\cM_{target~space})=0$,
    which is always
    the case for superconformal theories. (In case $U(1)_A$ is broken,
the symmetry can be `restored', however, by a shift in
    the $B_{NS}$-field \cite{Brunner:2003zm}.
    There is therefore exactly one B-boundary state for
    each value of $B_{NS}$. The `faulty' B-states do exist, but they do
    not preserve half of worldsheet susy.)}
We will only consider the $\alpha_A=0$ and $\alpha_B=0$ representatives.

The known consistent solutions to crosscap and boundary gluing
conditions are valid only for the trivial outer automorphism case (Type A).
However type B is mirror to type A, as it is obtained by flipping
the $U(1)$ R-charges. The mirror map can be constructed as a permutation
extension of the theory, as described in Appendix
\ref{sec:mirr-symm-extens}. 
Hence we can obtain
the Type-B crosscap and boundary states
as simple current extended states of the mirror type A theory \cite{dbranesquintic}.

\subsubsection{Crosscap states:}

Similarly to boundary states, crosscap states
will preserve some diagonal combination of the
left and right $\cN=2$ symmetries. The procedure to build the
orientifold theory is to mod out by
worldsheet parity and any other involutive global symmetry, $R$. We
insist on
involutive global symmetries
as we wish to preserve the geometry of the target manifold. This is
important to be able to
classify the theories consistently. Otherwise, one could argue that the
resulting theory is just the orientifold of
some orbifold of the target manifold.

Similarly to the boundary states there are two
possible {crosscap conditions} \cite{Brunner:2002em},
\begin{eqnarray}
   \label{eq:35}
   A{\rm -type}:&& G^{\pm} |C\rangle = e^{\pm i\bet_A} \bar{G}^{\mp} |C\rangle
~,~\non
&& J|C\rangle =\bar{J}|C\rangle ~.\non
B{\rm -type}:&& G^{\pm} |C\rangle = e^{\pm i\bet_B} \bar{G}^{\pm} |C\rangle
~,~~ \non
&& J|C\rangle = - \bar{J}|C\rangle~.
\end{eqnarray}
The phases $\bet_{A,B}$ are related to the $U(1)$-automorphism class. In
eq. (\ref{eq:glue}) we shall see that this determines the spacetime
susy properties of the O-plane. 
As also in the case of boundary states, the crosscaps for the trivial
automorphism class can be directly constructed, but
B-type crosscaps will be obtained via the mirror simple current map from
the A-type states.

\def\beq{\begin{equation}}
\def\bea{\begin{eqnarray}}
\def\eea{\end{eqnarray}}
\def\eeq{\end{equation}}
\def\C{\Gamma}
\def\s{\sigma}
\def\e{\epsilon}
\def\[{[}
\def\]{]}
\def\bZ{\bar{Z}}
\def\bz{\bar{z}}
\def\btau{\bar{\tau}}
\def\a{\alpha}
\def\b{\beta}

\bigskip
\subsection{Orientifolds in Gepner Models}

In Gepner models an important additional step in the construction of
supersymmetric BPS like
crosscap states is the GSO projection. Previous studies 
indicated that it was possible to decouple the internal geometry of the
theory from the spacetime
via a ``pre-GSO'' projection (see e.g. \cite{Fuchs:2000gv}). However, in the
context of orientifolds this
separation of the theory into spacetime and internal sectors is very
awkward,
as without the complete GSO projection the resulting
crosscaps did not satisfy the BPS condition \cite{Govindarajan:2003vv, Govindarajan:2003vp,Huiszoon:2003ai}.
Hence, a careful study of the GSO projection is necessary. Specifically
the GSO projection involves the
simple current extension from the world sheet supersymmetric theory,
$\cA^{ws}$, to the spacetime supersymmetry
theory, $\cA^{ext}$. This extension is generated by the spectral flow field,
$S$, which is of even order, $N_S$.
As described in
\cite{Huiszoon:2003ai}, by requiring that the
crosscap
only couples to fields, $i\in\cA^{ext}$, i.e.
such that $Q_S(i)=0$, the resulting A-type crosscap has the following
general
description:
\beq \label{eq:Cr}
|\C \rangle^\s_{[K]} = \sqrt{N_S} \sum_{\{i|REP_{[i]},~Q_S(i)=0\}}
(\frac{\s_0 P_{i,K} + \s_1
P_{i,KS}}{2\sqrt{S_{0i}}})
  |[i]\rangle\rangle_{\e_S(K)\s} \;\;\;,
\eeq
where $\s_0,\s_1$ and $\s=\s_0/\s_1$ are signs and
\beq \label{eq:twist}
|[i]\rangle\rangle_{\e_S(K)\s} = \sum_{n=0}^{N-1} [\e_S(K)\s]^n \e_{S^n}(i),
|S^ni\rangle\rangle_{1,C}~,
\eeq
is the relation between Ishibashi states in $\cA^{ext}$ and in
$\cA^{ws}$ (which in turn are the sum of Ishibashi states
in the minimal model).
The Klein bottle currents are required to have monodromy
\beq
Q_S(K)=\frac{2p}{N_S},\,\, p\in\ZZ.
\eeq
The weights in this last sum are just phase factors,
\beq
\e_S(i)\equiv e^{\pi\im(h_i-h_{Si})}.
\eeq
Therefore, the crosscap states obey a twisted gluing
condition,
\beq \label{eq:glue}
[S_n - (-1)^{n + h_S} \e^{\ast}_S(K)\s \bar{S}_{-n}] |C \rangle^\s_{[K]}
= 0~.
\eeq
The term $\e^{\ast}_J(K)\s = (\e_J(K)\s)^{-1}$ is the automorphism type of
the state. From the monodromy restriction of $K$, one can see
that $(\e_S(K)\s)^{N_S}=1$, so that the automorphism type of the plane takes
values in $\ZZ_{N_S}$. This gluing condition implies that
the planes preserve the supersymmetry generated by the linear
combination $S_0+\epsilon^*_S(K)\bar{S}_0$ \cite{Huiszoon:2003ai}. As
$S_0$ generates a discrete $\ZZ_{2N_S}$ subgroup of the $U(1)_R$ symmetry, we
recognize that $\eps^{\ast}_S$ determines the phase $\beta_A$ from the previous
subsection.

\subsection{Crosscap and Boundary Mass and Charges}

By considering the overlap with different massless fields we can get the
tensions and charges of the boundary or crosscap state. These charges encode
geometric characteristics of the corresponding orientifold or D-brane.
The overlap with the massless NS-NS vacuum state,
$[[\vec{0};\vec{0};\vec{0}]o_D]$,
is proportional to the tension of these states.\footnote{ The fields are
labeled
according to tensor components:
$[[\vec{l};\vec{m};\vec{s}]f_D]=[[l_1\ldots l_r;m_1\ldots m_r;s_1 \ldots
s_r]f_D]$, for a tensor product of $r$ minimal models and a $D_{8-n,1}$
spacetime
component
   whose fields $f_D$ are either scalar ($o_D$), spinor ($s_D$),
vector ($v_D$) or conjugate spinor ($c_D$). The square brackets indicate
that it is an
equivalence class of fields
under the simple current extension.
Further details on notation can
be found
in the appendices. In addition, \cite{Huiszoon:2003ai} contains a short
review on the properties of
the spacetime algebra,  $D_{8-n,1}$.}
It is convenient to also compute the boundary state data, as at the end
of the day one has to make sure that
the D-brane tadpole divergences are canceled by O-plane charges.

  From the crosscap expansion equation  (\ref{eq:Cr}) we see that the
tension is
\bea
\label{eq:mass}
M^{\[K\]}_C=\langle O|C\rangle_{\[K\]}^\s =\sigma_0
\frac{\sqrt{N_S}}{2\sqrt{S_{O,O}}}P^{ws}_{O,K},
\eea
where the prefactor is just a Gepner model dependent overall
normalization. In the case of the boundary state, the overlap of the boundary state,
(\ref{eq:D-expansion}),
with the vacuum gives a mass of
\beq
\label{eq:bmass}
M^{[a]}_B=\langle O|B_{[a]}\rangle =\sqrt{N S_{0a}^{ws}}.
\eeq

The other characteristics
to determine are the charges with respect to RR ground states.
In the non linear sigma model realization of the theory, the
RR ground states of left and right $U(1)_R$charge, $(q,\tilde{q})$ can be
identified with holomorphic $(n/2-q,n/2+\tilde{q})$ forms
\cite{Brunner:2004zd}.\footnote{Note that the R-charges/harmonic form
identification
here is for the internal sector of the theory, so the R-charge should be
restricted to
the minimal models.}  One
could directly get the charges by computing the overlap of the
crosscap or boundary state
with the RR groundstate corresponding to primary field, $i_{{RR}_{gs}}$,
\begin{eqnarray}
Q_i^{crosscap}&=&\langle i_{{RR}_{gs}}| C\rangle=\Gamma_{i_{{RR}_{gs}}},\non
Q_i^{brane,a}&=&\langle i_{{RR}_{gs}}| B_a\rangle=B_{a{i_{{RR}_{gs}}}}.
\end{eqnarray}

However, the RR ground states are in one to one correspondence
with the NS-NS chiral-chiral or chiral-antichiral primaries by spectral
flow \cite{Greene, warner}, and it is more convenient to
use chiral/antichiral fields to compute the charges.
In a minimal model of level $k_i$ chirals (anti-chirals) are labeled by
$\chi_{\pm l_i}=(l_i, \pm l_i,0),~l_i\in
\{0,\ldots,k_i\}$, have charge $q_i=\frac{\pm l_i}{k_i+2}$ and weight
$h_i=|q_i|/2$. In constructing a Gepner model chiral we can
take the tensor of minimal model chirals as long as they satisfy the GSO
projection,
$Q_S(\chi)=\sum_{i=1}^r \frac{l_i}{2(k_i+2)} +Q_{s_D}(f_D)= 0$  mod 1,
where the contribution for the (chiral) spacetime fields is
$Q_{s_D}(v_D)=1/2$ or $Q_{s_D}(o_D)=0$.

A chiral-chiral primary, $i$, of charge $(q,\tilde{q}=q)$ corresponds, by
symmetric Left/Right spectral flow,
to a harmonic forms $(n-q,q)$ in the horizontal cohomology. The flow is generated
by $S$, so that $i_{{RR}_{gs}}=S i$. As they have equal
charges in the Left and Right moving fields they naturally couple to
A-type Ishibashi states giving us the A-type
charges
\begin{eqnarray}
Q_i^{crosscap}&=&\langle i_{{RR}_{gs}}| C\rangle=\langle S i| C\rangle\non
&=& \Gamma_i=\frac{\sqrt{N_S}}{2\sqrt{S_{0i}}}\sigma_1 P_{i,K}
\epsilon_{S}(K)\epsilon_S(i).
\end{eqnarray}

Chiral-antichiral primaries of charge $(q,\tilde{q}=-q)$, correspond by
asymmetric Left/Right spectral
flow, to harmonic forms in the vertical cohomology. These states naturally
couple to B-type Ishibashi states.
The B-type crosscap (and boundary) states are given by a non-trivial
automorphism of the susy algebra. As mentioned
earlier this is the same as flipping the $U(1)$ R-charges of the right
moving
chiral algebra. This is equivalent to working with the conjugate modular
invariant
instead of the diagonal modular invariant which we have been using as a
starting point for
A-type states. In particular, we can construct the charge conjugate
invariant partition
function as a simple current
(permutation) extension of the diagonal theory by the currents of the
Mirror Symmetry Extension group $G_{ms}$. The
$G_{ms}$ is the simple current description of the mirror map of Greene
and Plesser
generated by elements of the form
\bea
G_{ms}=\{v_D^{(n+r)\epsilon} \Pi_{r=1}^r p_i^{\alpha_i}, \sum_{i=1}^{r}
\frac{\alpha_i}{h_i}-\frac{(n+r)\epsilon}{2}=0\mbox{ mod }1,\,\,\,\,
\epsilon\in\{0,1\}\}\mbox{ mod } G_{ext}
\label{eq:ms}
\eea
(details on the construction of this simple current group are given in
Appendix \ref{sec:mirr-symm-extens}).
We can then describe B states as $G_{ms}$ and $G_S$ extension states of
the $\cA^{ws}$ theory.
In effect, the B-type states end up being a weighted sum of A-type states:
\begin{eqnarray}
\label{eq:mirrorC}
\Gamma^{K,B}_i&=&\sum_{J\in G_{ms}, S^n\in G_S} \frac{1}{\sqrt{N_S
N_{ms}S_{oi}}} \sigma (JS^n K) P^{ws}_{S^n J K,i} \nonumber \\
&=&\sum_{J\in G_{ms}} \sigma (J K) \frac{1}{\sqrt{N_{ms}}} \Gamma^{JK},
\end{eqnarray}
\begin{eqnarray}
\label{eq:mirrorD}
B^B_{a,i}&=&\frac{1}{\sqrt{N_{ms}}} \sum_{J\in G_{ms}, S^n\in G_S}
\sqrt{\frac{N_S}{S_{oi}}}  S_{S^n J K,i} \nonumber \\
&=&{\sqrt{N_{ms}}} B_{a,i},
\end{eqnarray}
where $N_{ms}$ is the order of the $G_{ms}$ and $\sigma(JK)$ are signs
that have to satisfy some consistency conditions:
\begin{equation}
\beta_{M J}=\beta_ M \beta_ J e^{\im 2 \pi X(M,J)},
\end{equation}
where
\begin{eqnarray}
\beta_J&=& \sigma(JK) \epsilon_J(K),\\
\epsilon_J(K)&=&e^{\im \pi(h_K-h_{JK})},\\
X(J^m,J^n)&=&-nm h_J.
\end{eqnarray}
Since the extension current has a non-integer spin, we have had to
generalize the constraints on the $\beta$'s by the
$X$ matrix which encodes data for making the simple current extension
\cite{thesis}. Details on the $X$ matrix and
its properties are found in the appendix. The $\beta$'s depend on $K$ by
the requirement that the $\sigma$'s are signs.
These constraints only uniquely determine the signs of $\sigma(MK)$
terms where $M=L^2$ for some $L$. The remaining
signs provide additional degrees of freedom in constructing alternative
crosscaps. The exact details of the mirror symmetry
map and the required signs depend on the particular model in question,
so the calculations
will be done in a case by case basis. An illustrative example is
presented in Appendix \ref{sec:sample-mirr-symm}.
It is evident that the charges are then directly computable once the
chiral labels and the $P$-matrix are known.
The $P$-matrix for the Gepner model has been computed and is presented in Appendix \ref{sec:p-matrix-entries}.
The listing of chiral labels, although simple is tedious and model dependent.
Given this information it is possible to compute the charges for any Gepner model. The
goal of this paper is to examine one dimensional Calabi-Yaus - tori. The specifics of
these geometries will be addressed shortly.

\section{Orientifold planes from Crosscap states in $c=3$
Gepners}
\label{sec:orient-plan-from}
\setcounter{equation}{0}

\subsection{A torus simplification}

In our study, $c=3$ Gepner models corresponding to $T^2$s, 
the low dimensionality of the torus results in an important simplification
in the computation of the charges. Geometrically this
simplification
occurs because the only non-trivial complex harmonic forms are of type
$h^{0,0}$, $h^{1,0}$, $h^{0,1}$ and $h^{1,1}$.
These
cohomology rings are in one-to-one correspondence with RR-ground
states of the CFT. Spectral flow which maps the $U(1)_R$ charges as,
$(q,\tilde{q})\rightarrow (q\pm c/6, \tilde{q} \pm c/6)$ in turn
relates this to the ring of (NS-NS) chiral primaries.
The torus has central charge $c=3$, so the maximum chiral charge is $q=c/3=1$.
For the Left-Right symmetric theory, corresponding to A-type states, one
makes the identification between the $(c,c)$ ring, the RR-ground
states and the non-trivial harmonic forms as (see e.g. \cite{Greene})
  \begin{eqnarray*}
  (0,0)_{NSNS} &\longrightarrow & (-1/2,-1/2)_{RR_{gs}}
  ~~~\Leftrightarrow ~~~h^{1,0},  \\
  (1,1)_{NSNS} &\longrightarrow& (1/2,1/2)_{RR_{gs}}~~~\Leftrightarrow
  ~~~ h^{0,1},
\end{eqnarray*}
where the spectral flow has signs $(-,-)$. However, it is possible to
identify
the chiral (0,0) state with both RR ground states by applying spectral
flow in the
opposite direction:
\begin{equation}
(1/2,1/2)_{RR_{gs}} \leftarrow (0,0)_{NSNS} \rightarrow
(-1/2,-1/2)_{RR_{gs}}.
\end{equation}
The left ground state is generated by spectral flow of (+,+). Flow of
type (-,-) is generated by
$S$, while the (+,+) is made by the conjugate field, $S^c$.
A representative element of $h^{1,0}$ is then given by $|S\rangle$. A
direct overlap
calculation using the crosscap expansion, (\ref{eq:Cr}), thus yields the
$dZ$ charge\footnote{$Q_s(K)$ is usually defined mod 1, however to make sense of
this definition one must use the true conformal weights of the fields.}
\bea
\label{eq:qz}
Q_z^{\[K\]}&=&\langle dZ|\Gamma\rangle^s_{\[K\]}=\frac{\sigma_1
\sqrt{N_S}}{2\sqrt{S_{O,O}}}e^{\im\pi Q_s(K)} P^{ws}_{O,K},\\
Q_{\bz}^{\[K\]}&=&\langle d\bZ|\Gamma\rangle^s_{\[K\]}=\frac{\sigma_1
\sqrt{N_S}}{2\sqrt{S_{O,O}}}e^{-\im\pi Q_s(K)} P^{ws}_{O,K},
\eea
while the second equation, corresponding to $d\bar{Z}$ charge, is given by the overlap
with $|S^c\rangle$.
Up to a phase this agrees with the expression for the tension,
($\ref{eq:mass}$). This should not be surprising,
as the central charge of the theory is given by the overlap with the top
form ($dZ$ for the torus), and
supersymmetry equates the mass and magnitude of the central charge for
BPS objects.

Using the D-brane expansion, (\ref{eq:D-expansion}), one can similarly
get the A-type D-brane charges
\beq
\label{eq:bqz}
Q^{brane}_z(K)=e^{2\pi \im Q_s(K)} \sqrt{NS_{0K}^{ws}}.
\eeq
In the case of the torus the central charge can be easily understood
geometrically. Consider a torus whose
lattice is made of two vectors of length $R_1$ and $R_2$ meeting at an
angle $\theta$. This can be
transformed to the standard torus lattice spanned by  $\{1, \tau\}\in\CC$,
where $\tau=e^{\im \theta}R_2/R_1$.
These lattice vectors define the $\hat{\alpha}$
and $\hat{\beta}$ cycles of the torus. Integration of the holomorphic top
form, $\Omega=d Z$,
with respect to these cycles yields the period vector of the torus:
\bea
\int_{\hat{\alpha}} \Omega=1, & \int_{\hat{\beta}} \Omega= \tau,
\eea
so that,
\beq
d Z = d X + \tau d Y,
\eeq
where $X, Y\in(0,1)$ describe two real axis aligned with the lattice.
At low energy the A-type states are related to special lagrangian
manifolds. In the torus these are straight line
loci, which can be accordingly decomposed in terms of $\hat{\alpha}$ and $\hat{\beta}$
cycles. Correspondingly the A-type orientifold/brane
in the torus $T^2$ is expected to be characterized by a middle homology class
$\Gamma=n \hat{\alpha}+ m \hat{\beta} \in H_1(T^2, \ZZ)$. This class can be described
according to its charge under the RR gauge fields corresponding to the NLSM elements
$\{dZ, d\bZ\}$:
\bea
Q_z=\bra{dZ} \Gamma \rangle =\int_\Gamma dZ= n+m\tau,\\
Q_{\bz}= \bra{d\bZ} \Gamma \rangle =\int_\Gamma d\bZ= n+m\btau.
\eea
These equations can be inverted to yield the cycle content,
\bea
n&=&-\frac{\btau Q_z- \tau Q_{\bz}}{2 \im \tau_2},\\
m&=&\frac{Q_z- Q_{\bz}}{2 \im \tau_2},
\label{eq:cycles}
\eea
which encode the central charge.

\bigskip

The B-type states are built from conjugate modular invariants where the
Left and Right sectors have opposite chiral charges.
Anti-symmetric spectral
flow of signs (+-) can then be used to relate the chiral-antichiral
primaries
with the vertical cohomology:
  \begin{eqnarray*}
  h^{1,1}:  (0,0)_{NSNS} &\longrightarrow & (-1/2,1/2)_{RR_{gs}},  \\
  h^{0,0}: (1,-1)_{NSNS} &\longrightarrow& (1/2,-1/2)_{RR_{gs}}.
\end{eqnarray*}
Like in the A-type case, one can also flow the NS-NS vacuum to
the $h^{0,0}$
RR ground state by using a flow with signs (-+)
\begin{equation}
(1/2,-1/2)_{RR_{gs}} \leftarrow (0,0)_{NSNS} \rightarrow
(-1/2,1/2)_{RR_{gs}}.
\end{equation}
Thus we see that the these charges are
the result of overlap with B-type primaries, $|S\rangle_B$ and
$|S^c\rangle_B$.
Equation (\ref{eq:mirrorC}) allows us to express the B-type crosscap
coefficients
in terms of the A-type, so we can obtain the B-charges as combinations
of the
A-charges:
\begin{equation}
\label{eq:charges_B}
Q_z^{\[K\],B}=\frac{1}{\sqrt{N_{ms}}}\sum_{J\in G_{ms}} \sigma(J,K)
Q_z^{\[JK\]}.
\end{equation}
Similarly, one can use the Mirror Symmetry extension of the boundary states,
(\ref{eq:mirrorD}), to get the B-type D-brane charges:
\begin{equation}
\label{eq:Dcharges_B}
Q_z^{B-brane}(K)=\sqrt{N_{ms}} Q_z(JK).
\end{equation}
We will apply these equations to the different Torus Gepner models to
obtain the charges
and tensions of the O-planes and, for comparison, D-branes.

\subsection{The $(1,1,1)$ Gepner Model}
\label{sec:1-1-1}

The $(1,1,1)$ Gepner model consists of the product of three level $k=1$
minimal models with a $D_{7,1}$ spacetime component.
The internal geometry can be described by the variety
\begin{equation}
  x_1^3+x_2^3+x_3^3=0,~~~~ x_i\in\CC\PP^2,
\end{equation}
which corresponds to a torus with $\tau=e^\frac{\im 2\pi}{3}\sim
e^\frac{\im \pi}{3}$. The first step in constructing crosscaps is
determining
the set of Klein bottle currents (KBCs), $K$. Recall that these are
simple currents in the theory $\cA^{ws}$ which satisfy,
$Q_S(K)=2 p/N_S \mbox{ mod } 1,\, p\in\ZZ$.  The number of available KBC
labels is therefore $N_S/2$, where $N_S $ is the order
of the group generated by $S$. However, there is a sign freedom
associated to each Klein bottle label yielding a total of $N_S$
different crosscaps. In this Gepner model $N_S=6$ and there are three
different KBCs. For simplicity we will
choose the KBCs representatives to be of the form,
$K=[[\vec{0};\vec{m};\vec{0}]o_D]$ which will be denoted as $(m_1,m_2,
m_3)$. Each $m_i$ is either zero or $\pm 2$.
Using the equations (\ref{eq:qz}) and (\ref{eq:mass}) we can construct a
table of O-plane charges and tensions. In Table \ref{table:k111}
the KBC label is listed in the first column, and the corresponding
tensions and charges are in the last two columns. The results in the table
are all normalized by the mass of the D-branes, which is the same for
all $K$ labels.
Note that the tension and charge have sign degrees of freedom,
$\sigma_0$ and $\sigma_1$. One of these can be determined
by the tadpole cancellation condition \cite{thesis}, leaving free the
relative sign. The second column lists the monodromy
charge of the Klein bottle current. This distinguishes different KBC
orbits and is related to the automorphism class of the
crosscap, $\sigma \epsilon_S(K)$ which is $e^{\im \pi Q_S(K)}$.

\begin{table}
\begin{center}
\begin{tabular}{|c|c|cc|cc|}
\hline\rule{0pt}{5mm}
K & $Q_s(K)$ & $M^{brane}$& $Q_z^{brane}$& $\sigma_0 M^{crosscap}$ &$\sigma_1 Q_z^{crosscap}$\\[3pt]
\hline
$(-2,2,2)$ & $1/3$ & 1 &$\tau$& 1 &$1+\tau$ \\
$(-2,-2,2)$ & $2/3$ & 1  &$-1-\tau$& 1 &$\tau$ \\
$(0,0,0)$ & $1$ & 1 & $1$& -1 &-1 \\
$(-2,2,0)$ & $1$ & 1 &$1$ &1& 1 \\
\hline
\end{tabular}
\caption[something]{{\em Summary of results for D-branes and O-planes in
(1,1,1) Gepner model. Results are normalized by $M^{brane}$ which is the same for all $K$.}}
\label{table:k111}
\end{center}
\end{table}

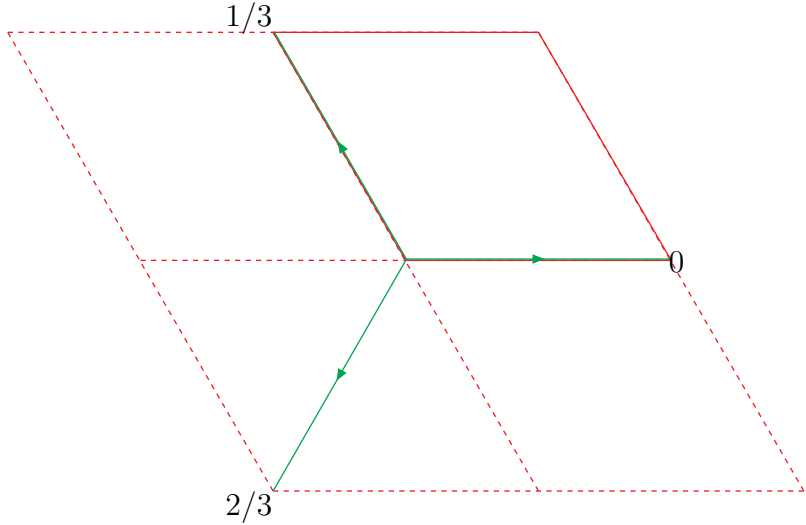
\begin{figure}[tbp]
\begin{center}
\begin{picture}(300,200)(0,0)
\SetColor{Red}
\Line(150,87)(250,87)
\Line(150,87)(100,173)
\Line(100,173)(200,173)
\Line(200,173)(250,87)

\DashLine(0,173)(200,173){2}
\DashLine(50,87)(250,87){2}
\DashLine(100,0)(300,0){2}
\DashLine(0,173)(100,0){2}
\DashLine(100,173)(200,0){2}
\DashLine(200,173)(300,0){2}
\SetColor{Green}
\ArrowLine(150,87.5)(250,87.5)
\ArrowLine(150.5,87)(100.5,173)
\ArrowLine(150,87)(100,0)

\SetColor{Black}
\Text(250,87)[l]{0}
\Text (100,173)[br]{1/3}
\Text (100,0)[tr]{2/3}

\end{picture}\\
\end{center}
\caption{{\em D-branes in $(1,1,1)$ Gepner model. The fundamental torus has
solid red outline, while the D-branes cycles appear as green arrows.}}
\label{fig:dbrane111}
\end{figure}

\begin{figure}[tpb]
\begin{center}
\begin{picture}(300,200)(0,0)
\SetColor{Red}
\Line(150,87)(250,87)
\Line(250,87)(200,173)
\Line(200,173)(100,173)
\Line(100,173)(150,87)

\SetColor{Dandelion}
\DashLine(150,87)(100,173){2}
\DashLine(150,87)(100,0){2}
\DashLine(100,173)(50,87){2}
\DashLine(50,87)(100,0){2}
\DashLine(100,0)(200,0){2}
\DashLine(200,0)(250,87){2}

\SetColor{Blue}
\ArrowLine(150,87.5)(250,87.5)
\ArrowLine(150.5,87)(100.5,173)
\ArrowLine(150,87)(100,0)

\SetColor{Black}
\Text(250,87)[l]{$\alpha=1$}
\Text (100,173)[br]{$\alpha=2\pi/3$}
\Text (100,0)[tr]{$\alpha=4\pi/3$}

\end{picture}
\end{center}
\caption{{\em Diagram of O-plane loci in the $(1,1,1)$ Gepner model. The
three different loci are automorphic and correspond to the
same involution class, $\alpha= 1\sim e^\frac{\im 2 \pi}{3}\sim
e^\frac{\im 4 \pi}{3} $. The fundamental torus is outlined by
solid red lines while the rotated (automorphic) ones are outlined with
orange hashes.}}
\label{fig:oplane111}
\end{figure}
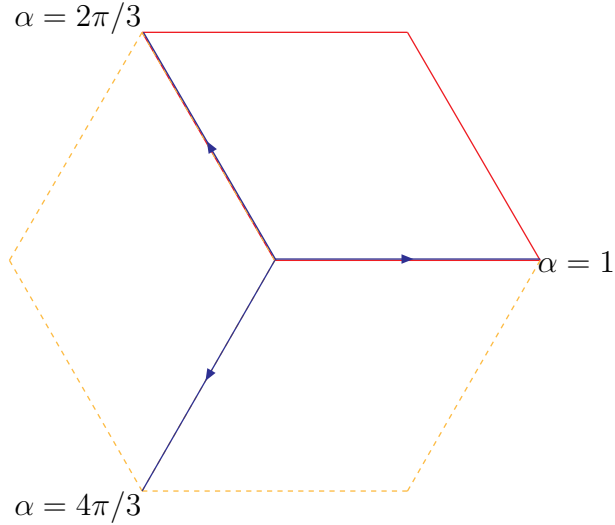

For comparison we can use equations (\ref{eq:bmass}) and (\ref{eq:bqz})
to also compute the data corresponding to boundary A-type states with
boundary labels equal to the KBCs. The boundary
automorphism class is given by $e^{\im 2 \pi Q_S(K)}$, where $2\pi
Q_S(K)$ has the nice geometric interpretation of being the angle which the
boundary locus makes with the torus lattice, as shown in Figure
\ref{fig:dbrane111}. Not surprisingly, this diagram is
compatible with the decomposition of the underlying Special Lagrangian
cycle approximation of the low energy D-brane in terms of the $\hat{\alpha}$
and $\hat{\beta}$
torus cycles (i.e. the 1 and $\tau$ coefficients in the D-brane charge).
The D-brane charges are dual to these homology elements and thus form a
lattice, which
in this case can be seen to be generated by $\{1, \tau\}$. Although the
O-planes are not expected to have a classical analogue and do not
obviously add to form a homology sublattice, we can formally combine their
charge vectors to span a sublattice of the D-brane charges. Examining the
O-plane charges we notice that this {\em pro forma}
charge lattice is also generated by
$\{1,\tau\}$.
Plotting the cycles dual to the O-plane charges yields
Figure \ref{fig:oplane111}, where it is seen that the
O-plane coincides with the fixed point loci of the involution Table
\ref{table:involutions}.
Both the D-brane and the O-plane diagram exhibit the $\ZZ_3\subset\ZZ_6$ rotational
symmetry of the
$\tau=e^\frac{\im\pi}{3}$ torus.
The (1,1,1) Gepner model only manifestly preserves this $\ZZ_3$ subgroup
rather than the full geometric $\ZZ_6$ toroidal symmetry.
This $\ZZ_3$ symmetry
maps the O-planes into each other. As a result, they are all equivalent
and correspond to the same involution class of Table \ref{table:involutions}
given by the parameter $\alpha=1 \sim e^\frac{\im 2\pi}{3} \sim
e^\frac{\im 4 \pi}{3}$. The Gepner $\ZZ_3$ symmetry generated by $g$
acts on this involution,
$I_\alpha$ parameter as $g I_\alpha g^{-1}=e^\frac{\im 2 \pi}{3}
\alpha$. 
This $\ZZ_3$ symmetry is displayed by the action on the periods:
\begin{eqnarray}
\label{eq:10}
1&\rightarrow& \tau,\non
\tau &\rightarrow& -1- \tau.
\end{eqnarray}
Note that the angle of inclination of a given O-plane is $\pi Q_S(K)$
with orientation given by the sign choice, $\sigma_1$ (i.e. a shift of
the angle by $\pi$), rather than by $2\pi Q_S(K)$ as in D-branes.

\bigskip
In order to study B-type states we use the mirror symmetry map, described by
equation (\ref{eq:ms}), which tells us that the group responsible for the mirror
map is $\ZZ_3\times\ZZ_3$ with generators $p_1p_2^2$ and $S^2$. To
obtain the B-type states we only need to extend $\cA^{ext}$ by the
mirror symmetry group
mod GSO, i.e. by the $\ZZ_3$ group generated by $J=p_1 p_2^2$. From
equation (\ref{eq:mirrorC}) it is apparent that one only needs to
obtain the
$\sigma(J,K)$ signs. This is dependent on the $K$ label. It so happens
that the signs generated by the mirror symmetry extension,
$\sigma(J,K)$
conspire with the $P^{ws}$ matrix signs to generated an overall scaling
of $\sqrt{3}$ for the crosscap, which is identical to the boundary state
scaling under mirror symmetry. Upon normalizing the results we get a charge table identical to
the A-type table, Table \ref{table:k111}. As expected this model is
self-mirror. The mirror to $K=(-2,2,2)$ is calculated as an illustrative
example in Appendix \ref{sec:sample-mirr-symm}.

\subsection{The $(1,4)$ Gepner Model}

The $(1,4)$ Gepner model corresponds to the product of a $k=1$ and a
$k=4$ minimal model with a $D_{7,1}$ spacetime component. The presence
of an even current in principle should alert us to fixed points in the
simple current extensions. We explained, however, that this is not
relevant for the computation of O-plane charges.

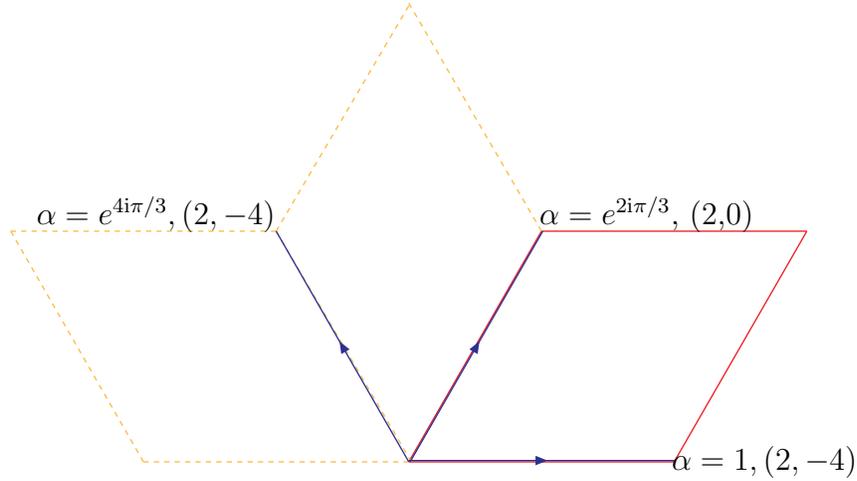
\begin{figure}[tbp]
\begin{center}
\begin{picture}(300,200)(0,0)
\SetColor{Red}
\Line(150,0)(250,0)
\Line(150,0)(200,87)
\Line(200,87)(300,87)
\Line(250,0)(300,87)

\SetColor{Dandelion}

\DashLine(200,87)(150,173){2}
\DashLine(150,172)(100,87){2}

\DashLine(50,0)(150,0){2}
\DashLine(50,0)(0,87){2}
\DashLine(0,87)(100,87){2}
\DashLine(100,87)(151,0){2}

\SetColor{Blue}
\ArrowLine(150,0.5)(250,0.5)
\ArrowLine(150.5,0)(200.5,87)
\ArrowLine(150,0)(100,87)
\SetColor{Black}
\Text(250,0)[l]{$\alpha=1, (2,-4)$}
\Text (200,87)[bl]{$\alpha=e^{2\im\pi/3}$, (2,0)}
\Text (100,87)[br]{$\alpha=e^{4\im\pi/3},(2,-4)$}
%\PText (50,87)(0)[r]{3/6}
%\PText (100,0)(0)[tr]{4/6}
%\PText (200,0)(0)[tl]{5/6}
\end{picture}\\
\end{center}
\caption{{\em O-planes for (1,4) Gepner model $\alpha=1 \sim e^\frac{\im 2
\pi}{3} \sim e^\frac{\im 4 \pi}{3}$ involution conjugacy.
The three different labeled cycles are equivalent under the automorphism of this
torus and correspond to different representative elements
of the same involution conjugacy class. The Gepner symmetry on the covering plane
is a rotation by $\pi/3$. This manifests itself as an ST action on the complex
structure. The fundamental torus is outlined with solid red lines. Hashed orange
lines are the rotated tori.}}
\label{fig:oplane14a}
\end{figure}

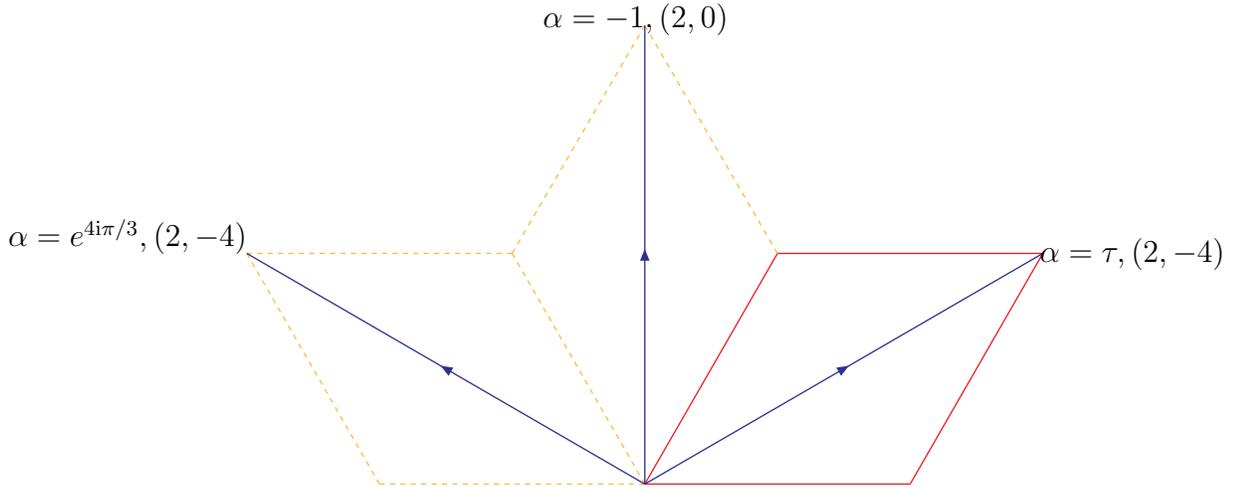
\begin{figure}[tbp]
\begin{center}
\begin{picture}(300,200)(0,0)
\SetColor{Red}
\Line(150,0)(250,0)
\Line(250,0)(300,87)
\Line(300,87)(200,87)
\Line(200,87)(150,0)

\SetColor{Dandelion}
\DashLine(200,87)(150,173){2}
\DashLine(150,173)(100,87){2}
\DashLine(100,87)(150,0){2}

\DashLine(150,0)(50,0){2}
\DashLine(50,0)(0,87){2}
\DashLine(0,87)(100,87){2}

\SetColor{Blue}
\ArrowLine(150,0)(300,87)
\ArrowLine(150,0)(150,173)
\ArrowLine(150,0)(0,87)
\SetColor{Black}
\Text(300,87)[l]{$\alpha=\tau, (2,-4)$}
\Text (147,170)[b]{$\alpha=-1, (2,0)$}
\Text (0,87)[br]{$\alpha=e^{4\im\pi/3},(2,-4)$}
%\PText (50,87)(0)[r]{3/6}
%\PText (100,0)(0)[tr]{4/6}
%\PText (200,0)(0)[tl]{5/6}
\end{picture}\\
\end{center}
\caption{{\em The (1,4) Gepner model O-planes for the $\alpha=-1 \sim
e^\frac{\im \pi}{3} \sim e^\frac{\im 5\pi}{3}$ involution
conjugacy class. The three different cycles are equivalent under the
automorphism of this torus. This Gepner symmetry acts as a rotation by
$\pi/3$ on the covering space. This rotation manifests itself as an ST
action on the complex structure. The
fundamental torus is outlined with solid red lines, while the rotated
(automorphic) ones are outlined with orange hashes.}}
\label{fig:oplane14b}
\end{figure}

The low energy underlying geometry of this model has a representation as
the solution set of
\begin{equation}
x_1^6+x_2^3+x_3^2=0,~~~~x_i\in W\CC\PP_{1,2,3},
\end{equation}
which again corresponds to a torus with $\tau=e^{\frac{\im \pi}{3}}$.
Although this torus is $PSL(2,\ZZ)$ equivalent to that of the $(1,1,1)$
Gepner
model the symmetries of the (1,4) Gepner model are richer, spanning the
full geometric $\ZZ_6$ . To classify the KBCs we note that the
spectral flow group has order $N_S=12$, so we can choose six KBCs as
representatives
of the possible crosscaps. As before, one can use the equations
(\ref{eq:mass}) and (\ref{eq:qz}) to compute a table, Table
\ref{table:k14}, of tensions
and charges for the A-type states. The charges are normalized by the
D-brane mass, which is independent of $K$.
  As in the previous case one can interpret the charges in terms of the
$\hat{\alpha}$
and $\hat{\beta}$ torus cycles. When diagrammed (see Figures
\ref{fig:oplane14a} and \ref{fig:oplane14b}) on the torus lattice the
crosscap
cycles are seen to correspond to the fixed point loci listed  in Table
\ref{table:involutions}. As emphasized, the richer Gepner symmetry, $\ZZ_6$,
now allows for the two distinct involution classes listed in Table
\ref{table:involutions}.
We can obtain all involution classes by $g_6^{n}I_\alpha
g_6^{-n}\rightarrow e^\frac{\im n}{3} \alpha$.
The O-planes transform as $\ZZ_6$ orbits whose action on the periods is
\begin{eqnarray}
1&\rightarrow& \tau,\non
\tau &\rightarrow& \tau-1 .
\end{eqnarray}
Naturally there are two separate O-plane orbits depending on the
involution class. For the involution class
$\alpha=1\sim e^\frac{\im 2 \pi}{3} \sim e^\frac{\im 4 \pi}{3}$ the
O-planes are shown in Figure \ref{fig:oplane14a}.
Those compatible with the second involution class, $\alpha=e^\frac{\im
\pi}{3} \sim -1  \sim e^\frac{\im 5 \pi}{3}$ are diagrammed
in Figure \ref{fig:oplane14b}. The two involution classes produce two
separate charge lattices:
\begin{itemize}
\item{} $ \alpha=1\sim e^\frac{\im 2 \pi}{3} \sim e^\frac{\im 4 \pi}{3} \mbox{
generated by }\{\tau+1, 2\tau-1\}$,
\item{}  $\alpha=e^\frac{\im \pi}{3} \sim -1  \sim e^\frac{\im 5 \pi}{3} \mbox{
generated by }\{1, \tau\},\mbox{ like the Gepner (1,1,1) Model}$.
\end{itemize}
Let us re-emphasize that for O-planes this charge lattice is pro forma
only; see the comment above eq. (\ref{eq:10}). 

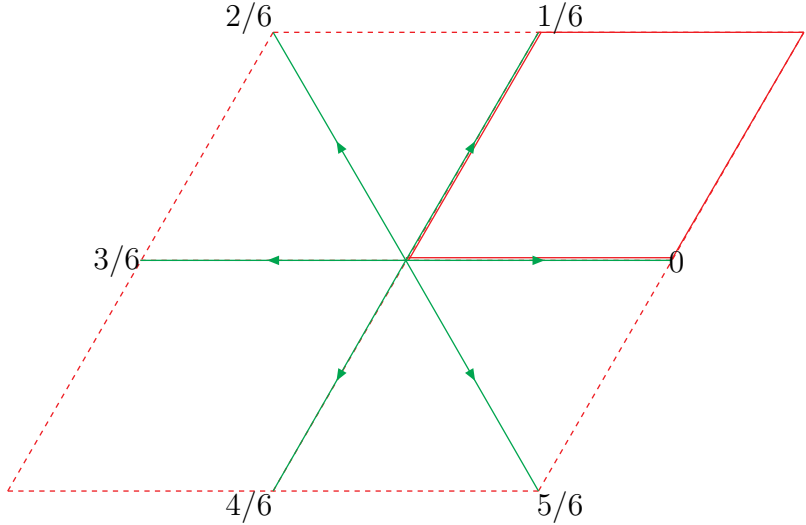
\begin{figure}[tbp]
\begin{center}
\begin{picture}(300,200)(0,0)
\SetColor{Red}
\DashLine(0,0)(200,0){2}
\DashLine(0,0)(100,173){2}
\DashLine(100,173)(300,173){2}
\DashLine(200,0)(300,173){2}
\DashLine(100,0)(200,173){2}
\DashLine(50,87)(250,87){2}
\Line(150,88)(250,88)
\Line(151,87)(201,173)
\Line(200,173)(300,173)
\Line(250,87)(300,173)
\SetColor{Green}
\ArrowLine(150,87)(250,87)
\ArrowLine(150,87)(200,173)
\ArrowLine(150,87)(100,173)
\ArrowLine(150,87)(50,87)
\ArrowLine(150,87)(100,0)
\ArrowLine(150,87)(200,0)
\SetColor{Black}
\Text(250,87)[l]{0}
\Text (200,173)[bl]{1/6}
\Text (100,173)[br]{2/6}
\Text (50,87)[r]{3/6}
\Text (100,0)[tr]{4/6}
\Text (200,0)[tl]{5/6}
\end{picture}\\
\end{center}
\caption{{\em D-branes for the (1,4) Gepner model seen as straight lines in
the covering space.
Dashed delimit tori domains. Arrows mark the D-branes, which are
labeled by $Q_s(K)$.
The fundamental torus is outlined with solid red lines.}}
\label{fig:dbrane14}
\end{figure}

For comparison we can repeat the analysis for A-type D-branes with
boundary label $K$ by reading of the tensions and charges from
equations (\ref{eq:bmass}) and (\ref{eq:bqz}). The results for these
boundary states are included in the first two columns of
Table \ref{table:k14}. Figure \ref{fig:dbrane14} displays these D-branes
as straight lines in the covering space of the torus.
As expected the generators of the D-brane charge lattice are $\{1,\tau\}$.

\begin{table}
\begin{center}
\begin{tabular}{|cc|c|cc|cc|}
\hline \rule{0pt}{5mm}  & K & $Q_S(K)$ & $M^{brane}$ & $Q^{brane}_z$ & $\sigma_0 M^{crosscap}$ & $\sigma_1 Q^{crosscap}_z$ \\[3pt]
\hline
  & $(2,-2)$ & 1/6 & 1 & $\tau$ & $\sqrt{3}$ & 1+$\tau$ \\
  & $(2,0)$ & 1/3 & 1 & $\tau$-1 & -1 & $-\tau$ \\
  & $(2,2)$ & 1/2 & 1 & -1 & $\sqrt{3}$ & -2$\tau$+1 \\
  & $(2,4)$ & 2/3 & 1 & -$\tau$ & 1 & $-\tau$+1 \\
  & $(2,6)$ & 5/6 & 1 & -$\tau$+1 & $\sqrt{3}$ & 2-$\tau$ \\
  & $(2,-4)$ & 0 & 1 & 1 & 1 & 1 \\
\hline
\end{tabular}
\caption[$(1,4)$ Gepner model brane and plane charges]{{\em Gepner (1,4)
Model Charges and Tensions for D-branes and O-planes labeled by $K$ and
normalized by the D-brane tension, which is constant for all $K$.}}
\label{table:k14}
\end{center}
\end{table}
\bigskip
An other example of the richness of the symmetries of the $(1,4)$ Gepner
model is that the mirror symmetry group is
actually a subgroup of the GSO simple current extension group. Hence,
this torus is manifestly self mirror. This
is apparent from the CFT side  as the GSO extended theory is
self conjugate (i.e. the conjugation matrix is the diagonal).
As a result the table of A-type charges and tensions (Table
\ref{table:k14}) is also a valid summary for
B-type states.

\subsection{The $(2,2)$ Gepner Model}

The (2,2) Gepner model is generated by tensoring two level $k=2$ Minimal
models with a $D_{7,1}$ spacetime part. It has the algebro-geometric
description
\begin{equation}
x_1^4+x_2^4+x_3^2=0,~~~~x_i\in W\CC\PP_{1,1,2}
\end{equation}
and corresponds to a torus with $\tau=\im$. The GSO extension is by a
group of order $N_S=8$ so we need four Klein bottle currents.
Proceeding as in the previous case we can use (\ref{eq:mass}) and
(\ref{eq:qz}) to compile a table of charges and tensions for
the crosscaps of this model (see Table \ref{table:k22}, note that the
data is normalized by the D-brane mass, which is constant).
Some new features arise when examining these results. First we can
see that the O-planes indexed by KBC (0,0) and (0,4) have no mass nor
charge. These are real states with physical interactions,
as the coupling to other fields (given by other P-matrix elements) are
not identically zero. They seem to correspond to orientifolds
without O-planes - as these are states which have no mass or charge
singularity. Interestingly enough, our involution classification
for this torus, Table \ref{table:involutions}, has an involution
conjugacy class $\alpha=1\sim -1$ with $\beta=1/2$ and $s=0$, i.e.
with no fixed points. That involutions with a 'shift' generate
orientifolds without O-planes is in fact well-known.\footnote{We thank
  C. Bachas for pointing this out.}

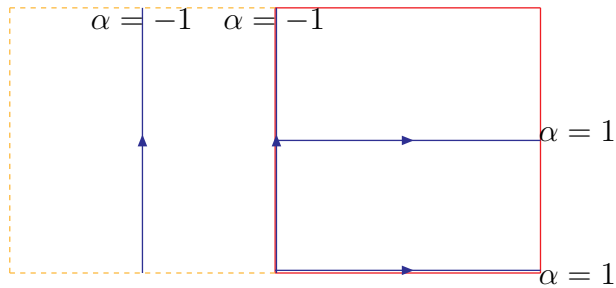
\begin{figure}[tbp]
\begin{center}
\begin{picture}(300,200)(0,0)
\SetColor{Red}
\Line(100,100)(200,100)
\Line(200,100)(200,0)
\Line(200,0)(100,0)
\Line(100,0)(100,100)

\SetColor{Dandelion}
\DashLine(100,100)(0,100){2}
\DashLine(0,100)(0,0){2}
\DashLine(0,0)(100,0){2}

\SetColor{Blue}

\ArrowLine(100,1)(200,1)
\ArrowLine(100,50)(200,50)
\ArrowLine(100.5,0)(100.5,100)
\ArrowLine(50,0)(50,100)

\SetColor{Black}
\Text(200,0)[l]{$\alpha=1$}
\Text(200,50)[lb]{$\alpha=1$}
\Text(100,100)[ct]{$\alpha=-1$}
\Text(50,100)[t]{$\alpha=-1$}
\end{picture}
\end{center}
\caption{{\em Gepner $(2,2)$ O-planes seen as straight lines. There are two
homologous cycles in the $\alpha=1 \sim -1$
involution class. The fundamental torus is outlined by solid red lines,
while the rotated automorphic
ones are outlined by hashed orange lines.} }
\label{fig:oplane22s2}
\end{figure}

\begin{figure}[tbp]
\begin{center}
\begin{picture}(300,200)(0,0)

\SetColor{Red}
\Line(100,100)(200,100)
\Line(200,100)(200,0)
\Line(200,0)(100,0)
\Line(100,0)(100,100)

\SetColor{Dandelion}
\DashLine(100,100)(0,100){2}
\DashLine(0,100)(0,0){2}
\DashLine(0,0)(100,0){2}

\SetColor{Blue}

\ArrowLine(100,0)(200,100)
\ArrowLine(100,0)(0,100)

\SetColor{Black}
\Text(200,100)[l]{$a=\im$}
\Text(0,100)[r]{$a=-\im$}

\end{picture}
\end{center}
\caption{{\em Gepner $(2,2)$ model O-planes seen as straight lines for the
$a=\im \sim -\im$ class involution. The fundamental torus is outlined
by solid red lines while the rotated automorphic one is outline by
hashed orange lines.}}
\label{fig:oplane22s1}
\end{figure}
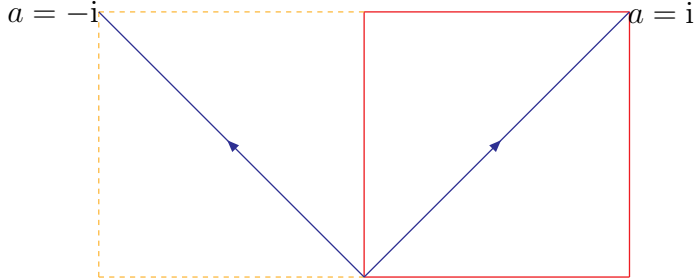

According to the involution table there should be two more involution
classes. This Gepner model manifestly has the full geometric symmetry,
$\ZZ_4$,
so we do expect to find O-planes corresponding to both involution
classes. The action on the periods is:
\bea
1 &\longrightarrow & \tau,\non
\tau &\longrightarrow & -1.
\eea
The $\alpha=\im \sim -\im$ involution can be seen to correspond to O-planes along
the diagonal of the torus, as diagrammed in Figure \ref{fig:oplane22s1}.
As in the previous cases, the different diagonal O-planes are related to
each other by the Gepner symmetry ($\ZZ_4$). Another peculiarity of
this model is that the other involution class, $\alpha= 1 \sim -1$, has
species $s=2$. This is manifest in the O-plane diagram Fig. \ref{fig:oplane22s2}
where we can see that this involution results in two disconnected
homologous loci. As before there are several different O-planes which
fill a Gepner
$\ZZ_4$ multiplet. When examining the formal sublattice generated by O-plane
charge vectors 
we come
upon 
a surprise. The lattice is now generated by $\{2, 2 \tau\}$
(which
is a proper sublattic of $\{1+\tau, \tau -1\}$):
\begin{itemize}
\item{}  $\alpha=\im \sim -\im  \mbox{ generated by }\{1+\tau, 1-\tau\}$,
\item{}  $\alpha=1\sim -1  \mbox{ generated by }\{2, 2\tau\}, \mbox{ (actually
a proper sublattice of the }\alpha=\im\mbox{ case})$.
\end{itemize}

\begin{figure}[tpb]
\begin{center}
\begin{picture}(300,200)(0,0)
\SetColor{Red}
\Line(100,100)(200,100)
\Line(200,100)(200,200)
\Line(200,200)(100,200)
\Line(100,200)(100,100)

\DashLine(0,0)(200,0){2}
\DashLine(200,0)(200,200){2}
\DashLine(200,200)(0,200){2}
\DashLine(0,200)(0,0){2}
\DashLine(0,100)(200,100){2}
\DashLine(100,0)(100,200){2}

\SetColor{Green}

\ArrowLine(100,101)(200,101)
\ArrowLine(101,100)(101,200)
\ArrowLine(100,100)(0,100)
\ArrowLine(100,100)(100,0)

\SetColor{Black}
\Text(200,100)[l]{0}
\Text(100,200)[b]{1/4}
\Text(0,100)[r]{2/4}
\Text(100,0)[t]{3/4}

\end{picture}
\end{center}
\caption{{\em Gepner $(2,2)$ model D-Branes seen as straight lines in the
torus covering space. The fundamental torus is outlined in solid red lines.}}
\label{fig:dbrane22}
\end{figure}
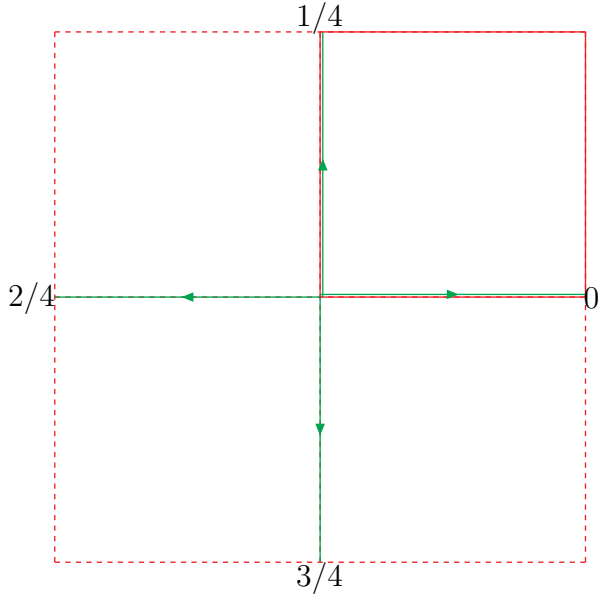

For comparison the D-brane results are listed in the same table, Table
\ref{table:k22}. The D-brane charge lattice is generated by $\{1, \tau \}$.
The D-branes are diagrammed in Figure \ref{fig:dbrane22}.

\begin{table}
\begin{center}
\begin{tabular}{|cc|c|cc|cc|}
 \hline \rule{0pt}{5mm} & K & $Q_s(K)$ & $M^{brane}$ & $Q_z^{brane}$ & $\sigma_0 M^{crosscap}$ & $\sigma_1 Q_z^{crosscap}$ \\[3pt]
\hline
  & $(-2,-2)$ & 1/2 & 1 & -1 & 2 & $2\tau$ \\
  & $(-2,2)$ & 0 & 1 & 1 & 2 & 2 \\
  & $(-2,4)$ & 1/4 & 1 & $\tau$ & $\sqrt{2}$ & $1+\tau$ \\
  & $(2,4)$ & -1/4 & 1 & -$\tau$ & $\sqrt{2}$ & $1-\tau$ \\
  & $(0,0)$ & 0 & 1 & 1 & 0 & 0 \\
  & $(0,4)$ & 1/2 & 1 & -1 & 0 & 0 \\
\hline
\end{tabular}
\caption[$(2,2)$ charges]{{\em Gepner $(2,2)$ Model charges and tensions for
O-planes and D-branes labeled by $K$. The table is normalized by the D-brane masses, which are independent of $K$.}}
\label{table:k22}
\end{center}
\end{table}

\bigskip

One can generate the geometric mirror theory by a simple current
extension with $J=p_1p_2^3$ mod $S$, which is order 2 \cite{walcher}. As
this mirror symmetry extension is even 
the signs cannot be determined
completely resulting in an extra sign degree of freedom, $\eta$, in the mirror
charge description:
\begin{eqnarray}
Q^{\[K\]B,crosscap}_z&=&\frac{1}{\sqrt{2}} (Q^{\[K\]}_z+\eta Q^{\[J K\]}_z), \\
Q^{B,brane}_z(K)&=&\sqrt{2}Q_z(K).
\end{eqnarray}
The results for the B-side are presented in Table \ref{table:mirrork22},
where a + or - subscript to the $K$ label
distinguish the sign choice for $\eta$. Note that the results in the
table are normalized by the mass of the D-brane - which
is constant for all $K$. There are two possible charge lattices for the
mirror side:
\begin{itemize}
\item{} the lattice generated by $\{1+\tau, 1-\tau\}$,
\item{} and the lattice generated by $\{1,\tau\}$.
\end{itemize}
The mirror D-brane lattice is generated, as usual, by $\{1,\tau\}$.

\begin{table}
\begin{center}
\begin{tabular}{|c c| c c c c|}
\hline \rule{0pt}{5mm}
  & K  & $M^{B,brane}$ & $Q_z^{B,brane}$ & $\sigma_0 M^{B,crosscap}$ & $\sigma_1 Q_z^{B,crosscap}$ \\[3pt]
\hline
  & $(-2,-2)_\pm$& 1 & -1 & 1 & $\tau$ \\
  & $(-2,2)_\pm$ & 1 & 1 & 1 & 1 \\
  \hline
  & $(-2,4)_+$ & 1 & $\tau$ & $\sqrt{2}$ & 1+$\tau$ \\
  & $(-2,4)_-$ & 1 & $\tau$ & $0$ & 0 \\
  \hline
  & $(2,4)_+$  & 1 & -$\tau$ & $\sqrt{2}$ & 1-$\tau$ \\
  & $(2,4)_-$  & 1 & -$\tau$ & 0 & 0 \\
\hline
\end{tabular}
\caption[$(2,2)$ charges]{{\em Gepner $(2,2)$ Model charges and tensions for
B-type O-planes and D-branes labeled by $K$. The entries are normalized by
the brane mass which is independent of $K$.}}
\label{table:mirrork22}
\end{center}
\end{table}

\subsection{Summary}

From studying these examples we can see that there is a difference in
presentation of the charge lattices of the orientifolds for the different models.
We expect that this will impose restrictions on the moduli space of the orientifolds.
This is clearly different from the D-brane case where the charge lattice has the same generators
throughout. What is particularly interesting is how the $\tau=\im$
torus (the (2,2) Gepner model) has a O-plane charge vectors form a
basis of a
sublattice of index 2. 
This is related to this torus having species $s=2$
which is topologically disconnected from other species of orientifolds. Particularly, this lattice
disparity between the $\tau=\im$ and the $\tau=e^\frac{\im \pi}{3}$ demonstrates that
the orientifold theories with Kalb-Ramond anti-symmetric fields valued at $B=1/2$ and $B=0$
exist in separate branches in moduli space. In section \ref{sec:disc-outl-towards},
we will explore constructions of toroidal families where these discontinuities in species
will require the degeneration of the cycles where the O-planes lie.

\section{Free Field Formulation}
\setcounter{equation}{0}

The torus, the simplest Calabi-Yau manifold, is special in that it has
a constant flat  metric.
This fact allows us to explicitly describe the
physics at all points in moduli space using the worldsheet sigma model. We
will use the NSR
free field formulation to obtain the charges and tensions of the O-planes
and D-branes and compare these results with the data garnered from the CFT
perspective. Much of what follows is based on the work of Di Vecchia et
al, (eg see \cite{DiVecchia1, DiVecchia2} and references therein) and
Walcher \cite{walcher}. We could also have used the GLSM description
of the torus. However, the general connection between this description,
and orientifolds at large volume is already addressed in
\cite{Brunner:2004zd}. As far as CY-spaces are concerned, $T^2$s are special in that we can also use the free-field
formulation.

The worldsheet action of a string compactification on a two
dimensional torus is 
\begin{equation}
S=-\frac{1}{4 \pi\alpha'}\int_\Sigma \left((G_{\alpha \beta}\eta^{a b}+
  B_ {\alpha \beta}\epsilon ^{a b}) \partial_a X^\alpha\partial_b X^\beta
  -\im G_{\alpha\beta}\bar{\psi}^\alpha\rho^c\partial_c\psi^\beta
  -\im B_{\alpha\beta}\bar{\psi}^\alpha\rho^3\rho^c\partial_c\psi^\beta
  \right)d^2\sigma.
\label{eq:Taction}
\end{equation}
We are using the notational conventions of GSW for the fermions, with $\rho$ being the two dimensional
Dirac matrices with $\{\rho^a, \rho^b\}=2\eta^{ab}$.
The worldsheet metric is Minkowski with a negative time component and $\epsilon^{01}=1$.
The spacetime variables, $X^\alpha$, are real valued and periodically identified with period 1.
The $T^2$ target space metric is \cite{walcher}
\bea G_{\alpha\beta} = \left( \begin{array}{cc}
R_1 ^2 & R_1 R_2\cos \alpha\\
R_1 R_2 \cos \alpha & R_2 ^2 \end{array} \right).
\label{eq:Tmetric}
\eea
This corresponds to a torus with periods of length $R_1$ and $R_2$
meeting at an angle $\alpha$, with complex structure $\tau=R_2 e^{\im \alpha}/R_1$.
Additionally there is a constant 
antisymmetric background field, $B_{12}=-B_{21}=B$.
The presence of the $B$ field can be absorbed as a total derivative in the
action. This does not affect the Lagrangian locally, so the equations of
motion remain the same. However, it does have global or topological effects.
 From the point of view of the open string, this modifies the boundary
conditions, while the closed string sees this as modification of the
zero modes.

The variation of this action results in the open string boundary conditions
\beq
\delta X^J(G_{IJ}\partial_\sigma X^I+\cB_{IJ}\partial_\tau
X^I)|_{\sigma=0,\pi}=0~,~~~~~~\cB=B+2\pi\alpha'F~,
\label{eq:boundary_terms}
\eeq
for the compact directions, $I=1,2$ (and similarly for the fermionic
variables). 
Setting $\delta X^I=0$  corresponds to Dirichlet boundary
conditions. A non-zero $\cB_{12}\neq 0$ only has consequences when both $\delta X^1\neq0$ {\em and}
$\delta X^2\neq 0$, i.e. for D2 branes or O2 planes, as the
antisymmetry of $\cB$ 
otherwise ensures that the term in parenthesis reduces to the standard
Neumann term. 
 This directly implies that the
$\cB$ field has no effect for D1, O1, D0 or O0 branes and planes.

Using world sheet duality, these open string 
boundary conditions define a state in the Hilbert
space of the closed string. On the boundary state, $|B\rangle$,
the boundary conditions become constraints. Expressed in terms of the
standard oscillator expansions, they are
\bea
\matrix{ (\alpha^\mu_n+ R^\mu_{~\nu}\tilde{\alpha}^\mu_{-n})|B\rangle=0, \cr
(\psi_r^\mu-\im \eta
R^\mu_{~\nu}\tilde{\psi}_{-r}^\nu)|B,\eta\rangle=0,} 
&& \left\{\matrix{ {\rm NS:} & r \in \ZZ+\hlf \cr {\rm R:} & r \in \ZZ} \right. ~.
\label{eq:boundary}
\eea
with
\begin{eqnarray}
  \label{eq:11}
  R = \left(\matrix{ R_{(N)} & 0 \cr 0 & R_{(D)}}\right)~,~
\left\{\matrix{ R_{(N)}= (G-\cB)^{-1}(G+\cB)~,&~ {\rm Neumann
    ~directions} \cr R_{(D)} =-\one~,&~{\rm Dirichlet ~directions}}\right.
\end{eqnarray}
Here $\alpha_n^\mu$ is the {\em n}th mode in the $\mu$ direction of the
bosonic coordinate $X^\mu$. The $\psi^\mu_r$'s are the oscillator
modes of the fermionic coordinates, whose boundary conditions can be
obtained by supersymmetry from the bosonic ones. 
The GSO projection will sum over the
two spin structures $\eta =\pm 1$. All the boundary information is
thus really encoded in the matrix $R$
which is derived from the choice of boundary conditions imposed on
(\ref{eq:boundary_terms}).

We can define the crosscap by an equation analogous to (\ref{eq:boundary}),
\beq
(\alpha^\mu_n+(-1)^n R^{\mu}_{~\nu}\tilde{\alpha}^\nu_{-n}|C\rangle=0\\
\label{eq:crosscapboson}
\eeq
\beq
(\psi_n^\mu-\im \eta (-1)^n R^{\mu}_{~\nu}\tilde{\psi}_{-n}^\nu)|C,\eta\rangle=0.
\label{eq:crosscapfermion}
\eeq
In this case the additional $(-1)^n$ factor accounts for the
parity action on the crosscap, $\sigma\rightarrow\pi-\sigma$.
The boundary and crosscap state constraints, (\ref{eq:boundary}),
(\ref{eq:crosscapboson}) and (\ref{eq:crosscapfermion}) are very similar.
For the purposes of our study we are interested in the charges of these
states with respect to the Ramond Ramond ground states. As a result we really
only need the solutions to these constraint equations for the zero modes
($n=0$) of the RR sector.  Now, note that the zero modes of both the crosscap and the
boundary state satisfy identical defining equations. This implies that
the crosscap charges should be a subset of the possible D-brane charges.
We will use this to think of the crosscap charges as generated by
a set of D-branes (with the correct charges) placed at the point of involution of
the spacetime theory. Consequently this short study applies equally well to
O-planes and D-branes, which we will occasionally denote collectively as defect states, $|D\rangle.$

For the torus these boundary and crosscap expressions, (\ref{eq:boundary}), (\ref{eq:crosscapboson}) and
(\ref{eq:crosscapfermion}) can be explicitly solved yielding these states
as coherent sums of oscillators. In the Ramond-Ramond sector the fermion crosscap equation
(\ref{eq:crosscapfermion}) is solved by
\beq
|C\rangle'=-\exp \left[\im\eta \sum_{n=1}^\infty(-1)^n
\psi^\mu_n R_{\mu\nu}\tilde{\psi}^\nu_{-n}\right]|C,\eta\rangle_0,
\eeq
where the zero mode factor $|C,\eta\rangle_0$ must obey
\beq
(\psi_0^\mu-\im \eta R^{\mu}_{~\nu}\tilde{\psi}_0^\nu)|C,\eta\rangle_0=0.
\label{eq:crosscapfermionzero}
\eeq
Following Di Vecchia et al. \cite{DiVecchia} (particularly Appendix A), we
can reformulate this requirement in terms of a matrix
\beq
|C,\eta\rangle_0=M_{A\tilde{B}}|A\rangle|\tilde{B}\rangle,
\eeq
where the $|A\rangle$, and $|\tilde{B}\rangle$ are basis ground state kets 
transforming under two copies the 32 dimensional SO(1,9) spinor representation 
of the Ramond oscillator zero modes. Then equation
(\ref{eq:crosscapfermion}) applied to the zero modes, reduces to
\beq
     (\Gamma^\mu)^TM-i\eta(\Gamma_{11})^TM R^{\mu}_{~\nu} \Gamma^\nu=0,
\label{eq:RRgs}
\eeq
where the fermion zero mode action on the RR ground states is
represented by two copies of the $32\times 32$ $\Gamma$-matrices of
SO(1,9): $\psi^\mu_0 = \Gam^{\mu}\otimes \one,~\tilde{\psi}^{\mu}_0 =
\Gam_{11}\otimes \Gam^{\mu}$~\cite{DiVecchia}. One finds that the
solution to $M$ depends on the number of positive and negative
eigenvalues of $R$, corresponding to the number of Neumann and
Dirichlet directions respectively. For $D-p$ Dirichlet directions in
$\cB=0$ Minkowski space, $M$
equals
\begin{eqnarray}
  \label{eq:12}
  M = C\Gam^0\Gam^1\ldots\Gam^p\frac{1-i\eta\Gam_{11}}{1-i\eta} 
\end{eqnarray}
with $C$ the charge conjugation matrix.

Finally the GSO projection in the Ramond sector amounts to averaging over the two
$\eta=\pm$ spin structures resulting in,
\beq
|C\rangle_R=\frac{1}{2}(|C,+\rangle_R+|C,-\rangle_R).
\eeq
The full fermionic Linear Sigma Model Crosscap is thus
     \beq
     |C\rangle_R=\cos{[\sum_{r=1}^\infty(-1)^r \psi_{-r}\cdot
     R \cdot \tilde{\psi}_{-r}]}
|a\rangle M_{a\tilde{b}}|\tilde{b}\rangle+
\sin{[\sum_{r=1}^\infty(-1)^r \psi_{-r}\cdot
     R \cdot \tilde{\psi}_{-r}]} |a'\rangle
     M_{a'\tilde{b}'}|\tilde{b}'\rangle.
     \eeq
where the spin basis ${A,\tilde{B}}$ has been broken into chiral
blocks ${a,\tilde{b},a',\tilde{b}'}$ after the GSO projection.
Of course, since we are interested in measuring the overlap with the
RR ground state zero modes, only
the zero mode part encoded in the matrix $M_{a\tilde{b}}$ will be relevant for our study.

Similar to the RR-charges of the O-plane, one can ask about their
tensions. We will use the shortcut that the tension can be directly
found from the Born-Infeld action, rather than using
the NS-NS-sector defect state. It is of course factually the same computation \cite{Schmidhuber:1996fy}.

\subsection{A-planes: O1}

As mentioned earlier under equation (\ref{eq:boundary_terms}) the $\cB$-field only
affects the boundary conditions when none of the directions are Dirichlet.
In this section we will examine the behavior when one of the directions satisfies
Dirichlet and the other Neumann boundary conditions. From the point of
view of the two torus such a brane wraps a middle-dimensional
cycle. Such cycles must corresponds to a special Lagrangian submanifold.
As they are one dimensional they appear as lines in the torus covering space. The special
Lagrangian property means that they have minimal volume - so that they are straight
lines. Different submanifolds can thus be parametrized by the angle, $\theta$, that they
make with the $R_1$ axis. Generally one can also shift the position of the cycle, but
since we want to apply the results to O-planes we will assume that the cycle starts at the
origin of the lattice. Using the argument that in the Linear Sigma Model the O-plane
charges can be thought of as a stack of D-branes, we can start our O1-plane study by
examining D1-branes at different angles $\th$. O-planes are thus also
expected to 'wrap' a special Lagrangian submanifold. Indeed the most well known way to construct
SLAGs is as the fixed point of an antiholomorphic involution. However, the O-planes
only exist for those SLAGs which truly are the fixed points of involutions.

As discussed by Walcher \cite{walcher} the angle $\theta$ with which
D-branes wrap is not arbitrary, but has to be compatible with the
description of the torus in the $\RR^2$ covering plane. In other
words
the cycle, $\hat{\gamma}$, upon which the brane wraps has to close upon
itself.
This implies that there exist integers $n,m$ such that
\beq
\tan\theta=\frac{n R_2 \sin\alpha}{m R_1+n R_2 \cos\alpha},
\label{eq:theta_quantisation}
\eeq
which corresponds to $\hat{\gamma}$ wrapping the $R_1$ cycle $m$ times
and the $R_2$ cycle $n$ times. This cycle can be described by a
displacement
vector, $L^I=(m,n)$, of length,
$|L|^2=L\cdot G\cdot L=(mR_1)^2+(n R_2)^2+2 m n R_1 R_2 \cos\alp$.
We will see that these $(m,n)$ are the charges of the D-brane.

The $R$ matrix for these one dimensional rotated defect states is:
\begin{eqnarray}
  \label{eq:15}
  R_{IJ} = \left(\matrix{ R_1^2\cos(2\th) & R_1R_2\cos(2\th-\alp) \cr
                          R_1R_2\cos(2\th-\alp) & R_2^2\cos(2\th-2\alp)}\right)~.
\end{eqnarray}
The easiest way to obtain this matrix is by starting in an orthogonal frame $x^i$. A suitable change of basis is given
by,  $
e_I^{\,\,\,i}= \left( \begin{array}{cc} R_1 & 0 \\ R_2\cos\alp
    &R_2\sin\alp
\end{array} \right)$, where the orthogonal coordinates are lowercase
and the indices likewise lowercase roman letters. One can then start out in an orthogonal frame where
$x^1, x^2$ are chosen to satisfy Neumann and Dirichlet boundary
conditions, i.e. $R_{11}=-R_{22}=1~,R_{12}=0$, rotate to obtain a generic
inclined defect state, and then finally change basis to the natural
torus coordinates. The choice of boundary conditions encoded in the
$R$ matrix identifies the defect state as being A-type. The
(left-moving) $U(1)$
charge is measured by the operator
$J=\im\eps_{\alp\bet}\psi^{\alp}\psi^{\bet}$ with $\eps_{\alp\bet}$
the Levi-Civita tensor of the spacetime torus. Its eigenvalue
on the
defect state returns equals minus the product of $R$ eigenvalues. Hence, one has $J|D\rangle=\bar{J}|D\rangle$, identifying the boundary state
as being A-type.

The corresponding RR ground state solution to equation (\ref{eq:RRgs})
is thus given by\footnote{We are implicitly choosing the remaining 7
  spatial directions to be Dirichlet.}
\beq M_{ab}=C\Gamma^0(\Gamma^1 R_1 \cos\theta+R_2
\cos(\alpha-\theta)\Gamma^2).
\eeq
Here $\Gam^1$, $\Gam^2$ are in the basis (\ref{eq:Tmetric}); the
metric is therefore not diagonal with regard to the anticommutation
relations for the $\Gamma$ matrices. The defect state couples to RR 2-form fields with polarizations
in the $01$ and the $02$ directions. Overlap computations with the
Ramond vertex operators \footnote{Note the normalization,
\beq\langle C_n|C_n\rangle=
\frac{ \mathcal{A}^{\mu_1...\mu_n}\mathcal{A}_{\mu_1...\mu_n}}{n!}.
\eeq }
\beq
\langle C_n|=\frac{1}{ n!4\sqrt{2}}\langle A|
\mathcal{A}_{\mu_1...\mu_n}(C\Gamma^{\mu_1...\mu_n})_{A\tilde{B}}\langle \tilde{B}|,
\eeq
results in,
\beq
C_{(1)}=\mu_1\frac{ 2 \sqrt{2} V_{1+1} \sin(\alpha-\theta)}{R_1 \sin
\alpha}\mathcal{A}_{01}\epsilon^{01}
+\mu_1\frac{ 2 \sqrt{2} V_{1+1} \sin(\theta)}{R_2 \sin
\alpha}\mathcal{A}_{02}\epsilon^{02}\eeq
where $V_{1+1}$ is the world volume of the defect and arises from the delta functions
in the bosonic sector of the defect state \cite{DiVecchia1}.
These boundary state techniques do not yield the overall normalization of the state \cite{DiVecchia2}.
Moreover this normalization is different for the Ramond and the NS sectors of the theory.
Different techniques can be used to explicitly obtain these normalizations. However, for our purposes it
will be sufficient to absorb this ambiguity in the charge density, $\mu_1$. We will absorb the $2 \sqrt{2}$
factor as well. Later we will check the consistency of our choice by examining the BPS condition.

We have thus found the O1-plane charges
\begin{eqnarray}
  \label{eq:16}
  Q^{01} = V_{1+1} \frac{\sin(\alp-\th)}{R_1\sin\alp}~,~Q^{02} = V_{1+1} \frac{\sin(\th)}{R_2\sin\alp}
\end{eqnarray}.
One can rewrite this in more recognizable form
by replacing the worldvolume, $V_{1+1}$, by the cycle length, $L$. 
After some simplification using $L=|(m,n)|$, and
(\ref{eq:theta_quantisation}),
one obtains the charges,
\bea Q^{01}=m, & Q^{02}=n, & 
\eea
which precisely corresponds to a decomposition of the plane cycle,
$\hat{\gamma}=m\hat{\alpha}+n \hat{\beta}$, in terms of the standard
torus homology basis (that is the $R_1$ and $R_2$ cycles.) The charges, $(m,n)$, then count
the number of cycles in the two different torus homology
classes around which the O1-plane/D1-brane wraps. These are the same charges which were
obtained in section \ref{sec:orient-plan-from}, whereupon the CFT calculation identified the charges with the
cohomology elements dual to these cycles.

It is now a trivial check on our computations to verify that the answer BPS. 
The absolute value of the charge is proportial to the length of the
D-brane \begin{eqnarray}
  \label{eq:18}
  Q.G.Q= L^2~,
\end{eqnarray}
and the latter is the mass divided by the
tension.

\subsection{B-planes: O2/O0}

By choosing the modified Neumann boundary conditions in equation (\ref{eq:boundary_terms})
for both directions one can construct two dimensional boundary states. By the same reasoning as above we will
assume that the O-plane can be considered a superposition of boundary states at the point of
involution. 
The boundary conditions are now encoded in the matrix,
    \begin{eqnarray}
     R_{ij} = \frac{1}{1+b^2}\left( \begin{array}{cc}
1-b^2  & 2b \\
2b & 1-b^2 \\
 \end{array} \right)~,
\label{eq:S2}
 \end{eqnarray}
where orthogonal coordinates have been used to produce a simpler expression. In this basis
 $\cB=\cB_{12}= bR_1R_2\sin\alp$. The $R$ matrix
now has two positive eigenvalues, so the defect state is of B-type.

The corresponding Ramond ground state information is then contained in
the matrix \cite{DiVecchia2}
\begin{equation}
M_{ab}=C\Gamma^0(\Gamma^1\Gamma^2 R_1R_2 \sin \alpha + \cot\alpha +\frac{\cB}{R_1R_2\sin\alpha}),
\end{equation}
where we have switched to natural torus coordinates (\ref{eq:Tmetric}). From this expression it is evident
that the O2-plane/D2-brane can couple to a Ramond 1 form, $C_{(0)}$ and to a three form, $C_{(2)}$. Computing the overlap
with the corresponding vertex operators gives the coupling
\begin{equation}
C_{(2)}=\frac{2 \sqrt{2}\mu_2 V_{2+1}}{R_1 R_2 \sin\alpha} \mathcal{A}_{012}\epsilon^{012},
\end{equation}
for the three form and
\begin{eqnarray}
C_{(0)}&=&\frac{2 \sqrt{2}\mu_2 V_{2+1}\cB}{R_1 R_2 \sin\alpha} \mathcal{A}_{0} \\
&=& \frac{\mu_2 V_{2+1}}{R_1 R_2 \sin\alpha} \mathcal{A}_{0}\frac{\cB_{ij}\epsilon^{ij}}{2!},
\end{eqnarray}
for the 1-form. Note that $\mu_2$ will be used to absorb the $2 \sqrt{2}$ factor as before.
If we replace the world volume term, $V_{2+1}$ by the size of the D2-plane/O2-plane, $N R_1 R_2 \sin \alpha$,
where $N$ counts the number of wrappings around the torus, we can extract the charges
\begin{eqnarray}
q_2&=& N~,\\
q_0&=& N \cB~.
\end{eqnarray}
Intuitively we can relate $q_2$ to the number of 2-cycles of the torus, which by Poincare duality, counts the
charge in the dual cohomology class of $dZ\wedge d\bar{Z}$. The $q_0$ number counts the amount of magnetic flux. This can
be attributed to the Wess-Zumino term \cite{DiVecchia2,callan}
\begin{equation}
\mu_p\int_{\cM_{p+1}} \left(\sum_p C_{p+1}\right)\wedge e^{2\pi \alpha' F+B}
\label{eq:WZ}
\end{equation}
in the generalized defect action which shows how the magnetic flux can manifest itself as
dissolved D0-branes/O0-planes.

The role of the induced D0/O0 charge can be understood by T-dualizing the theory along an axis orthogonal
to $X^1$. If we then make the identification $\cB\rightarrow -R_1 R_2 \sin \alpha \tan \theta$, or equivalently
$q_0/(R_1R_2\sin\alpha)\rightarrow-\tan\theta$, we obtain the $R$ and $M$ matrices corresponding to
an O1-plane/D1-brane inclined at angle $\theta$, quantized as described in the previous section. In other words, for a given
inclination of the D1-brane/O1-plane, the relative number of
$\hat{\beta}$ to $\hat{\alpha}$ cycles manifests itself as magnetic
flux or D0/O0-plane charge in the T-dual picture. We have thus implicitly
captured the pure D0-brane description as well. Recalling also that
T-duality on a torus is mirror symmetry, the above has simultaneously
verified the connection with the CFT results in the previous section.

As a quick check we can compute the BPS condition. Using the natural
inner product on vertical cohomology charge vectors,
\begin{equation}
q_0^2+ q_2^2 \det G =M^2,
\end{equation}
we find agreement with the Born-Infeld factor for $N$ D-branes or O-planes
\begin{equation}
N\sqrt{\det (G+B+2\pi\alpha' F)}=N\sqrt{(R_1 R_2 \sin\alpha)^2+\cB^2}.
\end{equation}

\section{Discussion and Outlook: towards elliptic fibrations}
\label{sec:disc-outl-towards}
\setcounter{equation}{0}

The tori we studied in section \ref{sec:orient-plan-from} 
are naturally described as elements
of different algebraic families. In this section we wish to examine how the
involutive maps which give rise to the $T^2$ orientifolds can be compatibly
extended to such families. In particular this demands some action on the base which
we will discuss later. We will restrict our study to antiholomorphic
maps so as to take advantage of existing mathematical results on antianalytic
maps on Riemann surfaces.  Since we are ultimately interested in applications to
string compactifications, we will require that the total space of the algebraic family be
an elliptically fibered K3 surface.

Geometrically we expect that the K3 O-planes associated with each orientifold can be
characterized, to some degree, in terms of their RR charges by  D-branes which
wrap the fixed point locus.  We will use known mathematical results on real submanifolds of K3's to validate
our results \cite{Nikulin, Silhol, Degtyarev}. The analysis is by no means exhaustive: our goal is
only to illustrate how the simple results of tori O-planes may be generalized to richer geometric
constructions in higher dimensions.

In this paper we deal with two elliptically fibered K3's in detail:
\begin{itemize}
\item a simple example - the ``linear model'',
\item the sextic.
\end{itemize}
The first item is a simple model which exemplifies the techniques and ideas used. The next one
is a pencil family containing the (1,4) Gepner model which extends the analysis of the linear
toy example to a more realistic one and introduces some interesting points in the construction
of K3 involutions. Two families, the quartic and the cubic, which contain the other two Gepner model
are technically more complex and will be briefly discussed in the appendices.

Each torus of the family is characterized by its  $J$-invariant.
The $J$-invariant may take an infinite value in the limit
that the elliptic curves become singular.  Singular fibers in
elliptic fibrations have been classified by Kodaira, and
correspond in a natural way to the ADE classification of extended
Dynkin diagrams.  In his notation a smooth elliptic curve is of
type $I_0$, a nodal rational curve is of type $I_1$, a cycle of
$N$ smooth rational curves is of type $I_N$ (Dynkin type
$A_{N-1}$), a cuspidal rational curve is of type $II$, a
configuration of two tangent rational curves are of type $III$
(type $A_1$), and three concurrent rational curves are of type
$IV$ (type $A_2$).  There are also ``quadratic twisted'' versions
of all of these: $I_0^*$ (Dynkin type $D_4$), $I_N^*$ (type
$D_{N>4}$), $IV^*$ (type $E_6$), $III^*$ (type $E_7$), and $II^*$
(type $E_8$).  The $J$-invariant takes infinite values at fibers
of types $I_N$ and $I_N^*$ with $N \geq 1$, and finite values at
all other singular fibers.  The $J$-invariant vanishes at singular
fibers of types $II$, $IV$, $IV^*$, and $II^*$.  It takes value
$1$ at fibers of types $III$ and $III^*$.  The $J$-invariant of a
singular fiber of type $I_0^*$ can take any finite value
whatsoever.

In particular this means that one can determine the singularity
type, if any, of an elliptic curve occurring in a family
parametrized by a complex variable $w$ by the ramification behavior
of Kodaira's  functional invariant ${\mathcal{J}}(w)$.  The
function ${\mathcal{J}}(w)$ maps from base-coordinate $w$ to the
$J$-line, and for an elliptic curve over $w_0$ to be singular it
is necessary that ${\mathcal{J}}(w_0) \in \{0, 1, \infty\}$. If
${\mathcal{J}}(w_0) = \infty$ then this is also sufficient.  More
generally, however this is not the case (see, for example, Lemma 4.5 in
\cite{DoranPFU}).  The full story is given in Table \ref{table:kodaira}
which correlates the multiplicity $\mu({\mathcal{J}})$ of
${\mathcal{J}}(w)$ at $w_0$ with the Kodaira singularity types.
\begin{table}
\begin{center}
\begin{tabular}{||c|c|c|c||}
\hline \rule{0pt}{5mm} ${\mathcal{J}}(w_0)$ & $\mu({\mathcal{J}})$
& Kodaira types  & Contribution to Euler Number \\[3pt]
\hline \hline \rule{0pt}{5mm} $0$ & $0 (\bmod{3})$ & $I_0$ \
\mbox{or} \ $I_0^*$ & 0 \ \mbox{or} \ 6 \\[3pt]
\cline{2-3} \rule{0pt}{5mm}  & $1 (\bmod{3})$ & $II$ \ \mbox{or} \
$IV^*$ & 2 \ \mbox{or} \ 8 \\[3pt]
\cline{2-3} \rule{0pt}{5mm}  & $2 (\bmod{3})$ & $IV$ \ \mbox{or} \
$II^*$ & 4 \ \mbox{or} \ 10 \\[3pt]
\hline \rule{0pt}{5mm}
  $1$ & $0 (\bmod{2})$ & $I_0$ \ \mbox{or} \ $I_0^*$ & 0 \ \mbox{or} \ 6 \\[3pt]
\cline{2-3} \rule{0pt}{5mm}  & $1 (\bmod{2})$ & $III$ \ \mbox{or} \
$III^*$ & 3 \ \mbox{or} \ 9 \\[3pt]
\hline \rule{0pt}{5mm}
  $\infty$ & $\mbox{pole of order} \ N$ & $I_N$ \ \mbox{or} \ $I_N^*$ & $N$ \ \mbox{or}\ $N+6$
\\[3pt]
     \hline
\end{tabular}
\end{center} \caption{{\em Table of Kodaira Types.}}
\label{table:kodaira}
\end{table}
Thus given the functional $\cJ$-invariant of a family of elliptic
curves, we can determine which members are singular and have multiple
representatives for the same value of $J$. The last column in Table \ref{table:kodaira}
lists the contribution of each singularity to the Euler number of the total space
formed by the family of elliptic curves. For K3 surfaces this must be equal to $\chi=24$.

To construct a type A K3-orientifold from an elliptically fibered K3 surface through an antiholomorphic
involution (both on the base and the fiber), recall from section
\ref{sec:geom-t2-orient} that the requirement
for the existence of a torus with antiholomorphic map is that the $J$-invariant be real.
The orientifold family associated to the torus family is then the pullback
of the $\cJ$-section to the real numbers combined with the datum on the choice of involution
(the choice of involution type within each species branch). This choice may be constrained
by compatibility requirements over the family.
In the case of the standard $w\rightarrow\bar{w}$ action on the base, the discriminant, given by equation \ref{eq:dis},
identifies the choice on each branch. Regardless, it is then noteworthy that there is a global disconnect in the choice
of species of involution at $\cJ=1$. The species number is a topological property of the
fixed point locus corresponding to the
O-plane. Within a family the species could therefore be affected by the
singular fiber structure of the family, which is reflected in the $\cJ$-invariant
global choice that the overall space is a K3 and the choice of involution on the base.
As we will show now, in an illustration of the
power of geometrical insight, this is generally the case.

\subsection{The Linear Model}
Consider the elliptic family whose J-invariant simply given by
\begin{equation}
\cJ(t)=t.
\end{equation}
This family has singular fibers $IV^*$, $III^*$ and  $I^*_1$ located at $t=0$, $1$ and $\infty$ respectively, with
Weierstrass form parametrized by
\begin{eqnarray}
g_2=3t^3(t-1)^3,\\
g_3=t^4(t-1)^5.
\end{eqnarray}
These and the other families were constructed using the methods described in \cite{Herfurtner}.
As mentioned earlier a torus admits antiholomorphic maps whenever its $J$-invariant is real.
In this simple case the pullback of the $J$-invariant is just the restriction that $\mbox{Im }t=0$.
This follows from the restriction of the standard antiholomorphic involution on
the ambient $\PP^2$ bundle over $\PP^1$ to the subspace described by Weierstrass equation. The restriction
to the base is then the standard antiholomorphic map $t\rightarrow \bar{t}$, with fixed point set given by $t$ real.
Since the pullback of the $\cJ$-invariant is a single cover of $\PP^1$ we call this model the ``linear'' one.
Given that the $\cJ$-invariant pullback is simply connected it would seem naively that the orientifold moduli space
would also be connected. However, in order to describe the orientifold, in addition to the $J$-invariant, we need
to specify the type of involution. The space of involutions has a topological disconnect at $\cJ=1$
(see Figure \ref{fig:2a}): for $\cJ<0$ there only exists species $s=1$ involutions
while for $\cJ>0$ one has either $s=2$ or fixed point free $s=0$ involutions. This seems to indicate that there
must be some sort of degeneration of cycles along the path $\mbox{Im }t=0$ as this otherwise smooth path traverses the
topological discontinuity at $\cJ=0$. This degeneration of fixed point cycles is seen to occur at
some of the Kodaira singular fibers as described by Silhol \cite{Silhol} and appears to be a generic phenomenon in these toroidal families.

\begin{figure}[tbp]
   \begin{center}
     \begin{picture}(300,100)(0,0)
       % Axes

       \SetColor{Gray}

       \Line(20,0)(280,0)
       \Text(100,-10)[b]{$0$}
       \Text(200,-10)[b]{$1$}
       \Text(280,-10)[b]{$\infty$}
    \Vertex(20,0){1}
    \Vertex(100,0){1}
    \Vertex(200,0){1}
    \Vertex(280,0){1}
        \Text(150,-25)[b]{$\cJ(t)$}

       \Text(100,80)[b]{$IV^*$}
       \Text(200,80)[b] {$III^*$}
       \Text(295,80)[b]{$I_1^*$}

       \SetColor{Black}
    \Curve{(20,75)(280,75)}

        \Vertex(100,75){2}
        \Vertex(200,75){2}
        \Vertex(280,75){2}

    \Text(102,70)[l]{$t=0$}
    \Text(202,70)[l]{$t=1$}
    \Text(282,70)[l]{$t=\infty$}

    \SetColor{Gray}
       \Text(20,-10)[b]{$-\infty$}
       \Text(20,80)[b]{$I_1^*$}
        \Text(20,70)[l]{$t=\infty$}
        \Vertex(20,75){2}

  %starting at J=0 normal behaviour and going in a negative direction
  %work up to branch point J from 0 to -infinity
   \COval(100,40)(8,4)(0){Black}{White}
   \COval(88,40)(7.5,4)(0){Black}{White}
   \COval(76,40)(7,4)(0){Black}{White}
   \COval(63,40)(7,3.5)(0){Black}{White}
   %\COval(55,40)(7.5,3)(0){Black}{White}
   \COval(51,40)(7,3)(0){Black}{White}
   %\COval(180,40)(12,5)(0){Black}{White}
   \COval(40,40)(8,3)(0){Black}{White}

   %branching at J=0. get figure 8 with overlapping ovals
   \COval(30,45)(5,2.75)(0){Black}{White}
   \COval(30,37)(5,2.75)(0){Black}{White}
   \COval(30,45)(4.6,2.45)(0){White}{White}
%   \COval(190,40)(1,2.15)(0){Black  }{White}
    %limit point. touching cycles. identified with +infinity
   \COval(20,46)(6,3)(0){Gray}{White}
   \COval(20,34)(6,3)(0){Gray}{White}

    %wrap around. now two cycles at infinity. one shrinks to a poinr at J=1
   \COval(280,46)(6,3)(0){Black}{White}
   \COval(280,34)(6,3)(0){Black}{White}

   \COval(270,48)(6,3)(0){Black}{White}
   \COval(270,32)(6,3)(0){Black}{White}

    \COval(260,48)(6,3.5)(0){Black}{White}
   \COval(260,32)(6,3.5)(0){Black}{White}

    \COval(250,48)(6,3.5)(0){Black}{White}
   \COval(250,32)(6,3.5)(0){Black}{White}

    \COval(240,48)(6,3.5)(0){Black}{White}
   \COval(240,32)(6,3.5)(0){Black}{White}

    \COval(230,48)(6,3.5)(0){Black}{White}
   \COval(230,32)(6,4)(0){Black}{White}

   \COval(220,48)(5,3.3)(0){Black}{White}
   \COval(220,32)(6,4)(0){Black}{White}

   \COval(210,48)(3,3)(0){Black}{White}
   \COval(210,32)(6,4)(0){Black}{White}

   \COval(200,48)(1,1)(0){Black}{White}
   \COval(200,32)(6,4)(0){Black}{White}

   %cycle surviving shrinkage shifts up to horizontal

   %\COval(20,48)(1,1)(0){Gray}{White}
   %\COval(20,32)(6,4)(0){Gray}{White}
   %J>-infinity and J<0 normal one cycle
   \COval(190,32.5)(6,4)(0){Black}{White}

   \COval(178,33)(6,4)(0){Black}{White}

   \COval(166,33.5)(6,4)(0){Black}{White}

   \COval(154,33)(6,4)(0){Black}{White}

   \COval(141,34)(6.5,4.5)(0){Black}{White}

   \COval(129,36)(6.5,4.5)(0){Black}{White}

   \COval(114,38)(7,4.5)(0){Black}{White}

     \end{picture}
     \end{center}
 \caption{{\em The linear model $J$-line pullback with a cartoon depicting the degeneration of the real loci.}}
   \label{fig:Jist}
\end{figure}
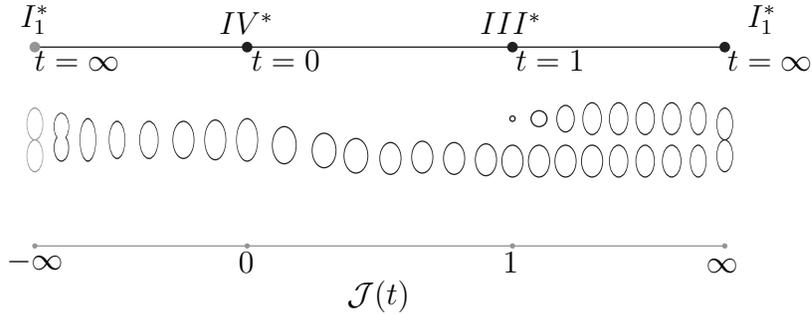

Silhol \cite{Silhol} analyzed the possible transitions of cycles along paths given by real parameters
for real elliptic curves. His conclusions were that the species number could jump only over
singularities of type $I^*_n$, for $n$ odd, or over those of type $III^*$. By plotting the real loci of the Weierstrass
equation, $Y^2=4 X^3-g_2(t) X-g_3(t)$ it is apparent that there is a single fixed point locus for $\cJ<0$ which splits
into two branches at the $I^*_1$ fiber located at minus infinity. Wrapping around and progressing from positive infinity
towards $\cJ=1$ we see that one of the cycles begins to shrink until it finally ``caps off'' at $\cJ=1$. We can continue uninterrupted
from $\cJ=1$ until $\cJ<0$ with no further exceptional behavior. There is no degeneration of
the real locus over the $IV^*$ fiber located at $\cJ=0$. This behavior is shown schematically in Figure \ref{fig:Jist}.

We can connect these results with the known properties of antiholomorphic
involutions on the full K3. In particular, the antianalytic fixed point loci of
K3 surfaces have been studied \cite{Nikulin,Degtyarev}. These
loci can be classified according to some topological characteristics  relating to the action of the involution on
the K3 lattice.  It was found that (with two exceptions) they
correspond to the union of a genus $g$ surface with $k$ copies of $S^2$ (the two sphere).
The Figure \ref{fig:JistK3} of the fixed point locus on the K3 shows how the base fixed
point locus supports a genus 1 curve, so $g=1$. The two-cycles represented by copies of $S^2$ are not necessarily purely
transcendental or algebraic in nature and we should not expect to see these additional fixed point locus components in the
Weierstrass presentation. The precise relationship between the lattice-theoretic description of
antiholomorphic involutions on K3 surfaces and elliptic fibrations with section on K3 surfaces will be studied
carefully in future work.

Generically we have found that as the standard involution on the base results in a base fixed point locus
which is topologically a circle, the fixed point locus in the fiber sweeps out a curve which is at least of genus 1.
This is compatible with the Nikulin classification \cite{Nikulin}. The trivial base involution arises
from restriction of the standard antiholomorphic map on the ambient space. However, it may be possible to
find other antiholomorphic maps and the corresponding orientifolds. This is exemplified by the
Sextic model considered below.

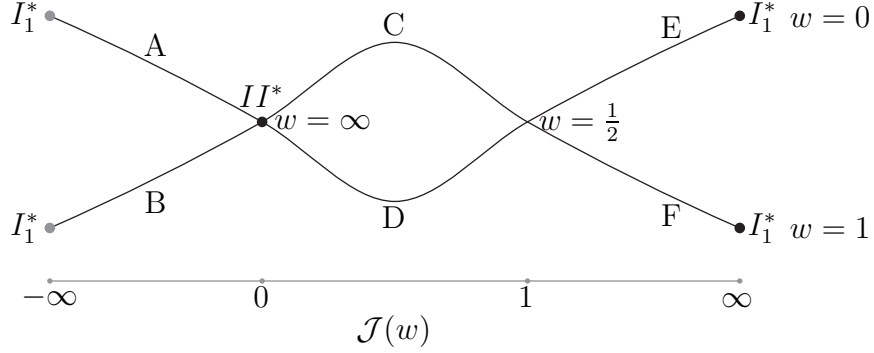
\begin{figure}[tbp]
   \begin{center}
     \begin{picture}(200,200)(0,0)

       \CArc(100,100)(60,0,360)
        \CArc(100,100)(75,0,345)
        \CArc(100,100)(80,0,90)
        \CArc(100,100)(95,0,90)
        \Curve{(175,100)(178,95)(180,100)}
        \Curve{(172.44,80.59)(190,87)(195,100)}
        \CArc(100,187.5)(7.5,90,270)
        \Text(182,90)[u]{$I^*_1$}
        \Text(90,187.5)[r]{$III^*$}
     \end{picture}
   \end{center}
 \caption{{\em Schematic diagram of the topological structure of the involution fixed point locus for the $\cJ=t$ model.}}
   \label{fig:JistK3}
\end{figure}
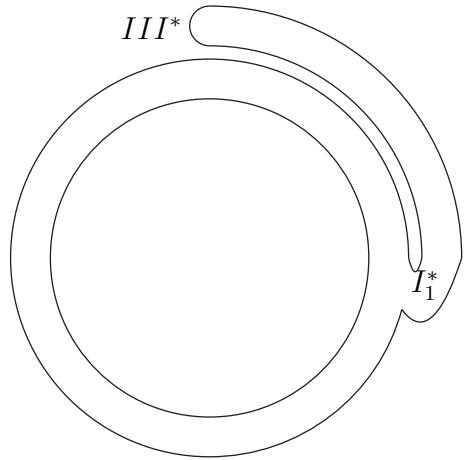

\subsection{The Sextic Pencil}

\begin{figure}[tbp]
   \begin{center}
     \begin{picture}(300,150)(0,0)
       % Axes

       \SetColor{Gray}

       \Line(20,0)(280,0)
       \Text(100,-10)[b]{$0$}
       \Text(200,-10)[b]{$1$}
       \Text(280,-10)[b]{$\infty$}
    \Vertex(20,0){1}
    \Vertex(100,0){1}
    \Vertex(200,0){1}
    \Vertex(280,0){1}
        \Text(150,-25)[b]{$\cJ(w)$}

       \Text(100,70)[u]{$II^*$}
       \Text(295,20)[r]{$I^*_1$}
       \Text(295,100)[r]{$I^*_1$}

       \SetColor{Black}

    \Curve{(20,20)(100,60)(150,90)(200,60)(280,20)}
    \Curve{(20,100)(100,60)(150,30)(200,60)(280,100)}

        \Vertex(100,60){2}
        \Vertex(280,20){2}
        \Vertex(280,100){2}

    \Text(105,60)[l]{$w=\infty$}
    \Text(205,60)[l]{$w=\frac{1}{2}$}
    \Text(300,100)[l]{$w=0$}
    \Text(300,20)[l]{$w=1$}

    \SetColor{Red}
    \Text(60,85)[b]{A}
    \Text(60,30)[u]{B}
    \Text(150,93)[b]{C}
    \Text(150,24)[u]{D}
    \Text(255,92)[b]{E}
    \Text(255,25)[u]{F}

    \SetColor{Gray}
       \Text(20,-10)[b]{$-\infty$}
       \Text(5,20)[l]{$I^*_1$}
       \Text(5,100)[l]{$I^*_1$}
        \Vertex(20,20){2}
        \Vertex(20,100){2}

     \end{picture}
     \end{center}

\bigskip
\bigskip
\begin{tabular}{|c|c|c|c|c|c|c|c|}
  \hline
\rule{0pt}{5mm}
  Segment & $-\infty$ &  & 0 &  & 1 &  & $\infty$ \\[3pt]
  \hline
  A &  & real 0 to $-\infty$  &  &  &  &  &  \\
  B &  & real 1 to $\infty$ &  &  &  &  &  \\
  C &  &  &  & complex $\infty$ to $\frac{1}{2}$ constant real part &  &  &  \\
  D &  &  &  & conjugate $\infty$ to $\frac{1}{2}$ &  &  &  \\
  E &  &  &  &  &  & real $\frac{1}{2}$ to 0 &  \\
  F &  &  &  &  &  & real $\frac{1}{2}$ to 1 &  \\
  \hline
\end{tabular}

\caption{{\em The pullback of the real $\cJ$-invariant for the sextic family.}}
 \label{fig:sextic}
\end{figure}

\begin{figure}[tbp]
   \begin{center}
     \begin{picture}(200,200)(0,0)

       \CArc(100,100)(60,0,360)
       \CArc(100,100)(95,0,360)
       \CArc(100,100)(70,-90,90)
       \CArc(100,100)(85,-90,90)
       \CArc(100,177.5)(7.5,90,270)
       \CArc(100,22.5)(7.5,90,270)

%        \Curve{(83.50,193.56)(85,193)(100,175)}
 %       \Curve{(86.11,178.78)(89,177.3)(92,174)}
  %      \Curve{(92,174)(89.5,170)(87.84,168.94)}

   %     \Curve{(86.11,21.22)(89,22.7)(92,26)}
    %    \Curve{(92,26)(89.5,30)(87.84,31.06)}
    %   \Curve{(83.50,6.44)(85,7)(100,25)}

      % \CArc(230,100)(30,0,360)
  % \Text(178,100)[l]{ $II^*$}
   \Text(82,22)[cl]{$I^*_1$}
   \Text(82,177)[cl]{$I^*_1$}
  %  \Text(230,100)[c]{CD}
%        \CArc(100,100)(80,0,90)
 %       \CArc(100,100)(95,0,90)
   %     \Curve{(172.44,80.59)(190,87)(195,100)}
    %    \CArc(100,187.5)(7.5,90,270)
     %   \Text(182,90)[u]{$I^*_1$}
      %  \Text(90,187.5)[r]{$III^*$}
     \end{picture}
   \end{center}
 \caption{{\em Schematic diagram of the topological structure of the involution fixed point locus for the Sextic. The handle on the left hand side
 corresponds to the AB branch on the base parameter locus, while the double handle on the right is the EF fixed point locus (of species type 2).
 The fixed point curve is of type $g=2$.}}
   \label{fig:sexticK3}
\end{figure}
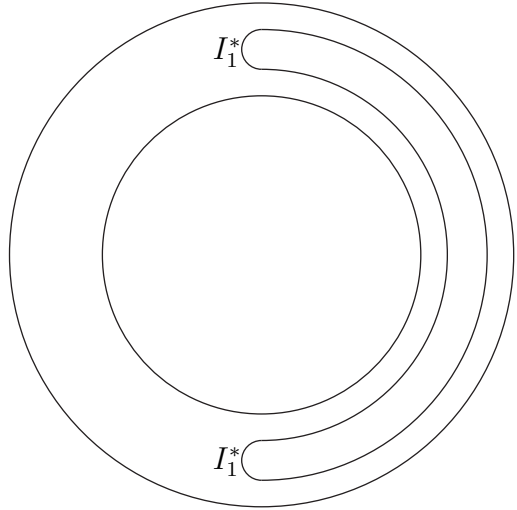

The next example is the family corresponding to the pencil containing the sextic hypersurface
\begin{eqnarray}
\label{eq:25}
x_1^6 + x_2^3 + x_3^2 - z^{-1/6} x_1 x_2 x_3 = 0 \ \mbox{in} \
{\mathbb{WP}}^2_{1,2,3}.
\end{eqnarray}
This family contains the Gepner model (1,4) at $z=\infty$ \cite{lian-yau}.\footnote{This paper lists the families containing the Gepner models we studied, however
they seem to have a sign difference in the $z$ term.} Although the natural parameter describing this family is
$w = -2^4 3^3 z$, it is advantageous to make the $PSL(2,\ZZ)$ transformation $t=\frac{w-1}{w}$ to obtain the
Weierstrass parameters in polynomial form,
\begin{eqnarray}
g_2(t)&=&3(t-1)^4t^2,\\
g_3(t)&=&(t-1)^5t^3(t+1).
\end{eqnarray}
The $\cJ$-invariant is given by
\begin{equation}
\cJ(t)=-\frac{1}{4}\frac{(t-1)^2}{t},
\end{equation}
or as
\begin{equation}
\cJ(t)=-\frac{1}{4w(w-1)},
\end{equation}
in terms of the natural pencil parameter.  This family has two $I^*_1$ singularities at $w=0$ and $w=1$ and a type $II^*$
fiber at $w=\infty$.  The $\cJ$-invariant is a little more complicated than the previous case.
The pullback in this case is double cover of $\PP^1$ with ramification points at $w=0$ and at $w=\infty$. The pullback
is illustrated in Figure \ref{fig:sextic} and a summary of the behavior of the $w$ parameter on each path is given in the accompanying
table.

An analysis similar to that of the linear model can be made for the regions where the pencil parameter, $w$
(or equivalently $t$), is real. As explained earlier, this base locus arises by simply considering the restriction of
standard antiholomorphic map on the ambient space ($x_i\rightarrow \bar{x_i}$) to the K3.
This real $w$-locus corresponds to the region where $\cJ<0$ and $\cJ>1$. In Figure \ref{fig:sextic} this is a loop delimited by the legs
labeled A,E,F and B. Along the leg A one has a single cycle that splits at the $I^*_1$ fiber as $\cJ\rightarrow -\infty$
($w\rightarrow 0$). The two resulting cycles then travel along the legs E and F to recombine at the other $I^*_1$ fiber as
$\cJ\rightarrow \infty$ ($w=1$). These cycles do not transition to the $\cJ<1$ region at the ramification point $w=\frac{1}{2}$ as
there is the aforementioned topological discontinuity dividing the J-line into regions solely admitting $s=1$ involutions
(for $\cJ<1$) and those allowing $s=0,2$ loci for $\cJ>1$. Back along the leg B the single cycle rejoins that of path A at the
ramification point $w=\infty$. This ramification point does contain a singular fiber of type $II^*$, though as shown by Silhol
\cite{Silhol} this does not result in a degeneration of the real locus, or change of species. This can be verified by simply plotting the Weierstrass equation along the path.
The path ABEF where the pencil parameter is real hence contributes a genus $g=2$
to the fixed point locus of the K3. The resulting fixed point curve is shown schematically in Figure
\ref{fig:sexticK3}.

One major difference with the simple model studied earlier is the existence of a central loop CD (i.e. for $0<\cJ<1$)
where $w$ is complex valued. The values of $w$ along one branch are conjugate to the those of the other branch.
Since $w$ is complex the Weierstrass equation has complex coefficients
and one cannot directly plot it in $\RR^2$ to investigate the fixed point loci (as in the linear case). However, we can
directly analyze the Weierstrass $X=\wp(u;g_2,g_3)$ and $Y=\wp'(u;g_2,g_3)$ functions. If we let the auxiliary variable $u$,
valued in the torus covering space, lie along the fixed point locus, we find that $\wp$ and $\wp'$ can be
independently rotated by an $w$ dependent phase to a real plane. Plotting the rotated variables, we see that the
fixed point locus is an $s=1$ curve. Given a choice of involution on one branch, we need to select the
other type involution for the complex branch in order to have matching curves at the ramification points.
As this $w$-locus is complex it is clearly not fixed by the standard antiholomorphic map inherited from the
ambient space. The standard involution instead maps one complex branch to its conjugate. In terms of the $w$ parameter the complex branches
are described by $w=1/2\pm \im v$ with $v\in (0, \infty)$. We can make a general linear transformation
to reparametrize the C and D branches so that they are described by the real parameter $v=\pm \im (w-1/2)$. As this new parameter
is real, it is fixed by some canonical extended antiholomorphic map. The linear transform also results in the legs A, B,E and F being now described
by complex parameters which consequentially are not fixed under the new involution.
Since the base is $\PP^1$, the restriction of the antiholomorphic involution on the ambient space should restrict to an involution on the base,
which is $w\rightarrow \pm \bar{w}$ modulo $PGL(2,\CC)$. The more suitable question is whether one can find the K3 involution from which this base
involution emerges. Regardless of the specific details of the involutive map, we can determine that the complex CD loop supports a genus $g=1$ curve.

The linear case has exemplified how the real locus on the base supports a genus $g$ curve, consistent with Nikulin's classification of fixed
point loci in K3 surfaces. The reality of the base parameter can be seen to be the result of restricting the standard ambient antiholomorphic
involution to the base of the K3. The sextic, a more realistic and relevant example, demonstrates that the parameter space where the elliptic fibers
admit involutions is typically larger than the space where the base parameter is real. Such loci ought to correspond to other antiholomorphic
involutions on the K3 with non-trivial action on the base.  The question then is to find and classify these other involutions.
The other families relevant to our paper, the quartic and cubic, are more complicated in the branching and ramification of the pullback of the real $\cJ$-line.
Like the linear and the sextic model the real locus on the base supports a genus $g$ curve. However, there are more complex and complex conjugate branches
which might be fixed under other non-standard antiholomorphic involutions. A brief account of the real base parameter locus
for these other families is presented in the adjoining appendices. It is our hope to extend this study in a subsequent
paper where we may be able to more systematically analyze the possible involutions and
orientifolds in K3's.

\subsection{Conclusion and Outlook}

Orientifold compactifications in string theory provide a very rich and
promising phenomenological playground. 
Toroidal compactifications are still the most well
understood starting point for orientifold
compactifications. The goal is to understand them more
generally. One road of approach is to deform away from the
toroidal point in moduli space \cite{Lust:2006zh}. Another is the
worldsheet gauged linear sigma model approach to better understand the
quantum geometry \cite{Brunner:2002em,Brunner:2003zm,Brunner:2004zd}.
The 
preview towards orientifolds of elliptically fibered
compactifications in this section points to a third way.
It is one of the main motivations for the study
undertaken here. In any of these $T^2$-compactifications 
play an essential role, either as building block
or example. They stand out due to
their simplicity. As we illustrated in this article, this allows a
deep connected and complementary 
understanding of various geometric or worldsheet
approaches to $T^2$-orientifolds, which should aid us in developing a
general understanding of these mysterious quantum objects.

\bigskip
\noindent
{\bf Acknowledgments:} We are very grateful to the hosts and
participants of the Simons Workshop in Mathematical Physics (Stony
Brook 2003). In particular we wish to thank Christian R\"omelsberger and
Cumrun Vafa. We also thank Jan de Boer, Kentaro Hori and Bert
Schel\-le\-kens
for useful comments, and especially Lennaert Huiszoon and Brian
Greene who were originally part
of this collaboration.
A number of calculations were checked with Bert Schellekens' program
{\tt kac} (http://www.nikhef.nl/$\sim$t58/kac.html), an indispensable
tool.
KS is grateful for partial support from DOE
grant DE-FG-02-92ER40699 and from a VIDI Innovational Research
Incentives Award of the Netherlands Organisation for
Scientific Research (NWO).

\appendix

\section{Families of the (2,2) and (1,1,1) Gepner Orientifolds}

For completeness we briefly analyze the elliptic K3-orientifolds based
on families of the (2,2) and the (1,1,1) Gepner orientifolds.

\subsection{The Quartic Pencil}
\setcounter{equation} {0}
The family containing the quartic hypersurface is described by
\begin{eqnarray}
     \label{eq:21}
  x_1^4 + x_2^4 + x_3^2 - z^{-1/4} x_1 x_2 x_3 = 0 \ \mbox{in} \
{\mathbb{WP}}^2_{1,1,2}.
   \end{eqnarray}
The variable $w=2^6z$ parametrizes the complex structure
moduli of this weighted quartic. In the limit that $w\rightarrow \infty$ we get
the Gepner model (2,2) studied in section 4. As before, we can change variables to
$t=\frac{w-1}{w}$ to obtain the Weierstrass parameters in polynomial form:
\begin{eqnarray*}
g_2(t)&=& 3 t^2(t-1)^3(t-4),\\
g_3(t)&=& t^3 (t-1)^5(t+8).
\end{eqnarray*}
In terms of this new parametrization the $\cJ$-invariant reads
\begin{equation}
\cJ(t)=-\frac{1}{27}\frac{(t-4)^3}{t^2},
\end{equation}
or in terms of $w$ we have,
\begin{equation}
\cJ(w) = \frac{(1+3w)^3}{27 w (1-w)^2} \ .
\end{equation}
This family has singularities of types $I^*_1$, $I^*_2$ and $III^*$ at $w=0$, 1 and $\infty$.
The locus with $\cJ$ real is visibly a triple cover of the real $J$-line, as indicated in
Figure \ref{fig:quartic}.

\begin{figure}[htb]
   \begin{center}
     \begin{picture}(300,120)(0,00)
       % Axes

       \SetColor{Gray}

       \Line(20,0)(280,0)
       \Text(100,-10)[b]{$0$}
       \Text(200,-10)[b]{$1$}
       \Text(280,-10)[b]{$\infty$}
    \Vertex(20,0){1}
    \Vertex(100,0){1}
    \Vertex(200,0){1}
    \Vertex(280,0){1}
        \Text(150,-25)[b]{$\cJ(w)$}

        \Text(200,110)[u]{$III^*$}
       \Text(295,20)[r]{$I^*_1$}
       \Text(295,100)[r]{$I^*_2$}

       \SetColor{Black}
       %\Line(20,20)(280,20)
       %\Line(20,100)(280,100)

       \Curve{(20,20)(100,60)(200,100)(280,100)}
        \Curve{(20,100)(60,70)(100,60)(150,30)(200,20)(280,20)}
        \Curve{(20,100)(60,85)(100,60)}
        \Curve{(100,60)(150,45)(200,20)}
        \Curve{(200,20)(210,21)(240,50)(275,99)(280,100)}

        \Vertex(200,100){2}
        \Vertex(280,20){2}
        \Vertex(280,100){2}

    \Text(100,70)[b]{$w=-\frac{1}{3}$}
    \Text(200,90)[u]{$w=\infty$}
    \Text(200,12)[u]{$w=\frac{1}{9}$}
    \Text(315,100)[u]{$w=1$}
    \Text(315,20)[u]{$w=0$}

    \SetColor{Red}
    \Text(60,90)[b]{A}
    \Text(60,70)[u]{B}
    \Text(60,35)[u]{C}
    \Text(150,85)[b]{D}
    \Text(150,45)[b]{E}
    \Text(150,20)[u]{F}
    \Text(245,97)[u]{G}
    \Text(245,45)[u]{H}
    \Text(245,20)[b]{K}

    \SetColor{Gray}
       \Text(20,-10)[b]{-$\infty$}
        \Vertex(20,20){2}
        \Vertex(20,100){2}
       \Text(5,20)[l]{$I^*_1$}
       \Text(5,100)[l]{$I^*_2$}
     \end{picture}
   \end{center}

\bigskip
\bigskip
\begin{tabular}{|c|c|c|c|c|c|c|c|}
  \hline
\rule{0pt}{5mm}
  Segment & $-\infty$ &  & 0 &  & 1 &  & $\infty$ \\[3pt]
 \hline
  A &  & complex 1 to $-\frac{1}{3}$ &  &  &  &  &  \\
  B &  & conjugate 1 to $-\frac{1}{3}$ &  &  &  &  &  \\
  C &  & real 0 to $-\frac{1}{3}$ &  &  &  &  &  \\
  D &  &  &  & real $-\frac{1}{3}$ to $-\infty$&  &  &  \\
  E &  &  &  & complex $-\frac{1}{3}$ to $\frac{1}{9}$ &  &  &  \\
  F &  &  &  & conjugate $-\frac{1}{3}$ to $\frac{1}{9}$ &  &  &  \\
  G &  &  &  &  &  & real $\infty$ to 1 &  \\
  H &  &  &  &  &  & real $\frac{1}{9}$ to 1 &  \\
  K &  &  &  &  &  & real $\frac{1}{9}$ to 0 &  \\
  \hline
\end{tabular}
 \caption{{\em The quartic.}}
   \label{fig:quartic}
\end{figure}

There are two type of paths: paths parametrized by real $w$ which can
be directly analyzed as in the linear model, and pairs of paths corresponding to complex $w$ and its
conjugate. Focusing on the real path we can start at $w=-\infty$ and follow the progression of
increasing $w$. This is schematically illustrated in Figure \ref{fig:quarticK3}. As described by Silhol \cite{Silhol}
along the CD branch the locus  varies smoothly until it splits at the $I_1^*$ singularity located at $w=0$
($\cJ\rightarrow-\infty$) in agreement with the topological distinction between $s=1$ involutions for $\cJ<1$ and
$s=2,0$ for $\cJ>1$. The cycle pair then vary along legs KHG until one of them is capped off at the $III^*$ fiber
located at $w=\infty$. The change in fixed point topology here has been mentioned by Silhol \cite{Silhol} and
is the same mechanism that allows us to align the topological discontinuity of involution over the $\cJ$-line with the
smooth family picture. This real locus is very similar to the linear model case and contributes a handle to the fixed point
locus.

Presumably one could find other involutions on the K3 which would fix the complex $w$ base branches. It would be
interesting to know whether both complex loops, AB and EF, could be fixed simultaneously or not. Since path EF does
not cross any intersections and supports $s=1$ fixed loci on its fibers it clearly yields a contribution of 1 to the
K3 fixed curve genus. More analysis is required.

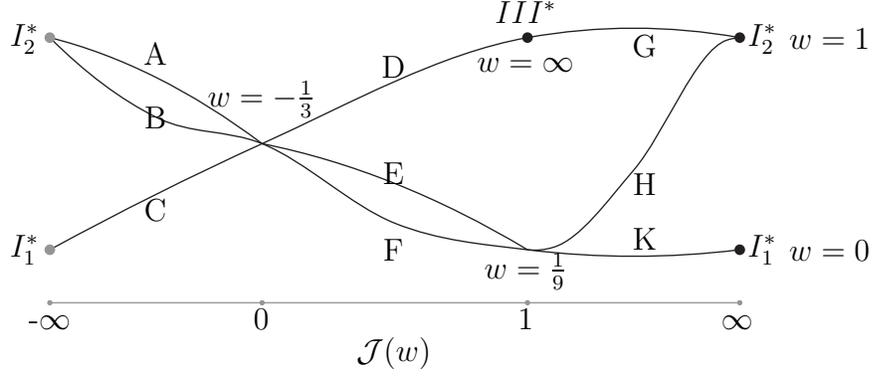
\begin{figure}[tbp]
   \begin{center}
     \begin{picture}(300,200)(0,0)
    %comment bottom axis
     \Curve{(0,50)(65,50)}
     \Curve{(95,50)(280,50)}
    %top axis up to ramification point
     \Curve{(0,65)(65,65)}
     %filled in ramification
     \Curve{(65,65)(95,65)}
     \Curve{(65,50)(95,50)}
     %end filling
     \Curve{(95,65)(155,65)}
     \Curve{(180,65)(280,65)}
  %comment split cycle
    \Curve{(180,80)(280,80)}
    \Curve{(180,90)(280,90)}
   %comment lower loop EF
 %   \CArc(80,30)(25,0,53.13)
 %   \CArc(80,30)(25,126.87,360)
 %   \CArc(80,30)(15,0,360)
%comment upper loop up to caps AB
%    \CArc(80,85)(25,-53.13,60)
%    \CArc(80,85)(25,120, 233.13)
%    \CArc(80,85)(15,-240, 60)
    %comments AB caps
%    \CArc(90,102.3)(5,60,240)
%    \CArc(70,102.3)(5,-60,120)
%splitting
    \CArc(180,72.5)(7.5,90,270)
    \CArc(180,65)(25,90, 180)
%cap off second cycle
    \CArc(280,85)(5,-90,90)
%text describing path
    \BText(-20,150){w}
    \Text(0,150)[c]{$-\infty$}
    \Text(80,150)[c]{$-\frac{1}{3}$}
    \Text(160,150)[c]{$0$}
    \Text(200,150)[c]{$\frac{1}{9}$}
    \Text(240,150)[c]{$1$}
    \Text(280,150)[c]{$\infty$}
    \LongArrow(12,150)(60,150)
    \LongArrow(90,150)(150,150)
    \LongArrow(170,150)(190,150)
    \LongArrow(210,150)(230,150)
    \LongArrow(250,150)(270,150)
    \Text(40,160)[r]{D}
    \Text(120,160)[r]{C}
    \Text(180,160)[r]{K}
    \Text(220,160)[r]{H}
    \Text(260,160)[r]{G}

%    \Text(48,30)[c]{E}
%    \Text(112,30)[c]{F}

 %   \Text(48,85)[c]{A}
 %   \Text(112,85)[c]{B}

  %  \Text(80,110)[c]{$I^*_2$}
    \Text(169,72)[c]{$I^*_1$}
    \Text(280,72)[c]{$III^*$}

   %\CArc(80,85)(25,126.87,360)

     %   \Text(182,90)[u]{$I^*_1$}
      %  \Text(90,187.5)[r]{$III^*$}
     \end{picture}
   \end{center}
 \caption{{\em Schematic diagram of the topological structure of the involution fixed point locus of the Quartic for the
 standard K3 involution (i.e., with standard action on base). The ends of
 the figure are identified. The fixed point curve is of type $g=1$.}}
   \label{fig:quarticK3}
\end{figure}
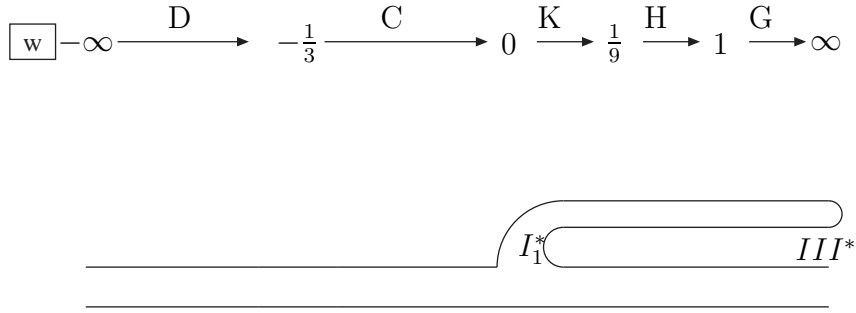

\subsection{The Cubic Pencil}

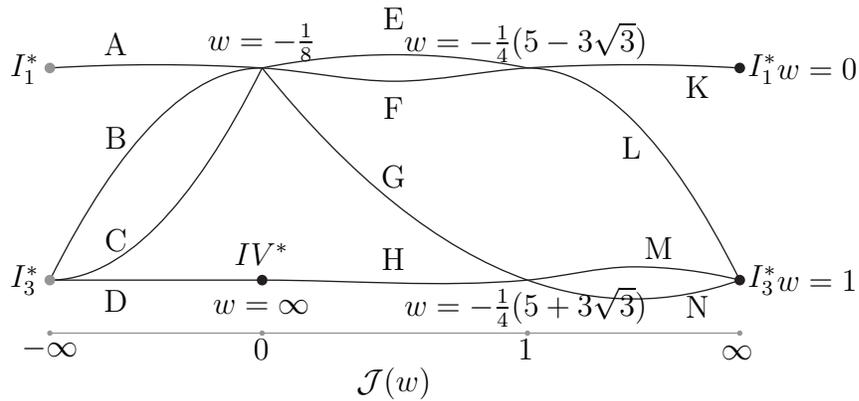
\begin{figure}[tbp]
   \begin{center}
     \begin{picture}(300,150)(0,0)
       % Axes

       \SetColor{Gray}

       \Line(20,0)(280,0)
       \Text(100,-10)[b]{$0$}
       \Text(200,-10)[b]{$1$}
       \Text(280,-10)[b]{$\infty$}
    \Vertex(20,0){1}
    \Vertex(100,0){1}
    \Vertex(200,0){1}
    \Vertex(280,0){1}
        \Text(150,-25)[b]{$\cJ(w)$}

         \Text(100,30)[u]{$IV^*$}
       \Text(295,20)[r]{$I^*_3$}
       \Text(295,100)[r]{$I^*_1$}

       \SetColor{Black}
       %\Line(20,20)(280,20)
       %\Line(20,100)(280,100)

       \Curve{(20,20)(100,20)(200,20)(240,25)(280,20)}
       \Curve{(20,100)(100,100)(150,95)(200,100)(280,100)}
       \Curve{(20,20)(60,80)(100,100)}
       \Curve{(20,20)(60,40)(100,100)}
       \Curve{(100,100)(150,105)(200,100)}
       \Curve{(100,100)(200,20)(280,20)}

       \Curve{(200,100)(240,80)(280,20)}

       \Vertex(280,20){2}
       \Vertex(280,100){2}
       \Vertex(100,20){2}

        \Text(100,10)[u]{$w=\infty$}
    \Text(100,110)[u]{$w=-\frac{1}{8}$}
    \Text(200,10)[u]{$w=-\frac{1}{4}(5+3\sqrt{3})$}
    \Text(200,110)[u]{$w=-\frac{1}{4}(5-3\sqrt{3})$}
        \Text(295,100)[l]{$w=0$}
    \Text(295,20)[l]{$w=1$}

    \SetColor{Gray}
    \Vertex(20,20){2}
       \Vertex(20,100){2}
\Text(5,20)[l]{$I^*_3$}
       \Text(5,100)[l]{$I^*_1$}
       \Text(20,-10)[b]{$-\infty$}
     \SetColor{Red}
     \Text(45,105)[b]{A}
     \Text(45,70)[b]{B}
     \Text(45,35)[u]{C}
     \Text(45,12)[u]{D}

    \Text(150,115)[b]{E}
    \Text(150,85)[u]{F}
    \Text(150,55)[b]{G}
    \Text(150,24)[b]{H}

    \Text(265,93)[u]{K}
    \Text(240,70)[u]{L}
    \Text(245,32)[l]{M}
    \Text(265,10)[u]{N}

     \end{picture}
   \end{center}

\bigskip
\bigskip
\begin{tabular}{|c|c|c|c|c|c|c|c|}
  \hline
  segment & $-\infty$ &  & 0 &  & 1 &  & $\infty$ \\
 \hline
  A &  & real 0 to -1/8 &  &  &  &  &  \\
  B &  & complex 1 to -1/8 &  &  &  &  &  \\
  C &  & conjugate 1 to -1/8 &  &  &  &  &  \\
  D &  & real 1 to $\infty$ &  &  &  &  &  \\
  E &  &  &  & complex  -1/8 to 0.49  &  &  &  \\
  F &  &  &  & conjugate  -1/8 to 0.49 &  &  &  \\
  G &  &  &  & real -1/8 to -2.55 &  &  &  \\
  H &  &  &  & real $-\infty$ to -2.55 &  &  &  \\
  K &  &  &  &  &  & real 0.49 to 0 &  \\
  L &  &  &  &  &  & real 0.49 to 1 &  \\
  M &  &  &  &  &  & complex  -2.55 to 1 &  \\
  N &  &  &  &  &  & conjugate -2.55 to 1 &  \\
  \hline
\end{tabular}

 \caption{{\em The cubic.}}
   \label{fig:cubic}
\end{figure}

The $T^2$ underlying the $(1,1,1)$ Gepner model, also known as the `cubic',
   is the hypersurface described by
\begin{eqnarray}
x_1^3 + x_2^3 + x_3^3 - z^{-1/3} x_1 x_2 x_3 = 0 \ \mbox{in} \
{\mathbb{P}}^2
\end{eqnarray}
The variable $w = 3^3 z$ parametrizes the complex
structure moduli of the cubic. There are $I^*_1$, $I^*_3$ and $IV^*$ singular fibers at the points $w=0$, 1 and
$\infty$. The change of variables $t=\frac{w-1}{w}$ yields
the Weierstrass parameters:
\begin{eqnarray*}
g_2(t)&=& 3t^2(t-1)^3(t-9),\\
g_3(t)&=& t^3(t-1)^4(t^2+18t-27),
\end{eqnarray*}
with $\cJ$-invariant
\begin{equation}
\cJ(t)=-\frac{1}{64}\frac{(t-1)(t-9)^3}{t^3}.
\end{equation}
In terms of the pencil parameter the $\cJ$-invariant is
\begin{equation}
   \cJ(w) = \frac{(1+8w)^3}{64 w (1-w)^3} \ .
\end{equation}
This time the restriction to real $J$-values yields a triple cover of the $J$-line. This
is shown in Figure \ref{fig:cubic}.

\begin{figure}[tbp]
   \begin{center}
     \begin{picture}(300,200)(0,0)
    %comment bottom axis
     \Curve{(-50,50)(65,50)}
     \Curve{(95,50)(320,50)}
    %top axis up to ramification point
     \Curve{(-50,65)(65,65)}
     \Curve{(95,65)(155,65)}
     \Curve{(180,65)(320,65)}
     %comment fill in ramification
     \Curve{(65,65)(95,65)}
     \Curve{(65,50)(95,50)}

  %comment split cycle
    \Curve{(180,80)(280,80)}
    \Curve{(180,90)(280,90)}
   %comment lower loop EF
 %   \CArc(80,30)(25,0,53.13)
  %  \CArc(80,30)(25,126.87,360)
  %  \CArc(80,30)(15,0,360)
%comment upper loop up to caps AB
  %  \CArc(80,85)(25,-53.13,60)
  %  \CArc(80,85)(25,120, 233.13)
  %  \CArc(80,85)(15,-240, 60)
   % %comments AB caps
   % \CArc(90,102.3)(5,60,240)
   % \CArc(70,102.3)(5,-60,120)
%splitting
    \CArc(180,72.5)(7.5,90,270)
    \CArc(180,65)(25,90, 180)
%cap off second cycle
    \CArc(280,85)(5,-90,90)

%%cap disjoint spheres at M and N
 %   \CArc(280,30)(15,0,360)
 %   \CArc(280,115)(15,0,360)

%text describing path
    \BText(-70,150){w}
    \Text(-50,150)[c]{$-\infty$}
    \Text(10,150)[c]{$-\frac{1(5+3\sqrt{3})}{4}$}
    \Text(80,150)[c]{$-\frac{1}{8}$}
    \Text(160,150)[c]{$0$}
    \Text(220,150)[c]{$\frac{1}{4}(3\sqrt{3}-5)$}
    \Text(280,150)[c]{$1$}
    \Text(320,150)[c]{$\infty$}
    \LongArrow(-38,150)(-15,150)
    \LongArrow(35,150)(65,150)
    \LongArrow(90,150)(145,150)
    \LongArrow(165,150)(185,150)
    \LongArrow(255,150)(270,150)
    \LongArrow(285,150)(305,150)

    \Text(-26,160)[r]{H}
    \Text(50,160)[r]{G}
   % \Text(63,160)[r]{A}
    \Text(118,160)[r]{A}
    \Text(175,160)[r]{K}
    \Text(262,160)[r]{L}
    \Text(295,160)[r]{D}

    %\Text(48,30)[c]{E}
    %\Text(112,30)[c]{F}

   % \Text(48,85)[c]{B}
   % \Text(112,85)[c]{C}

  %  \Text(280,115)[c]{M}
  %  \Text(280,30)[c]{N}

   % \Text(80,110)[c]{$I^*_3$}
    \Text(169,72)[c]{$I^*_1$}
    \Text(280,72)[c]{$I^*_3$}
    \Text(320,57)[c]{$IV^*$}

   %\CArc(80,85)(25,126.87,360)

     %   \Text(182,90)[u]{$I^*_1$}
      %  \Text(90,187.5)[r]{$III^*$}
     \end{picture}
   \end{center}
 \caption{{\em Schematic diagram of the topological structure of the standard (i.e., with standard base action) involution fixed point locus for the Cubic. The ends of
 the figure are identified. The fixed point curve is of type $g=1$.}}
   \label{fig:cubicK3}
\end{figure}
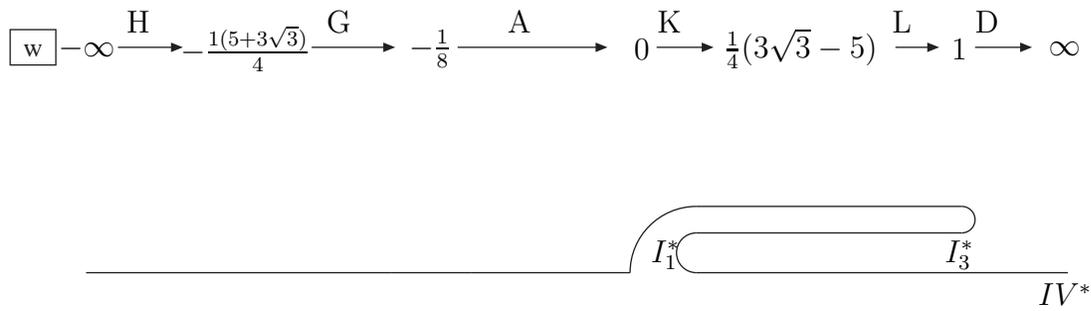

The overall result is similar to the quartic family. As in the sextic case the appearance of complex loops in the base
which can support fixed loci on their fibers indicates the possibility of involutions with non-trivial action on the
base. For simplicity we will focus on involutions which restrict to $w$ standard action on the base. Following the real
path of the parameter $w$, starting from $-\infty$ we have a single cycle along the H leg. As Silhol \cite{Silhol}
shows, the $IV^*$ fiber does not degenerate the real cycle. At the $w=-\frac{1}{8}$ ramification we continue along the real
path A reaching a splitting point induced by the fiber $I^*_1$ at $w=0$. The resulting cycle pairs travel along KL, in
accordance with this region having $s=2$, until one of them caps off a the $I^*_3$ fiber at $w=1$.  The remaining real cycle
then continues along D to return to its starting point. The resulting
K3 fixed locus component has genus $g=1$. This is
schematically shown in Figure \ref{fig:cubicK3}.
\\

\section{$P$-matrix entries in Gepner models}
\label{sec:p-matrix-entries}
\setcounter{equation}{0}

We seek explicit expressions for Gepner model $P$-matrix entries
with one index corresponding to that of a chiral field and the other
to a Klein-Bottle current of the $\cA^{ws}$ theory. The $\cA^{ws}$
model is obtained as a simple current extension from the tensor
product of $r$ minimal models and a $D_{8-n,1}$ WZW model. This
simple current extension, commonly referred to as ``fermion
alignment'', is responsible for ensuring that there is a well defined
world sheet supersymmetry.

  In terms
of the constituent $P$-matrices, the $P$-matrix of a simple current
extension is
\begin{eqnarray}
   \label{eq:1}
   P_{[a],[b]} = e^{\pi i (a_{[a]}+a_{[b]})}
\sum_{n=1}^N \eps_{J^n}(b) P_{a,J^nb}.
\end{eqnarray}
The phases $\eps_{J^n}(b) \equiv e^{\pi i (h_b-h_{J^nb})}$ and
$a_{[a]}= h_{[a]}- h_a$ (the difference between the true conformal
weight $h_{[a]}$ of the orbit $[a]$ and that of the chosen
representative element)
  ensure that expression for the $P$-matrix is invariant on
orbits $[a]$. For convenience the phase factor, $e^{\pi i
(a_{[a]}+a_{[b]})}$, will be dropped in the intermediate steps to
the final answer. With regard to its modular properties, each
minimal model, represented as a $SU(2)_k\times U(1)_4/U(1)_{2\mh}$
coset can be thought of as an $SU(2)\times U(1)^{\ast} \times U(1)$
tensor product extended by an order two simple current, whose orbits
impose {\em field identification} $(l,m,s) \Leftrightarrow
(k-l,m+\mh,s+2)$.

The NS Klein bottle currents of the $\cA^{ws}$ theory are the orbits
$[G_{8-d,1}^{NS}[0;\vec{n};\vec{m}]]$ with both $n_i$ and $m_i$ even.
Fermion alignment can be used to reduce all the $m_i=0$. The
(anti)-chiral fields of the minimal model are the fields $(l,l',0)$
where $l'=\pm l$.

Recall the forms of the $SU(2)_k$ and $U(1)_{2\mh}$ $P$-matrices
\begin{eqnarray}
   \label{eq:24b}
   P^{SU(2)_k}_{l_1,l_2}  &=&
   \ove{\sqrt{\mh}}\sin\left(\frac{\pi(l_1+1)(l_2+1)}{2\mh}\right)
   \sum_{u=0}^1 (-1)^{u(k+l_1+l_2)} ~,~~\mh=k+2,\non
   P^{U(1)_{2\mh}}_{m_1,m_2}
&=& \frac{e^{-\frac{\pi i m_1m_2}{2\mh}}}{2\sqrt{\mh}}
   \sum_{u=0}^1 (-1)^{u(\mh+m_1+m_2)}~.
\end{eqnarray}
Due to the selection rule $k+l_1+l_2 \in 2\ZZ$ for the $SU(2)$
$P$-matrix, odd and even $k$ minimal models behave differently. As
$l_2$ vanishes for {\em all} Klein bottle currents, this selection
rule reduces to $k+l_1 \in 2\ZZ$. For odd $k$ minimal models
therefore only {\em one} of the elements of the identification orbit
contributes, which depends on the nature of $l_1$. For even $k$
minimal models only $l_1 \in 2\ZZ$ $P$-matrix entries are nonzero,
but for each both elements of the identification orbit contribute.
It is straightforward to check that for (anti)chiral fields $(l,\pm
l,0)$ the $U(1)$ selection rules yield the same conditions.

Explicitly we find the following values for the minimal model
$P$-matrix
\begin{eqnarray}
   \label{eq:39}
   P^{min}_{(l,\pm l,0),(0,n,m)} &=&
   \frac{\sqrt{2}}{\mh}e^{\pm\frac{\pi i ln}{2\mh}}\left(e^{\pi i
(\frac{n}{2}-\frac{m}{2})}\sin\left(\frac{\pi(l+1)}{2\mh}\right)
\del^{(2)}_{k+l}\del^{(2)}_{m}+\cos\left(\frac{\pi(l+1)}{2\mh}\right)
\del^{(2)}_l\del^{(2)}_m\right).
\end{eqnarray}

Due to aforementioned difference in behavior of the $P$-matrix as a
function of whether $k$ is odd or even, we will first study the
cases where all the $k_i$ are purely odd or even. The mixed case can
then be built by fermion aligning a purely even and odd tensor
block. Finally we will tensor in the spacetime part.

\subsection{Models with all $k_i$ odd}

Tensoring $r$ odd $k_i$ minimal models and extending by the currents
$w_i$ we obtain the following $P$-matrix entries for the fermion
aligned tensor theory, $\cA^{r}$ theory. (Recall that a
representative of an orbit under the $w_i$ has all $s_{i\neq 1} =
s_1 \mod 2$, $s_1=-1,0,1,2$. We use below that $\eps_{v_i}(0,n_i,0)
= \im$) and the following useful relation
\begin{eqnarray}
   \label{eq:40}
   P^{min,~k~odd}_{(l,\pm l,0),(0,n,m)} &=& e^{-\frac{\pi i l m}{2}}
P^{min,~k~odd}_{(l,\pm l,0),(0,n,0)}.
\end{eqnarray}

\begin{eqnarray}
   \label{eq:41}
   P^{r}_{[\vec{l};\vec{l'};0],[0;\vec{n};0]} &=&
    \prod_{i=1}^{r}\left( P_{(l_i,\pm l_i, 0),(0,n_i,0)}^{min,k_i}\right)+
    \non &&
     ~+\sum_{i<j}\eps_{v_i}(0,n_i,0)P_{(l_i,\pm l_i,
    0),(0,n_i,2)}^{min,k_i}\eps_{v_j}(0,n_j,0)P_{(l_j,\pm l_j,
    0),(0,n_j,2)}^{min,k_j}
     \prod_{q\neq
     i,j}\left(P_{(l_q,\pm l_q, 0),(0,n_q,0)}^{k_q}\right)+ \non
&&~+\sum_{i<j<k<l}~
     \prod_{p=i,j,k,l}
     \left(\eps_{v_p}(0,n_p,0)P_{(l_p,\pm l_p,0),(0,n_p,2)}^{min,k_p}\right)
     \prod_{n \neq i,j,k,l} \left(P_{(l_q,\pm l_q,
0),(0,n_q,0)}^{min,k_q}\right)
     +\ldots
     \non
&=& \prod_{i=1}^{r}\left( P_{(l_i,\pm l_i,
0),(0,n_i,0)}^{min,k_i}\right)
    \left( 1 + \sum_{i<j} \im^2
    (-1)^{ l_i+l_j} + \sum_{i<j<k<l} \im^4
    (-1)^{l_i+l_j+l_k+l_l}+\ldots\right)
\non &=&{\rm Re}\left[ \prod_{p=1}^{r}(1+\im (-1)^{l_p})\right]
    \prod_{i=1}^r \left(P_{(l_i,\pm l_i, 0),(0,n_i,0)}^{min,k_i}\right)~\non
    &=&2^{\frac{r}{2}}\cos (\xi_r) \prod_{i=1}^r \left(P_{(l_i,\pm l_i,
    0),(0,n_i,0)}^{min,k_i}\right)~,\non
   P^{r}_{[\vec{l};\vec{l'}; 0],[v][0;\vec{n};0]} &=&
    \eps^{\ast}_{v_1}(0,n_1,0) \left[\sum_{i}
    \eps_{v_i}(0,n_i,0)P^{\min,k_i}_{(l_i,\pm l_i,0),(0,n_i,2)}\prod_{q
\neq i}\left( P_{(l_q,\pm l_q, 0),(0,n_q,0)}^{min,k_q}\right)+
    \right. \non
&&~~~~~~~~~~~\left.+\sum_{i<j<k}~
     \prod_{p=i,j,k}
     \left(\eps_{v_p}(0,n_p,0)P_{(l_p,\pm l_p,0),(0,n_p,2)}^{min,k_p}\right)
     \prod_{n \neq i,j,k} \left(P_{(l_q,\pm l_q,
0),(0,n_q,0)}^{min,k_q}\right)
     +\ldots \right]
     \non
&=&-\im\prod_{i=1}^{r}\left( P_{(l_i,\pm l_i,
0),(0,n_i,0)}^{min,k_i}\right)
    \left( \sum_{i} \im
    (-1)^{l_i} + \sum_{i<j<k} \im^3
    (-1)^{l_i+l_j+l_k}+\ldots\right)
\non &=& ~{\rm Im}\left[ \prod_{p=1}^{r}(1+\im(-1)^{l_p})\right]
    \prod_{i=1}^r \left(P_{(l_i,\pm l_i, 0),(0,n_i,0)}^{min,k_i}\right)\non
    &=& 2^{\frac{r}{2}}\sin (\xi_r)\prod_{i=1}^r \left(P_{(l_i,\pm l_i,
    0),(0,n_i,0)}^{min,k_i}\right)~,
\end{eqnarray}
where we have defined the phase angle,
\bea
     \xi_r\equiv \frac{\pi}{4}\sum_{i=1}^r (-1)^{l_i},
\eea
and used $[v][0;\vec{n};0]$ to denote the orbit resulting from the action of the vector current on the Klein Bottle current.

\subsection{Models with all $k_i$ even}

It can be shown by induction on $r$ that the relevant $P$-matrix
entries of the fermion aligned product of $r$ even $k$ minimal
models are:

\begin{equation}
     \label{eq:Pevenn}
     P^{r}_{[\vec{l};\vec{l'};0],[0;\vec{n};0]}\equiv  \alpha_r 2^r
\cos\left(\theta_r-\frac{\pi r}{4}\right),\\
\end{equation}
\begin{equation}
    \label{eq:Pevenv}
     P^{r}_{[\vec{l};\vec{l'};0],[v][0;\vec{n};0]}\equiv - \alpha_r
2^r \sin\left(\theta_r-\frac{\pi r}{4}\right),\\
\end{equation}
where we have defined,
\begin{eqnarray*}
     \theta_r&\equiv&\sum_{i=1}^{r}\im^{n_i}\frac{\pi
     (l_i+1)}{2(k_i+2)},\\
     \alpha_r&\equiv& \prod_{i=1}^r \frac{e^\frac{\pi l_i
     n_i}{2(k_i+2)}}{k_i+2}.
\end{eqnarray*}
%$\chi$ is shorthand for the product of chiral fields of the
%individual minimal model fields. $\vec{n}$ and $\vec{v n}$ is
%shorthand for the klein bottle current and its superpartner.
The induction proof relies on that fact that different theories
are fermion aligned by an order two simple current, $w$, which is
the product of the vector currents of each block. First, by
expanding out the trigonometric expressions in the $P$-matrix
expressions for the case $r=1$ (\ref{eq:Pevenn}) and (\ref{eq:Pevenv}),
we find agreement with (\ref{eq:39}). Then consider the vector extension of a
fermion aligned product of $r$ minimal models (again given by
equation (\ref{eq:Pevenn})) and one additional minimal model (given in
the form of (\ref{eq:Pevenn}) with $r=1$)
\begin{eqnarray}
\label{eq:pn_induction}
P^{r+1}_{[\vec{l};l_{r+1};\vec{l'},l'_{r+1};0],
[0;\vec{n},n_{r+1};0]}&=&P^r_{[\vec{l};\vec{l'};0],[0;\vec{n};0]}P_{(l;l';0),(0;n_{r+1};0)}^{min,k_{r+1}}+
\non
&& ~~~~+ \eps_{\vec{v}}([0;\vec{n};0])
\eps_{v_{r+1}}((0;n_{r+1};0))P^r_{[\vec{l};\vec{l'};0],[v][0;\vec{n};0]}P_{(l;l';0),(0;n_{r+1};2)}^{min,k_{r+1}}\non
&=&\alpha_r 2^r \cos \left( \theta_r -\frac{\pi r}{4}\right) \alpha_1 2
\cos \left( \theta_1 -\frac{\pi}{4}\right) \non
&& ~~~~+ (\im)^2
\alpha_r 2^r \sin \left( \theta_r -\frac{\pi r}{4}\right) \alpha_1 2
\sin \left( \theta_1 -\frac{\pi}{4}\right)\non
&=& 2^{r+1} \alpha_r\alpha_1 \cos
\left(\theta_1+\theta_r-\frac{\pi(r+1)}{4}\right)\non
&=& 2^{r+1} \alpha_{r+1} \cos \left(\theta_{r+1}-\frac{\pi(r+1)}{4}\right),
\end{eqnarray}
in agreement with (\ref{eq:Pevenn}).
Similarly,
\begin{eqnarray}
\label{eq:pv_induction}
P^{r+1}_{[\vec{l};l_{r+1};\vec{l'},l'_{r+1};0],
[v][0;\vec{n},n_{r+1};0]}&=&P^r_{[\vec{l};\vec{l'};0],[v][0;\vec{n};0]}P_{(l;l';0),(0;n_{r+1};0)}^{min,k_{r+1}}+
\non
&& ~~~~+ \eps_{[v]}([v][0;\vec{n};0])
\eps_{v_{r+1}}((0;n_{r+1};0))P^r_{[\vec{l};\vec{l'};0],[v][0;\vec{n};0]}P_{(l;l';0),(0;n_{m+1};2)}^{min,k_{r+1}}\non
&=&-\alpha_r 2^r \sin \left( \theta_r -\frac{\pi r}{4}\right) \alpha_1 2
\cos \left( \theta_1 -\frac{\pi}{4}\right) -
\non
&& ~~~~(-\im)(\im)\alpha_r 2^r \sin \left( \theta_r -\frac{\pi r}{4}\right) \alpha_1 2
\cos \left( \theta_1 -\frac{\pi}{4}\right)\non
&=& -2^{r+1} \alpha_r\alpha_1 \sin
\left(\theta_1+\theta_r-\frac{\pi(r+1)}{4}\right)\non
&=& -2^{r+1} \alpha_{r+1} \sin \left(\theta_{r+1}-\frac{\pi(r+1)}{4}\right),
\end{eqnarray}
again in agreement with the second expression of the (\ref{eq:Pevenv}).

\subsection{Models with even and odd $k$}

For the generic case of $r=r_e+r_o$ even and odd $k$ minimal
models we just take the tensor of the even and odd cases that we have studied
and fermion align via the order 2 current $w=v_{even block}v_{odd
block}$. The calculation is similar to (\ref{eq:pn_induction}) and
(\ref{eq:pv_induction}), even to the effect that the $\eps$ phase
factors are also -1 and +1. The result is
\begin{eqnarray}
     \label{eq:Pmix}
     P^{r}_{[\vec{l};\vec{l'};0],[0;\vec{n};0]}&\equiv&  \alpha_{r_e}
2^r \cos\left(\theta_{r_e}-\frac{\pi
r_e}{4}-\xi_{r_o}\right)
\prod_{i=0}^{r_o}\frac{1}{\sqrt{2}}P_{(l_i;l'_i;0),((0;n_i;0)}^{min,k_i},
     \\
     P^{r}_{[\vec{l};\vec{l'};0],[v][0;\vec{n};0]}&\equiv& - \alpha_{r_e}
2^r \sin\left(\theta_{r_e}-\frac{\pi
r_e}{4}-\xi_{r_o}\right)\prod_{i=0}^{r_o} \frac{1}{\sqrt{2}} P_{(l_i;l'_i;0),((0;n_i;0)}^{min,k_i}.
     %\theta_r&\equiv&\sum_{i=1}^{r}\im^{n_i}\frac{\pi
     %(l_i+1)}{2(k_i+2)}\\
     %\alpha_r&\equiv& \prod_{i=1}^r \frac{e^\frac{\pi l_i
     %n_i}{2(k_i+2)}}{k_i+2}
\end{eqnarray}

\subsection{Spacetime extension}

Finally we extend with the spacetime WZW theory. The
non-zero entries of the $P$-matrix of $D_{8-n,1}$ are readily
computed (see for instance \cite{sagnotti}):
\begin{eqnarray}
   \label{eq:2}
   P^{D}_{0,0} &=& - P^{D}_{v,v}
   = \cos(\frac{n\pi}{4}) ~,\non
  P^{D}_{0,v} &=&
    -\sin(\frac{n\pi}{4}) ~,\non
  P^{D}_{s,s} &=& P^{D}_{c,c} = e^{\frac{\im
n\pi}{4}}\cos(\frac{n\pi}{4}) ~,\non
  P^{D}_{s,c} &=& -\im e^{\frac{\im n\pi}{4}}\sin(\frac{n\pi}{4})~.
\end{eqnarray}
However, the nature of the extension depends on whether we need the
WZW part of the chiral field to be in the vector or in the scalar
representation. In particular the scalar representation is necessary
to find the mass/charge of O-planes. 
In case the WZW component of the chiral field
is the vector primary, we find
\bea
 P^{ws}_{[[\vec{l};\vec{l'};0]v],[[0;\vec{n};0]o]}=
P_{[\vec{l};\vec{l'};0],[0;\vec{n};0]}^rP^D_{v,o}+\epsilon_v(o)\epsilon_{[v]}([0;\vec{n};0])
     P_{[\vec{l};\vec{l'};0],[v][0;\vec{n};0]}^r P_{v,v}^D ~.
\eea
  If it is built on the scalar, then we obtain
  \bea
 P^{ws}_{[[\vec{l};\vec{l'};0]o],[[0;\vec{n};0]o]}=
P_{[\vec{l};\vec{l'};0],[0;\vec{n};0]}^rP^D_{o,o}+\epsilon_v(o)\epsilon_{[v]}([0;\vec{n};0])
     P_{[\vec{l};\vec{l'};0],[v][0;\vec{n};0]}^r P_{o,v}^D ~.
\eea
In both cases the product of the $\eps$ phase factors equals
+1. Explicitly we thus obtain
\bea P^{ws}_{[[\vec{l};\vec{l'};0]v],[[0,\vec{n},0]o]}=e^{\im\pi\left( a_{[[\vec{l};\vec{l'};0]v]}+a_{[[0,\vec{n},0]o]}\right)} 2^{r_e}
\alpha_{r_e}
\sin\left(\theta_{r_e}-\frac{\pi}{4}(r_e+n)-\xi_{r_o}\right)
\prod_{i=1}^{r_o}\frac{1}{\sqrt{2}}P^{min,k_i}_{(l_i;l'_i;0),(o;n_i;0)},
\eea
and
\bea P^{ws}_{[[\vec{l};\vec{l'};0]o],[[0;\vec{n};0]o]}=e^{\im\pi\left( a_{[[\vec{l};\vec{l'};0]o]}+a_{[[0,\vec{n},0]o]}\right)} 2^{r}
\alpha_{r_e}
\cos\left(\theta_{r_e}-\frac{\pi}{4}(r_e+n)-\xi_{r_o}\right)
\prod_{i=1}^{r_o}\frac{1}{\sqrt{2}}P^{min,k_i}_{(l_i;l'_i;0),(o;n_i;0)}.
\eea
Here we have reintroduced the phase prefactor, equal to $\pm1$, which
encodes the difference in conformal weights of the true orbit field and that
of the representative element chosen for the calculation. As mentioned
earlier in the paper, the true weight of the orbit is the smallest of
the weights of the representative elements modulo 2.

\section{Mirror Symmetry Extension Map}
\label{sec:mirr-symm-extens}
\setcounter{equation}{0}

From the CFT perspective the difference between the A type theory and its
mirror is that one is based on a diagonal bulk modular
invariant while the other is based on the conjugation one. Simple
current technology allows us to build all but the most exotic modular
invariants. In particular we can find a set of fields which can
supplement the GSO and fermion aligning currents to produce the mirror
symmetric theory. The Mirror Symmetry extension currents are essentially
the conjugating fields available after GSO projection.

For a single minimal model conjugation is done through the phase
symmetry maps, $p_i=(0,2,0)_i$ of order $k_i+2$ and weight
$h_{p_i}=-1/(k_i+2)\mbox{ mod }1$.  As the conformal weight times the order is
integer we can use the group generated by this field to construct a
modular extension invariant. In the more general Gepner case, the
simple current group, $G_{ms}$, of Mirror Symmetry Extension currents 
is generated by some product of phase symmetry maps
$\Pi_{i=1}^r p_i^{\alpha_i}$. We will find constraints on the vector
 $\vec\alpha$ based on the requirement that
we have a proper simple current modular invariant and that the simple
current preserves the GSO projection of the initial (pre-mirror)
theory. 

Generally, to construct a simple current modular invariant we need to choose
a pairing $X$ \cite{thesis,walcher,simple,simple2}. This is a bihomomorphism,
$X:G\times G\rightarrow\RR$, which must satisfy, $X+X^T
(g,f)=Q_g(f)$ mod 1, for non-diagonal elements $g,f\in G$. On diagonal
elements one requires that 
$X(g,g)=-h_g$ mod 1. Different solutions for $X$ differ in half integer
values in the off-diagonal components.

In this case we opt for the simplest solution which will yield a
conjugate modular invariant. Define $X(p_i, p_j)=\delta_{ij}/(k_i+2)$. The simple
current modular invariant is
\beq
Z_G=\sum_{\lambda,\kappa} Z_{\lambda,\kappa} \chi_\kappa \bar{\chi}_\lambda,
\eeq
where $Z_{\lambda,\kappa}$ is an integer that counts the number of
solutions to
\bea
\lambda&=&J \kappa, ~~~J\in G \label{eq:mic_1},\non
Q_g(\kappa)+X(g,J)&=&0, ~~~~\forall g\in G. \label{eq:mic_2}
\eea
Now let $J=\Pi_{i=1}^r p_i^{\alpha_i}$, $g=\Pi_{i=1}^r p_i^{\beta_i}$
and $\kappa=[\vec{l};\vec{m};\vec{s}]f_D$, with $\vec{s}, f_D\in$ NS,
then the second condition reads,
\beq
\sum_{i=1}^r \frac{\beta_i m_i}{h_i}+\frac{\beta_i \alpha_i }{h_i}=0
~~~\forall \vec{\beta},
\eeq
which can be solved by $\alpha_i=-m_i$. Applying this to
the Eq. (\ref{eq:mic_1}) then tells us that the field $\kappa$ with phases
$\vec{m}$ is paired up with a field $\lambda$ whose phases are
$J [\vec{l};\vec{m};\vec{s}]f_D=[\vec{l};\vec{m}-2\vec{\alpha};\vec{s}]f_D=[\vec{l};-\vec{m};\vec{s}]f_D$
as desired for a conjugate field in the NS sector.

This result is not surprising as orbifolding/simple current extending the
minimal model by the phase symmetry $p$ yields its conjugate.
However, in order to construct a spacetime supersymmetric
theory we need to GSO project. We must make sure that the currents we are
extending by are admissible in the GSO projected theory. The GSO
projection drops all fields with non-integer monodromy charge under the
spectral flow, $s$, so we require that admissible mirror extending currents satisfy
the compatibility condition $Q_s(J)=0$, or, explicitly,
\beq
\sum_{i=1}^r \frac{\alpha_i}{h_i}=0.
\eeq
In the orbifold language this is equivalent to requiring that the
twisted sectors (in the Gepner model) are uncharged with respect to
the GSO map.

For the NS case, no further conjugation is
necessary. In the Ramond sector we need to conjugate the values of
$s_i$ as well as the spacetime spinor component. 
This conjugation depends on the parity of $n$ and $r$ as
this determines whether the sector
is self conjugate or not. The spacetime factor $D_{8-n,1}$ is self
conjugate for $n$ even. For $n$ odd we can extend by $v_D$ to conjugate.
Because of fermion alignment the action of $v_D$ is equivalent to the
use of $v_1$. 
Overall this yields a factor of the form $v_1^n$. In the Gepner
sector, we need to flip the individual $s_i$ values from 1 to -1 and
vice versa. If $r$ is even then these two states are equivalent modulo fermion
alignment - so the Gepner sector is self conjugate.
If the Gepner sector is not self conjugate, then conjugation can be
achieved through $v_1$. Overall we see that the conjugation of the
internal sector requires an extensions by $v_1^r$. Combining the results
for the internal Gepner and the spacetime fermion sector we see that we
we need to extend by $v_D^n v_1^r$ which is equal to  $v_1^{n+r}$ modulo fermion alignment.

We conclude that the Mirror Symmetry extension map is of the form
\bea
J=v_1^{(n+r)\epsilon}\prod_{i=1}^r p_i^{\alpha_i},
\eea
where $\epsilon$ is either 0 or 1 and distinguishes between the NS and R
sectors.
However, as before, we must make sure that these currents are compatible with the
GSO projection. In particular we need,
\bea
Q_s(J)&=& 0\mbox{ mod }2\non
&=&\sum \alpha_i/h_i +(n+r)\epsilon /2,
\eea
so that the Mirror Symmetry extension currents are neutral with respect to the GSO current.

\subsection{Summary}

The Mirror Map is generated by elements of the form
\bea
G^{Mirror}=\{v_D^{(n+r)\epsilon} \Pi_{r=1}^r p_i^{\alpha_i}, \sum_{i=1}^{r}
\frac{\alpha_i}{h_i}-\frac{(n+r)\epsilon}{2}=0\mbox{ mod }1,\,\,\,\,
\epsilon\in\{0,1\}\},
\eea
with pairing
\bea
X(p_i,p_j)=\frac{\delta_{ij}}{k_i+2}.
\eea

\subsection{Sample Mirror Symmetry Extension Calculation}
\label{sec:sample-mirr-symm}

To illustrate the mirror symmetry extension and the calculation of the
sign constraints we will compute the
B-type crosscap coefficients for the Gepner (1,1,1) model $K=(-2,2,2)$.
The mirror symmetry is generated by
the order 3 current $J=p_1 p_2^3$ with conformal weight $h_J=1/3$. The
action of $J$ on the Klein bottle current is:
\begin{center}
\begin{tabular}{ccccccc}
   J:(-2,2,2) & $\longrightarrow$ & (0,0,2) & $\longrightarrow$ &
(2,-2,2) & $\longrightarrow$ & (-2,2,2) \\
   $h_{K}=1$ &  & $h_{JK}=2/3$ &  & $h_{J^2K}=1$ &  &  \\
   $\beta_0=\sigma_0$& &$\beta_J=\sigma (JK) e^{\frac{\im \pi}{3}}$&
&$\beta_{J^2}=\sigma (J^2K) e^{\frac{\im \pi}{3}}$&\\
  \end{tabular}
\end{center}

There are three constraint equations:
\begin{eqnarray}
\beta_0&=&\sigma_0=\beta_J\beta_{J^2}e^{\im 2\pi X(J,J^2)}=-\sigma
(JK)\sigma(J^2K),\\
\beta_J&=&\sigma(JK) e^{\frac{\im \pi}{3}} =\beta_{J^2}
\beta_{J^2}e^{\im 2\pi X(J^2,J^2)}=-\sigma (JK)^2=-1,\\
\beta_J^2&=&\sigma(J^2K) e^{\frac{\im \pi}{3}} =\beta_{J}
\beta_{J}e^{\im 2\pi X(J,J)}=\sigma (J^2K)^2=1,
\end{eqnarray}
where we have used that
\begin{eqnarray}
\beta_{J^{n+m}}&=&\beta_{J^n}\beta_{J^m}e^{\im 2\pi X(Jn2,J^m)},\\
e^{\im 2\pi X(Jn2,J^m)}&=&e^{-\im 2\pi n m h_J}.
\end{eqnarray}
Hence the signs are:
\begin{eqnarray}
\sigma_0&=& 1,\\
\sigma(JK) &=& -1,\\
\sigma(J^2K) &=& 1.
\end{eqnarray}
Note that $JK=(0,0,2)$ is equivalent to $(2,2,-2)$ under the
$S^2=(2,2,2)$ GSO identification. We would then expect
$\sigma(J,K)$ to equal $\sigma(0,K)$ as $JK$ has the same minimal model
components as $K$ modulo permutation. This
discrepancy is compensated by a sign difference between
$P^{ws}_{(0,0,2),\chi}$ and $P^{ws}_{(2,2,-2),\chi}$.
The difference in sign in the P-matrix is in turn given by the choice of
$(0,0,2)$ versus $(2,2,-2)$ as
representative elements of the $S$ orbit of $K$. Recall that the
prescription is to choose the element with lowest
conformal weight mod 1. Had we had done that, it is clear that the
computation of the signs would have been trivial,
as for this Klein bottle, $J$ maps identical Klein bottle currents $K$
(modulo minimal model
permutation) to themselves. As a result the
mirror map results in an overall scaling by $\sqrt{3}$.

\providecommand{\href}[2]{#2}\begingroup\raggedright

\endgroup
\end{document}